\documentclass[titlepage,a4paper,12pt]{amsart}

\usepackage[latin1]{inputenc}
\usepackage[T1]{fontenc}

\usepackage{amsthm}

\usepackage{indentfirst}
\usepackage[final]{graphicx}
\usepackage{hyperref}
\hypersetup{
    colorlinks=true,       
    linkcolor=blue,        
    citecolor=blue,       
    urlcolor=blue,         
    pdfborder={0 0 0}      
}

\usepackage{epstopdf}

\usepackage{maple2e}
 
\DefineParaStyle{Maple Output} \DefineParaStyle{Maple Output}
\DefineParaStyle{Normal} \DefineCharStyle{2D Math}
\DefineCharStyle{2D Output}

\newcommand{\abs}[1]{\left\vert#1\right\vert}
\newcommand{\set}[1]{\left\{#1\right\}}
\newcommand{\norm}[1]{\left\Vert#1\right\Vert}

\newcommand{\pd}[2]{\frac{\partial#1}{\partial#2}}
\newcommand{\od}[2]{\frac{d#1}{d#2}}
\newcommand{\eps}{\varepsilon}
\renewcommand{\phi}{\varphi}
\newcommand{\diag}[1]{\mathop{\mathrm{diag}}\left\{#1\right\}}
\newcommand{\grad}{\mathop{\mathrm{grad}\,}}

\newcommand{\fft}{\mathop{\mathrm{fft}\,}}
\newcommand{\ifft}{\mathop{\mathrm{ifft}\,}}

\renewcommand{\div}{\mathop{\mathrm{div}}}
\newcommand{\supp}{\mathop{\mathrm{supp}}}
\newcommand{\sign}{\mathop{\mathrm{sign}}}

\renewcommand{\Re}{\mathop{\mathrm{Re}}}
\renewcommand{\Im}{\mathop{\mathrm{Im}}}

\newtheorem{thm}{Theorem}[section]
\newtheorem{defn}[thm]{Definition}
\newtheorem{lemm}{Lemma}[section]

\title{Moving load on a floating ice layer}
\author{Denys Dutykh}

\date{June 28, 2005}
\email{Denys.Dutykh@crans.org}
\address{\'{E}cole Normale sup\'{e}rieure de Cachan\\
61, avenue du Pr\'{e}sident Wilson\\
Pavillon des Jardins, ch. 122\\
94235 Cachan Cedex France}

\begin{document}
\begin{titlepage}
\sc \pagestyle{empty} \frenchspacing

\vspace{8cm}

\begin{center}\Large
MOVING LOAD ON A FLOATING ICE LAYER
\end{center}

\vspace{1cm}

\begin{center}
DENYS DUTYKH
\end{center}

\vspace{3cm}

\begin{center}
Director: FREDERIC DIAS
\end{center}

\vspace{\stretch{1}}
\begin{center}
  July 6, 2005
\end{center}
\end{titlepage}

\tableofcontents \listoffigures

\section{Introduction}

At the beginning of the forties of the last century, a systematic study of displacements and stresses in the multilayer medium was begun because of the need for practice.

The multilayer model represents the properties of numerous real objects very well, such as roadway coverings, earth foundations, layered floors of industrial buildings, etc. An evident advantage of this model is that it is able to represent the stepwise transition of the mechanical properties of real bases. Moreover, the multilayer medium can be used to approximate the bases with continuous mechanical properties depending on the depth. These features explain the researchers' interest in the theory of layered medium, which has a number of applications in engineering.

By 1974, the number of articles devoted only to static elastic problems for layered medium achieved almost four thousand. See \cite{1,2,3}, and the references are given there. Actually, the number of papers in this field of the theory of elasticity appreciably increased.

Designing high-rise buildings, barrages, dams, mines, road coverings and others is connected with solving different boundary value problems of the theory of elasticity for layered bases. Many of these problems are new and demand the application of new methods. Let us analyze the main directions in which the static problems of multilayer medium have been developed. We restrict our attention only to works where the elastic bases with an arbitrary number of layers were considered (substantively multilayer medium). As well, we do not consider the papers devoted to purely numerical methods such as the finite element method.

We should mention here that there are several approximate theories based on additional hypotheses about the character of deformations. These theories have been developed in the works of V.V. Bolotin, N.N. Leontyev, Yu.N. Novichkov, V.V. Partsevsky, B.E. Pobedrya, R.M. Rapoport and others \cite{4,5,6,7,8,9,10,11}.

The majority of studies on the static problems for an elastic layered medium are based on the following assumptions:
\begin{itemize}
    \item each layer is solid-core (does not contain any cavities,
    cracks)
    \item during the deformation, the layers do not detach
    \item the deformation is due only to surface loads (we neglect
    the mass forces)
    \item the mechanical properties are constant in each layer
\end{itemize}
Henceforward, we name such layered bases the medium of simple structure. This term was proposed in \cite{12,13}.

I.G. Alperin published the first article devoted to substantively multilayer medium \cite{14} in 1938. In this work, the author considered the plane deformation of layered elastic half-space. Using the integral Fourier transform, the author expresses the general solution of elasticity equations in each layer through four arbitrary functions of transform parameters. Then, the recurrence relations between these unknown functions follow from deformations' compatibility relations. Finally, the problem is reduced to the determination of these four functions on the upper layer. They are determined from boundary conditions and from the condition of stress attenuation at infinity.

The method proposed by I.G. Alperin belongs to the family of recurrence relations methods. For methods of this group, it is characteristic that the problem is reduced to the determination of auxiliary functions only for the upper layer. Recurrence relations determine the solution for other layers. At the same time, the order of the system of linear algebraic equations relative auxiliary functions (in the case of the first or the second basic boundary value problems) does not depend on the number of layers in the base. The rapid development of these methods was begun in the sixties. See \cite{15,16,17,18,19,20} for more details.

Actually, many modifications of the recurrence relations method were proposed. They differ from one another by the choice of auxiliary functions.

In 1942, G.S.~Shapiro \cite{21} proposed another analytical method of an axisymmetric boundary value problem of the theory of elasticity for multilayer bases. He used integral Hankel transform \cite{22}. Later, D.M.~Burmister \cite{23} published analogical results with an explanation of how to use the obtained solution in order to calculate the stress state of aerodrome coverings. The method described in \cite{21,23} consists of obtaining a system of algebraic linear equations with $4n$ unknown functions in Hankel transforms space, where $n$ means the number of layers. These equations are obtained using boundary conditions and deformation compatibility relations. The matrix of this linear system depends on the Hankel transform parameter. Let us mention that if we use this method to solve three-dimensional problems, we will have $6n$ unknown functions. Later this method was developing in the works of L.G. Bulavko, B.I. Kogan, M.B. Koreunskiy, I.A. Rodzevich, V.S. Nikitin and G.S. Shapiro, H. Bufler, M.~Fremond and others. A detailed review of the research in this direction is given in \cite{24}.

The obvious disadvantage of the method proposed by G.S. Shapiro and later by D.M. Burmeister is that it is relatively expansive from a computational point of view because, at each step of inverse transformation, we have to store and solve a large system of linear algebraic equations. As well, there is always a problem of singularities for little or big values of transform parameters. All these particularities are described in detail in \cite{30}.

Obviously, the method of recurrence relations is more convenient for solving elastic problems in a simple structure medium. His advantage is that we do not need to solve linear systems of high order. In addition, an important simplification can be achieved if we use the Lamzyuk-Privarnikov functions method \cite{13,25,26,27,28,29}. These functions depend on the integral transform parameter and geometrical and mechanical characteristics of layers and do not depend on the loads applied on the base. That is why Lamzyuk-Privarnikov functions can be determined before solving a boundary-value problem. To determine these functions, simple recurrence relations were constructed \cite{13,29}. The integral transforms of stresses and displacements in any layer can be obtained from special recurrence relations containing Lamzyuk-Privarnikov functions. Analytical solution of principal boundary-value problems for simple structure medium is given by Fourier or Hankel type integrals.

From a computational point of view, the method of recurrence relations coupled with the Lamzyuk-Privarnikov functions method is preferable. This approach is very economical in memory use, and it is linear in complexity when we compute transforms of stresses and displacements. Then, to make inverse transforms, fast routines can be applied. In the case of DFFT, the total complexity is $O(n\log n)$. There is a disadvantage to this method. In fact, before performing an accurate computation, one has to eliminate the singularities for some values of transform parameters. Those values depend on the concrete variant of the method and, of course, on the problem. A detailed description of the numerical realization of this method is given in \cite{13,26,27}.

Lamzyuk-Privarnikov functions method was generalized for the case of layered in radial direction wedge \cite{31,32}, layered cylinder \cite{33} as well for the problems of thermoelasticity \cite{34}. Besides static problems, this method has been used to solve almost dynamic problems of elasticity and viscoelasticity, in particular, the problem about harmonic oscillations of multilayer base \cite{35}, about the action of moving load on a layered medium \cite{36,37}. We should mention here that the Lamzyuk-Privarnikov functions method can still be used in the presence of cracks, cavities or inclusions \cite{12,38}. As well this method was used to solve contact problems of elasticity and viscoelasticity \cite{13,38}.

Recently, the Lamzyuk-Privarnikov functions method was extended for the case of continuous dependence of mechanical characteristics upon the depth of the layer \cite{39,40}. After proceeding to the limit, recurrence relations become a system of differential equations. Therefore, principal boundary value problems of the theory of elasticity for the layered medium are reduced to the consequent solving of two Cauchy-type problems in transform space.

A major motivation for the study of a moving load on a flexible beam or plate has been its application to transport systems (rail tracks, roads or runways), originally in temperate lands and subsequently in cold regions, where, in particular, floating ice sheets may be exploited.

Wave propagation in a water-ice system attracted considerable attention in the literature both as a theoretical subject as well as due to its importance to ice engineering. From the theoretical point of view, models of a water-ice system allow for the convenient application of mathematical tools for studying wave phenomena under realistic physical conditions. At the same time, from the engineering standpoint, there are questions of great practical significance in which wave propagation is of importance. These include stress control of the ice cover in the neighbourhoods of facilities built upon the ice, the performance of ice-breaking ships, damage of offshore constructions by floating ice sheets, the spontaneous appearance of large-scale cracks in the ice cover of water basins, etc.

There exists a significant body of literature on linear as well as non-linear wave propagation in a water-ice system. One of the first theoretical studies of the response of floating ice to moving loads was given in \cite{green}. More recently, a number of authors: Kheisin (1963, 1971) \cite{kheisin1,kheisin2}, Nevel (1970) \cite{nevel}, Marchenko (1988) \cite{mar1}, Marchenko and Semenov (1994) \cite{mar2}, Milinazzo (1995, 2004) \cite{mil2,mil1} and others have studied the effects of a moving load on floating ice. In these investigations, the ice sheet is treated as a thin plate of infinite extent supported below by water of uniform, finite depth, and the load is assumed to move at a constant speed. Kheisin in \cite{kheisin1} examines the steady motion of a point and a line load.

We cannot omit the recent monograph by Squire et al. \cite{squire} that is devoted to the rich topic of moving loads on ice plates.

Recently, a stability analysis of a water-ice system, in which ice was modelled as an elastic homogeneous layer of finite thickness, was carried out by Brevdo and Il'ichev \cite{bred}. In that work, it was shown that any 2-D homogeneous floating ice layer of infinite horizontal extension is exponentially stable.

There is also an article \cite{dias} by E. Parau and F. Dias in which they modelled the ice as a Kirchhoff-Love plate, and the water was assumed to be an ideal incompressible fluid. Their study was restricted to the effects of nonlinearity when the load speed is close to the minimum phase speed. A weakly nonlinear analysis, based on dynamical systems theory and on normal forms, was performed.

The thin plate assumption for the ice, while producing a considerable simplification in the model, causes, at the same time, the exclusion of the effects related to the elastic deformations present inside the ice sheet; within this assumption, only flexural deformations are accounted for.

The compressibility of water has to be taken into account for consistency reasons. In fact, the compressibility of water cannot be neglected in treating the wave dynamics in the model because the infinite medium dilatational and equivoluminal wave velocities in ice are comparable with the speed of the acoustic waves in water.

In the present work, we consider the ice to be an elastic plate of finite thickness. We take into account the compressibility of the water and the moving load as a rigid block.

\section{Formulation of the problem}

\begin{figure}[htbp]
\begin{center}
\unitlength 1mm
\begin{picture}(90.50,40.00)(0,0)

\linethickness{0.40mm}
\put(4.00,29.00){\line(1,0){69.50}}

\linethickness{0.40mm}
\put(4.00,21.00){\line(1,0){69.50}}

\linethickness{0.20mm}
\put(73.50,21.00){\line(1,0){10.50}}
\put(84.00,21.00){\vector(1,0){0.12}}

\linethickness{0.15mm}
\put(41.00,21.00){\line(0,1){21.00}}
\put(41.00,42.00){\vector(0,1){0.12}}

\linethickness{0.40mm}
\put(5.00,5.00){\line(1,0){69.00}}

\linethickness{0.20mm}
\multiput(5.00,1.50)(0.16,0.12){29}{\line(1,0){0.16}}

\linethickness{0.20mm}
\multiput(11.00,1.50)(0.16,0.12){29}{\line(1,0){0.16}}

\linethickness{0.20mm}
\multiput(8.00,1.50)(0.16,0.12){29}{\line(1,0){0.16}}

\linethickness{0.20mm}
\multiput(14.00,1.50)(0.16,0.12){29}{\line(1,0){0.16}}

\linethickness{0.20mm}
\multiput(20.00,1.50)(0.16,0.12){29}{\line(1,0){0.16}}

\linethickness{0.20mm}
\multiput(17.00,1.50)(0.16,0.12){29}{\line(1,0){0.16}}

\linethickness{0.20mm}
\multiput(23.00,1.50)(0.16,0.12){29}{\line(1,0){0.16}}

\linethickness{0.20mm}
\multiput(29.00,1.50)(0.16,0.12){29}{\line(1,0){0.16}}

\linethickness{0.20mm}
\multiput(26.00,1.50)(0.16,0.12){29}{\line(1,0){0.16}}

\linethickness{0.20mm}
\multiput(32.00,1.50)(0.16,0.12){29}{\line(1,0){0.16}}

\linethickness{0.20mm}
\multiput(38.00,1.50)(0.16,0.12){29}{\line(1,0){0.16}}

\linethickness{0.20mm}
\multiput(35.00,1.50)(0.16,0.12){29}{\line(1,0){0.16}}

\linethickness{0.20mm}
\multiput(41.00,1.50)(0.16,0.12){29}{\line(1,0){0.16}}

\linethickness{0.20mm}
\multiput(47.00,1.50)(0.16,0.12){29}{\line(1,0){0.16}}

\linethickness{0.20mm}
\multiput(44.00,1.50)(0.16,0.12){29}{\line(1,0){0.16}}

\linethickness{0.20mm}
\multiput(50.00,1.50)(0.16,0.12){29}{\line(1,0){0.16}}

\linethickness{0.20mm}
\multiput(56.00,1.50)(0.16,0.12){29}{\line(1,0){0.16}}

\linethickness{0.20mm}
\multiput(53.00,1.50)(0.16,0.12){29}{\line(1,0){0.16}}

\linethickness{0.20mm}
\multiput(59.00,1.50)(0.16,0.12){29}{\line(1,0){0.16}}

\linethickness{0.20mm}
\multiput(65.00,1.50)(0.16,0.12){29}{\line(1,0){0.16}}

\linethickness{0.20mm}
\multiput(62.00,1.50)(0.16,0.12){29}{\line(1,0){0.16}}

\linethickness{0.20mm}
\multiput(68.00,1.50)(0.16,0.12){29}{\line(1,0){0.16}}

\linethickness{0.40mm}
\put(33.50,29.50){\line(1,0){16.00}}
\put(33.50,29.50){\line(0,1){8.00}}
\put(49.50,29.50){\line(0,1){8.00}}
\put(33.50,37.50){\line(1,0){16.00}}

\put(81.00,17.00){\makebox(0,0)[cc]{$x$}}

\put(123.50,64.00){\makebox(0,0)[cc]{}}

\put(39.00,42.00){\makebox(0,0)[cc]{$y$}}

\linethickness{0.20mm}
\put(49.50,33.00){\line(1,0){8.00}}
\put(57.50,33.00){\vector(1,0){0.12}}

\put(55.00,36.00){\makebox(0,0)[cc]{$c_0$}}

\put(56.50,6.00){\makebox(0,0)[cc]{}}

\linethickness{0.20mm}
\put(61.00,5.00){\line(0,1){16.00}}
\put(61.00,21.00){\vector(0,1){0.12}}
\put(61.00,5.00){\vector(0,-1){0.12}}

\linethickness{0.20mm}
\put(61.00,21.00){\line(0,1){8.00}}
\put(61.00,29.00){\vector(0,1){0.12}}
\put(61.00,21.00){\vector(0,-1){0.12}}

\put(64.00,14.00){\makebox(0,0)[cc]{$H$}}

\put(62.50,14.00){\makebox(0,0)[cc]{}}

\put(64.00,25.00){\makebox(0,0)[cc]{$h$}}

\linethickness{0.15mm}
\put(11.00,20.00){\line(1,0){4.00}}

\linethickness{0.15mm}
\put(31.00,15.00){\line(1,0){4.00}}

\linethickness{0.15mm}
\put(46.00,15.00){\line(1,0){4.00}}

\linethickness{0.15mm}
\put(16.00,11.50){\line(1,0){4.00}}

\linethickness{0.15mm}
\put(21.50,17.50){\line(1,0){4.00}}

\linethickness{0.15mm}
\put(5.50,8.50){\line(1,0){4.00}}

\linethickness{0.15mm}
\put(5.50,17.50){\line(1,0){4.00}}

\linethickness{0.15mm}
\put(25.50,7.50){\line(1,0){4.00}}

\linethickness{0.15mm}
\put(26.50,13.50){\line(1,0){4.00}}

\linethickness{0.15mm}
\put(34.00,9.00){\line(1,0){4.00}}

\linethickness{0.15mm}
\put(37.00,13.00){\line(1,0){4.00}}

\linethickness{0.15mm}
\put(15.00,8.00){\line(1,0){4.00}}

\linethickness{0.15mm}
\put(43.50,7.50){\line(1,0){4.00}}

\linethickness{0.15mm}
\put(52.00,17.50){\line(1,0){4.00}}

\linethickness{0.15mm}
\put(52.00,8.00){\line(1,0){4.00}}

\linethickness{0.15mm}
\put(63.50,8.50){\line(1,0){4.00}}

\linethickness{0.15mm}
\put(66.50,17.50){\line(1,0){4.00}}

\linethickness{0.15mm}
\put(68.00,14.00){\line(1,0){4.00}}

\linethickness{0.15mm}
\put(69.00,7.50){\line(1,0){4.00}}

\linethickness{0.15mm}
\put(38.50,15.00){\line(1,0){4.00}}

\linethickness{0.15mm}
\put(44.50,10.00){\line(1,0){4.00}}

\put(19.00,25.00){\makebox(0,0)[cc]{$\nu(x,y), E(x,y), \rho(x,y)$}}

\linethickness{0.15mm}
\put(33.00,20.00){\line(0,1){2.00}}

\linethickness{0.15mm}
\put(49.00,20.00){\line(0,1){2.00}}

\put(31.00,18.00){\makebox(0,0)[cc]{$-a$}}

\put(49.00,18.00){\makebox(0,0)[cc]{$a$}}

\put(41.00,18.00){\makebox(0,0)[cc]{$o$}}

\put(40.00,17.00){\makebox(0,0)[cc]{}}

\put(80.00,5.00){\makebox(0,0)[cc]{$\Gamma_1$}}

\put(75.00,5.00){\makebox(0,0)[cc]{}}

\put(80.00,25.00){\makebox(0,0)[cc]{$\Gamma_2$}}

\put(80.00,30.00){\makebox(0,0)[cc]{$\Gamma_3$}}

\put(10.00,35.00){\makebox(0,0)[cc]{$T=T_1$}}

\put(10.00,15.00){\makebox(0,0)[cc]{$T=T_0$}}

\linethickness{0.15mm}
\put(51.00,10.00){\line(1,0){4.00}}

\linethickness{0.15mm}
\put(5.00,10.00){\line(1,0){4.00}}

\linethickness{0.15mm}
\put(21.00,15.00){\line(1,0){4.00}}

\linethickness{0.15mm}
\put(55.00,15.00){\line(1,0){4.00}}

\end{picture}
\end{center}
  \caption{Moving load on an ice layer}\label{pict}
\end{figure}
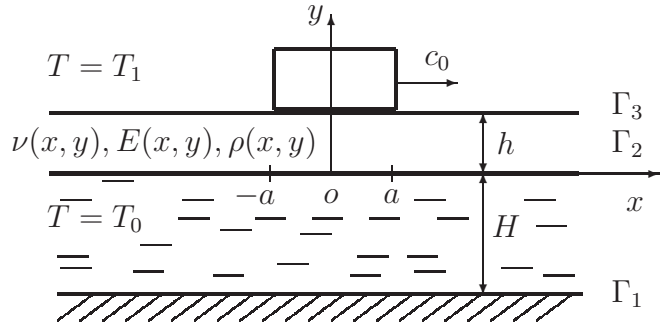

In the present work, we consider the problem of ice layer deformation under a moving block (see Figure \ref{pict}). We make the following assumptions. First of all, we neglect the gravitational forces. In other words, there are no exterior mass forces in this model\footnote{The validity of this assumption is discussed below.}. The problem is two-dimensional. The horizontal axis is denoted by $x$, while the vertical axis is denoted by $y$. The $x$-axis is along the interface between ice and water. The ice thickness is $h=const$. We suppose that the ice layer lies on compressible water of constant depth $H$. The water layer lies on a rigid half-space. In general, the mechanical properties (density, Poisson ratio, Young's modulus) of the ice layer are functions of $(x,y)\in\mathbb{R}\times[0,h]$. In the next section, we will discuss this point. A rigid block moves along the $x$-axis with constant velocity $c_0$. In this work, we take into account the friction between the block and the ice. The block load is modelled by a pressure patch applied to the ice upper boundary:
\begin{equation*}
  L(x,t) = \left\{%
\begin{array}{ll}
    l(x), & |x-c_0t|<a \\
    0, & \hbox{else} \\
\end{array}
\right.
\end{equation*}
A similar expression can be written for the friction between the ice layer and the block:
\begin{equation*}
  G(x,t) = \left\{%
\begin{array}{ll}
    g(x), & |x-c_0t|<a \\
    0, & \hbox{else} \\
\end{array}
\right.
\end{equation*}

Also, we suppose that the water temperature and the air temperature are $T_0$ and $T_1 = \mathrm{const}$, respectively. It is important to take into account heat transfer because the mechanical properties of ice depend on its temperature.

\section{Ice mechanical properties}

In this section, we will briefly review only a few aspects of this very complicated subject. The best general reference here is \cite{squire}.

Experiments show that Young's modulus depends on the fractional brine volume\footnote{To our knowledge ``fractional brine volume'' means the volume fraction of brine in the ice.} $\nu_b$:
\begin{equation*}
  \boxed{E=10-3.5\nu_b}
\end{equation*}
In this empirical formula, $E$ is expressed in GPa.

Next, the fractional brine volume depends on the water salinity and temperature through the following formula:
\begin{equation*}
\nu_b=\left\{%
\begin{array}{ll}
    \frac{S}{1000}\left(\frac{52.56}{|T|}-2.28\right), & -2.06^\circ C<T<-0.5^\circ C\\
    \frac{S}{1000}\left(\frac{45.917}{|T|}+0.93\right), & -8.2^\circ C<T<-2.06^\circ C\\
    \frac{S}{1000}\left(\frac{43.795}{|T|}+1.189\right), & -22.9^\circ C<T<-8.2^\circ C\\
\end{array}%
\right.
\end{equation*}
where $S$ is the ice salinity in $^\circ/_{\circ\circ}$.

\begin{figure}
  \includegraphics[width=1.0\linewidth]{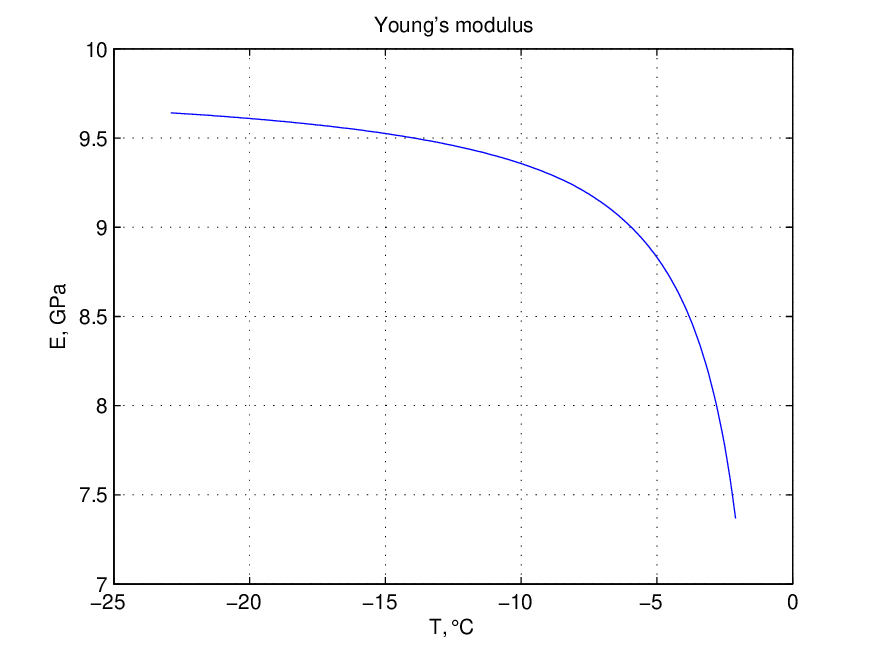}\\
  \caption{Young's modulus dependence on the ice temperature}\label{nub}
\end{figure}

The Young's modulus dependence on the ice temperature is shown in Figure \ref{nub}.

As for Poisson's ratio, experiments show that this parameter does not vary much in the ice with a typical value
\begin{equation*}
\boxed{\nu=0.33\pm0.03}
\end{equation*}
Anyhow, we will assume this parameter to be a function of $x$ and $y$ since it does not change our solution technique.

The density of bubble-free ice is also a function of temperature. Experiments show that the density $\rho$ is almost a linear function of temperature. Using this information and the fact that at $T=0^\circ C$ $\rho=916.5 kg/m^3$ and at $T=-30^\circ C$ $\rho=920.7 kg/m^3$, we can reconstruct this functional dependence:
\begin{equation*}
  \boxed{\rho(T)=-0.14T+916.5}
\end{equation*}
where $T$ in $^\circ C$, $\rho$ in $kg/m^3$.

\section{Lam\'{e}'s equations for an inhomogeneous medium} 

In this section, we review the well-known Lam\'{e}'s equations when the mechanical properties of the material are functions of space. The best general reference here is \cite{tim}.

First of all, we make some remarks on the notation. Let us denote by
$$
  \begin{pmatrix}
    \sigma_x & \tau_{xy} \\
    \tau_{yx} & \sigma_y \\
  \end{pmatrix}
$$
the stress tensor components and by
$$
  \begin{pmatrix}
    \eps_x & \gamma_{xy} \\
    \gamma_{yx} & \eps_y \\
  \end{pmatrix}
$$
the deformation tensor components.

The dynamic equations are
\begin{equation}\label{stat}
\left\{%
\begin{array}{ll}
  \pd{\sigma_x}{x}+\pd{\tau_{xy}}{y}+\rho X= & \rho\pd{^2u}{t^2},\\
  \pd{\tau_{yx}}{x}+\pd{\sigma_y}{y}+\rho Y= & \rho\pd{^2v}{t^2},\\
\end{array}%
\right.
\end{equation}
where $(u,v)$ is the displacement field and $(X,Y)$ is the volume force field.

The kinematic equations are
\begin{equation}\label{geom}
 \left\{%
 \begin{array}{lll}
   \eps_x=&\pd{u}{x},\qquad \gamma_{xy}=&\pd{u}{y}+\pd{v}{x},\\
   \eps_y=&\pd{v}{y},\qquad \gamma_{yx}=&\pd{v}{x}+\pd{u}{y}\\
 \end{array}%
 \right.
\end{equation}
and the physical equations are
\begin{equation}\label{hook}
 \left\{%
 \begin{array}{ll}
   \sigma_x=&2\mu\eps_x+\lambda\theta, \\
   \tau_{xy}=&\mu\gamma_{xy}, \\
   \sigma_y=&2\mu\eps_y+\lambda\theta
 \end{array}%
 \right.
\end{equation}
where $\theta=\eps_x+\eps_y$. Using (\ref{hook}), we obtain the following derivatives of the stress tensor components:
\begin{equation}\label{str2}
 \begin{array}{ll}
   \pd{\sigma_x}{x} =& 2\mu\pd{^2u}{x^2} + \lambda\left(
   \pd{^2u}{x^2}+\pd{^2v}{x\partial y}\right)+ 2\pd{\mu}{x}\pd{u}{x}
   +\pd{\lambda}{x}\left(\pd{u}{x}+\pd{v}{y}\right) \\
   \pd{\tau_{xy}}{y} =& \mu\left(\pd{^2u}{y^2}+\pd{^2v}{x\partial
   y}\right) + \pd{\mu}{y}\left(\pd{u}{y} + \pd{v}{x}\right) \\
   \pd{\tau_{yx}}{x} =& \mu\left(\pd{^2u}{x\partial y} +
   \pd{^2v}{x^2}\right) + \pd{\mu}{x}\left(\pd{u}{y} +
   \pd{v}{x}\right)\\
   \pd{\sigma_y}{y} =& 2\mu\pd{^2v}{y^2} + \lambda\left(\pd{^2u}{x\partial
   y} + \pd{^2v}{y^2}\right) + 2\pd{\mu}{y}\pd{v}{y} +
   \pd{\lambda}{y}\left(\pd{u}{x} + \pd{v}{y}\right)
 \end{array}
\end{equation}
Now we substitute equalities (\ref{str2}) into (\ref{stat}) and obtain Lam\'{e}'s equations:
\begin{equation}\label{lame}
 \begin{array}{rl}
   (\lambda+2\mu)\pd{^2u}{x^2} + \mu\pd{^2u}{y^2} + (\lambda+\mu)
   \pd{^2v}{x\partial y} +& \\ +\left(\pd{\lambda}{x}+2\pd{\mu}{x}\right)
   \pd{u}{x} + \pd{\lambda}{x}\pd{v}{y} + \pd{\mu}{y}\left(\pd{u}{y}
   + \pd{v}{x}\right) + \rho X =& \rho\pd{^2u}{t^2} \\
   \mu\pd{^2v}{x^2} + (\lambda+2\mu)\pd{^2v}{y^2} + (\lambda+\mu)
   \pd{^2u}{x\partial y} + \left(\pd{\lambda}{y} +
   2\pd{\mu}{y}\right)\pd{v}{y} +&\\+\pd{\lambda}{y}\pd{u}{x} +
   \pd{\mu}{x}\left(\pd{u}{y} + \pd{v}{x}\right) + \rho Y =&
   \rho\pd{^2v}{t^2}
 \end{array}
\end{equation}
If Lam\'{e}'s constants depend only on $y$ equations (\ref{lame}) become simpler:
\begin{equation*}
 \begin{array}{rl}
   (\lambda+2\mu)\pd{^2u}{x^2} + \mu\pd{^2u}{y^2} + (\lambda+\mu)
   \pd{^2v}{x\partial y} + \pd{\mu}{y}\left(\pd{u}{y}
   + \pd{v}{x}\right) + \rho X =& \rho\pd{^2u}{t^2} \\
   \mu\pd{^2v}{x^2} + (\lambda+2\mu)\pd{^2v}{y^2} + (\lambda+\mu)
   \pd{^2u}{x\partial y} + \left(\pd{\lambda}{y} +
   2\pd{\mu}{y}\right)\pd{v}{y} +&\\+\pd{\lambda}{y}\pd{u}{x} +
   \rho Y =& \rho\pd{^2v}{t^2}
 \end{array}
\end{equation*}
Finally one can rewrite equations (\ref{lame}) in vector form:
\begin{multline*}
  (\lambda+\mu)\underline{\grad}\div\underline{u} + \mu\Delta
  \underline{u} + \div\underline{u}\cdot\underline{\grad}\lambda
  +\\+
  (\underline{\underline{\grad}}\underline{u} +
  ^t\underline{\underline{\grad}}\underline{u}) \cdot\underline{\grad}
  \mu + \rho\underline{F} = \rho\pd{^2\underline{u}}{t^2}
\end{multline*}

If the mechanical properties are constant, then one can recognize the classical Lam\'{e}'s equations:
\begin{equation}\label{lame1}\boxed{
  (\lambda+\mu)\underline{\grad}\div\underline{u} + \mu\Delta
  \underline{u} + \rho\underline{F} = \rho\pd{^2\underline{u}}{t^2}}
\end{equation}

\section{International Equation of State of Seawater} 

In this work, we need to know the sound velocity in water. By definition, the velocity $\gamma$ of propagation of acoustic waves is given by
\begin{equation*}
  \gamma^2=\pd{p}{\rho}=\frac1{\beta\rho}
\end{equation*}
where $\beta$ is the compressibility factor of seawater and $p$ the pressure.

The speed of sound can be obtained from the International Equation of State of Seawater (IES80):
\begin{equation*}
  \rho(S,T,p) = \frac{\rho(S,T,0)}{1-p/K(S,T,p)}
\end{equation*}
with
\begin{multline*}
\rho(S,T,0) = 999.842594+6.793952\times10^{-2}T -
9.09529\times10^{-3}T^2 +\\ 1.001685\times10^{-4}T^3 -
1.120083\times10^{-6}T^4 + 6.536332\times10^{-9}T^5 +
\\8.24493\times10^{-1}S - 4.0899\times10^{-3}TS +
7.6438\times10^{-5}T^2S \\- 8.2467\times10^{-7}T^3S +
5.3875\times10^{-9}T^4S - 5.72466\times10{-3}S^{\frac32} \\+
1.0227\times10^{-4}TS^{\frac32} - 1.6546\times10^{-6}T^2S^{\frac32}
+ 4.8314\times10^{-4}S^2
\end{multline*}
and
\begin{multline*}
K(S,T,p) = 19652.21 + 148.4206T - 2.327105T^2 +
1.360447\times10^{-2}T^3\\ - 5.155288\times10^{-3}T^4 + 3.239908p +
1.43713\times10^{-3}Tp + 1.16092\times10^{-4}T^2p\\ -
5.77905\times10^{-7}T^3p + 8.50935\times10^{-5}p^2 -
6.12293\times10^{-6}Tp^2 +\\ 5.2787\times10^{-8}T^2p^2 + 54.6746S -
0.603459TS + 1.09987\times10^{-2}T^2S \\- 6.167\times10^{-5}T^3S +
7.944\times10^{-2}S^{\frac32} + 1.6483\times10^{-2}TS^{\frac32}\\ -
5.3009\times10^{-4}T^2S^{\frac32} + 2.2838\times10^{-3}pS -
1.0981\times10^{-5}TpS \\- 1.6078\times10^{-6}T^2pS +
1.91075\times10^{-4}pS^{\frac32} - 9.9348\times10^{-7}p^2S \\+
2.0816\times10^{-8}Tp^2S + 9.1697\times10^{-10}T^2p^2S
\end{multline*}
However, there are several simplified formulae. In the present work, we will make use of the following:
\begin{multline}\label{gamma}
  \gamma(S,T,p) = 1449.2 + 4.6T - 0.055T^2 + 0.00029T^3 \\+
  (1.34-0.01T)(S-35) + 1.58\times10^{-6}p
\end{multline}
where $\gamma$ is the speed of sound in $m/s$, $T$ is temperature in $^\circ C$, $S$ is salinity and $p$ is pressure in $Pa$.

In general, we consider the pressure to be hydrostatic\footnote{We can make this assumption since we deal with linear acoustics approximation in the water. Let us recall that this model implies pressure perturbation $p(x,t)$ to be such that $p(x,t) = \eps(x,t)p_0$ with $|\eps(x,t)|\ll 1$ and $p_0$ - hydrostatic pressure. For further discussion, the reader should see section \ref{acoustics} and more general reference \cite{sellier}.} and that is why we can replace the last term in (\ref{gamma}) by $-0.016z$ where $z$ is the water depth in $m$. Thus we have
\begin{multline*}\label{gamma2}
\gamma(z) = 1449.2 + 4.6T - 0.055T^2 + 0.00029T^3 \\+
(1.34-0.01T)(S-35) - 0.016z
\end{multline*}
See \cite{ocean} for more details.

In order to obtain a PDE with constant coefficients for water, we replace $\gamma(z)$ by its average value:
\begin{multline*}
  \gamma = \frac1H\int\limits_{-H}^0\gamma(z)dz =
  1449.2 + 4.6T - 0.055T^2 + 0.00029T^3 +\\
(1.34-0.01T)(S-35) + 0.008H
\end{multline*}
where $H$ is water depth.

\section{Linear acoustics in perfect fluids}\label{acoustics}

The author was inspired by \cite{bred,bred2} to use this model in the present work. In this section, we closely follow \cite{sellier}. We try to model seawater by considering it an ideal fluid.

\subsection{Basic hypotheses}

We consider the unsteady flow of a fluid under the following assumptions:
\begin{enumerate}
    \item the fluid is perfect (inviscid $\underline{\underline{\tau}} =
    \underline{\underline{0}}$),
    \item the fluid is not a heat conductor ($\underline{q} =
    \underline{0}$),
    \item there are no exterior mass forces,
    \item the flow is continuous.
\end{enumerate}

The flow is denoted by
\begin{equation}\label{pert}
   (\underline{V}+ \underline{v}, p_0 + p,\rho_0+\rho).
\end{equation}
It is written as a perturbation of the steady flow $(\underline{V},p_0,\rho_0)$ where $\underline{V}$ is the velocity vector, $p_0$ is the pressure and $\rho_0$ is the water density. It is assumed that the perturbation $(\underline{v}, p, \rho)$ is such that
\begin{eqnarray*}
  \underline{v}(x,t) &=& \underline{\underline{\eps_1}}(x,t)
  \underline{V},\qquad
  \text{where}\qquad 0\leq\norm{\eps_1(x,t)}\ll 1, \\
  p(x,t) &=& \eps_2(x,t)p_0,\qquad \text{where}\qquad
  0\leq|\eps_2(x,t)|\ll 1, \\
  \rho(x,t) &=& \eps_3(x,t)\rho_0,\qquad \text{where}\qquad
  0\leq|\eps_3(x,t)|\ll 1.
\end{eqnarray*}

\subsection{Linear acoustics approximation} 

Our perturbed flow (\ref{pert}) satisfies Euler's equations:

Conservation of mass:
\begin{equation}\label{mass}
  \pd{}{t}(\rho_0+\rho) + \div[(\rho_0+\rho)
  (\underline{V}+\underline{v})] = 0,
\end{equation}

Momentum conservation:
\begin{equation}\label{mouvt}
  (\rho_0+\rho)\left\{\pd{}{t} (\underline{V}+\underline{v}) +
  \underline{\grad}(\underline{V}+\underline{v})(\underline{V}+
  \underline{v})\right\} = -\underline{\grad}(p_0+p),
\end{equation}

Energy conservation:
\begin{multline*}
  \pd{}{t}\left\{
    (\rho_0+\rho)\left[
    e'+\frac12(\underline{V}+\underline{v})^2
    \right]\right\} +\\+ \div\left[(\rho_0 + \rho)
    (e'+\frac12(\underline{V}+\underline{v})^2)
    (\underline{V}+\underline{v})
    \right] = -\div[(p_0+p)(\underline{V}+\underline{v})]
\end{multline*}
where $e'$ is specific internal energy.

The fluid pressure is a function of density $\rho_0+\rho$ and entropy\footnote{In this work, we restrict ourselves to homoentropic flows} $s_0$. Linearizing the state equation around the basic state $(\rho_0,s_0)$ yields
\begin{equation}\label{state}
  p(x,t)=\gamma^2\rho(x,t),\qquad \gamma^2 := \left(
    \pd{p}{\rho}\right)_s (\rho_0,s_0).
\end{equation}
The basic flow $(p_0,\rho_0,\underline{V})$ being steady and uniform, the linearisation of equations (\ref{mass}), (\ref{mouvt}) is
\begin{equation}\label{linearm}
  \left(\pd{}{t}+\underline{V}\cdot
  \underline{\grad}\right)\rho + \rho_0
  \div\underline{v} = 0,
\end{equation}
\begin{equation}\label{linearmt}
  \pd{\underline{v}}{t} + \underline{\underline{\grad}}
  \underline{v}\cdot
  \underline{V} = -\frac{1}{\rho_0}\underline{\grad} p.
\end{equation}

From equation (\ref{linearm}) one finds
\begin{equation}\label{tmp}
  \div\underline{v}= -\frac{1}{\rho_0} \left(
    \pd{}{t}+\underline{V}\cdot\underline{\grad}
  \right)\rho.
\end{equation}
Taking the divergence of equation (\ref{linearmt}) and substituting into (\ref{tmp}) yields
\begin{equation*}
  \frac1{\gamma^2}\left\{
    \pd{}{t} + \underline{V}\cdot\underline{\grad}
  \right\}\left(
    \pd{\rho}{t} + \underline{V}\cdot\underline{\grad}\rho
  \right) = \Delta\rho
\end{equation*}
If we multiply the last equality by $\gamma^4$ and use (\ref{state}) we obtain
\begin{equation*}
  \boxed{\left\{
    \pd{}{t} + \underline{V}\cdot\underline{\grad}
  \right\}\left(
    \pd{p}{t} + \underline{V}\cdot\underline{\grad}p
  \right) = \gamma^2\Delta p}
\end{equation*}

If we assume that before perturbation, the fluid was motionless ($\underline{V} = \underline{0}$) we obtain the equation that we will work with:
\begin{equation}\label{linacoustic}
  \boxed{
    \frac1{\gamma^2}\pd{^2p}{t^2} = \Delta p
  }
\end{equation}

To this equation, we should add the boundary condition that the accelerations of solid and fluid coincide at the interface:
\begin{equation}\label{bc_ice}
  \rho_g\pd{^2v}{t^2} = \pd{p}{y}
\end{equation}
where $\rho_g$ is the ice density at the bottom, and $v$ is the vertical displacement.

\section{Complete model}\label{complete}

In this section, we put together all the equations obtained above and discuss initial and boundary conditions.

First of all, we start with the homogeneous problem: all mechanical properties of the ice are the same everywhere. Here, we recall different relations among mechanical constants. These relations were taken from \cite{sned}.
\begin{equation}\label{lambda}
  \lambda = \frac{\nu E}{(1+\nu)(1-2\nu)},
\end{equation}
\begin{equation}\label{mu}
  \mu = \frac{E}{2(1+\nu)},
\end{equation}
\begin{equation*}
  E = \frac{(3\lambda+2\mu)\mu}{\lambda+\mu},
\end{equation*}
\begin{equation*}
  \nu = \frac{\lambda}{2(\lambda+\mu)}.
\end{equation*}

The displacement field in the ice induced by the moving load satisfies Lam\'e-Clapeyron equations in $\mathbb{R}\times(0,h)$:
\begin{equation}\label{compl1}
  (\lambda+2\mu)\pd{^2u}{x^2} + \mu\pd{^2u}{y^2} + (\lambda+\mu)
  \pd{^2v}{x\partial y} = \rho_g\pd{^2u}{t^2} + \alpha\pd{T}{x},
\end{equation}
\begin{equation}\label{compl2}
  \mu\pd{^2v}{x^2} + (\lambda+2\mu)\pd{^2v}{y^2} + (\lambda+\mu)
  \pd{^2u}{x\partial y} = \rho_g\pd{^2v}{t^2} +\rho_g g + \alpha\pd{T}{y}
\end{equation}
where $\alpha$ is the linear expansion coefficient. The temperature $T(x,y)$ in the ice layer satisfies the heat conduction equation in the same region $\mathbb{R}\times(0,h)$:
\begin{equation}\label{heat}
  \frac1{\kappa}\pd{T}{t} + \eta\pd{}{t}\div\underline{u} = \Delta T +
  \frac1{\kappa}Q.
\end{equation}
where $\kappa$ is the thermal conductivity coefficient, $\eta=\gamma T_0/k$ with $T_0$ the temperature of undeformed body, $\gamma := \frac{E}{1-2\nu}\alpha$ and $k$ is the heat conductivity coefficient. Equation (\ref{heat}) is needed if we consider thermoelastic effects. Later, we will make further simplifications and take into account thermo-effects only in the ice's mechanical properties. There is a technical reason: the terms due to thermal expansion and gravitation do not belong to $L_2(\mathbb{R})$ and prevent the author from using integral transform methods or, more exactly, Fourier integral transforms. Let us now see how to avoid this difficulty.

In water, the pressure below the ice is given by the linear acoustic equation (\ref{linacoustic}):
\begin{equation*}
  \frac1{\gamma^2}\pd{^2p}{t^2} = \Delta p
\end{equation*}

Next, we describe initial and boundary conditions. Let us start with the initial conditions. Since we look for travelling-wave type solutions, this point can be omitted.

To describe the boundary conditions, we use the notation
$$\Gamma_1 = \{(x,y)\in\mathbb{R}^2|y=-H\},$$ $$\Gamma_2 =
\{(x,y)\in\mathbb{R}^2|y=0\},$$ $$\Gamma_3 =
\{(x,y)\in\mathbb{R}^2|y=h\}.$$

Along $\Gamma_3$, we have the following boundary conditions:
\begin{eqnarray}
 \sigma_y(x,h) &=& l(x), \\
 \tau_{xy}(x,h) &=& g(x), \\
 T(x,h) &=& T_1.
\end{eqnarray}
The physical meaning of the first equation is to impose the load on the upper ice boundary. The second one corresponds to the given friction between the ice layer and the moving load. The third condition imposes air temperature along the ice-air interface. In fact, we suppose this temperature to be constant along $\Gamma_3$.

On $\Gamma_2$ we have:
\begin{eqnarray}
  \sigma_y(x,0) &=& -p(x,0) + \rho_w g v(x,0), \\
  \tau_{xy}(x,0) &=& 0, \\
  \rho_g\pd{^2v}{t^2} &=& \pd{p}{y}, \\
  T(x,0) &=& T_0,
\end{eqnarray}
where $g$ is the gravity acceleration and $v(x,0)$ is the vertical displacement. The first condition is the pressure balance at the ice-water interface. The second condition means that we neglect the friction between ice and water. The third one was already discussed (\ref{bc_ice}). The last one is clear.

On $\Gamma_1$, we just have one condition:
\begin{equation}\label{dpdy}
  \pd{p}{y} = 0
\end{equation}
In fact, this is the same condition as (\ref{bc_ice}) in the case where the boundary is fixed.

Now, we simplify our model. First, we start with equations (\ref{compl1}),(\ref{compl2}). In these equations, we neglect thermal expansion terms and gravitational forces. To justify this simplification, we have to look at dimensionless equations. For example, we consider equation (\ref{compl2}), which is more interesting because it contains both terms. In non-dimensional form this equation is
\begin{equation*}
  \pd{^2v}{x^2} + \frac{\lambda+2\mu}{\mu}\pd{^2v}{y^2} +
  \frac{\lambda+\mu}{\mu}\pd{^2u}{x\partial y} = \frac{\rho_g
  H^2}{\mu\tau^2}\pd{^2v}{t^2} + \frac{\rho_g g H}{\mu} + \frac{\alpha
  T_0}{\mu H}\pd{T}{y}
\end{equation*}
where $H$, $\tau$, and $T_0$ are characteristic length (water depth), time and temperature, respectively.

The gravitational term is of order $10^{-6}H/m$. Thus, for physically relevant values\footnote{$H\leq1000m$} of $H$, this coefficient is small. Consequently, we can neglect the effect of gravity. In \cite{bred}, the authors explain this fact by the smallness of the characteristic velocity $\sqrt{g(H+h)}$ of free surface gravity waves in a water layer of depth $H<5000m$ compared to the infinite medium dilatational wave velocity in ice $\sqrt{\frac{\lambda+2\mu}{\rho_g}}$, or, in other words the water wave velocity is much smaller than the body wave velocity. Finally, the thermal expansion term is of order $10^{-13}$ because for ice $\alpha\approx 44(\pm9)\times10^{-6}K^{-1}$. This value was taken from \cite{expan}.

Then, we simplify the heat-conduction equation (\ref{heat}). First, in ice, there are no internal heat sources; thus, $Q=0$. The next term to be neglected is the thermoelastic one. In fact, we decouple the thermoelastic problem, and temperature is only needed to determine the ice's mechanical properties at each point. Then, we find the steady solution of this equation. This can be justified by the fact that boundary conditions are time-independent and that this process has taken infinite time before our consideration. The last assumption concerning this equation is that the temperature $T$ depends only on the variable $y$: $T=T(y)$. It is natural to make this assumption because boundary conditions are invariant by translations along the $x$-axis.

With all these assumptions, one can write the solution of the simplified heat-conduction equation (\ref{heat}):
\begin{equation}\label{temper}
  \boxed{
  T(y) = T_0\left(1-\frac{y}{h}\right) + T_1\frac{y}{h}}
\end{equation}

Now we come to the problem that we are going to solve in the next section:
\begin{eqnarray}\label{lam1}
  (\lambda+2\mu)\pd{^2u}{x^2} + \mu\pd{^2u}{y^2} + (\lambda+\mu)
  \pd{^2v}{x\partial y} &=& \rho_g\pd{^2u}{t^2}, \\
  \label{lam2}
  \mu\pd{^2v}{x^2} + (\lambda+2\mu)\pd{^2v}{y^2} + (\lambda+\mu)
  \pd{^2u}{x\partial y} &=& \rho_g\pd{^2v}{t^2}, \\
  \label{lam3}
  \frac1{\gamma^2}\pd{^2p}{t^2} &=& \Delta p.
\end{eqnarray}

To these equations, we have to add the boundary conditions discussed earlier. We repeat one more time that the absence of initial conditions is explained by the fact that we look for travelling wave solutions.

The general plan followed by the author is to:
\begin{enumerate}
    \item Solve problem (\ref{lam1}), (\ref{lam2}), (\ref{lam3}) for a homogeneous ice layer
    \item Generalize this solution to a multi-layer case as a special case of inhomogeneity
    \item Take the limit of an infinite number of layers (consequently, at this stage, we will obtain the solution of an inhomogeneous problem)
    \item If time permits, solve the quasi-dynamic contact problem
\end{enumerate}

\section{Fourier transform}

In this work, we often use integral Fourier transforms. For a good practical reference on this subject, we invite the reader to look at \cite{transforms}. If the reader is interested in theoretical aspects of Fourier transforms, we suggest the reading of \cite{princeton}. We use the following definition:
\begin{defn}
  If $f(x)$ is an absolutely integrable function on $\mathbb{R}$ (i.e.,
  $\int_{-\infty}^{+\infty}|f(x)|dx<\infty$), then the direct
  Fourier transform of $f(x)$, $\mathfrak{F}[f]$, and the
  inverse Fourier transform of $f(x)$, $\mathfrak{F}^{-1}[f]$ are thee functions given by
  \begin{equation}\label{direct}
    \mathfrak{F}[f](s) = \int\limits_{-\infty}^{+\infty} f(x)
    e^{-ixs}dx
  \end{equation}
  and
  \begin{equation}\label{inv}
    \mathfrak{F}^{-1}[f](s) = \frac1{2\pi}\int\limits_{-\infty}
    ^{+\infty}f(x)e^{ixs}dx
  \end{equation}
\end{defn}

\begin{defn}
  A function $f:\mathbb{R}\rightarrow\mathbb{R}(\mathbb{C})$ is said to be "classically transformable" if either
  \begin{enumerate}
    \item $f$ is absolutely integrable on $\mathbb{R}$, or
    \item $f$ is the Fourier transform (or inverse Fourier transform) of an absolutely integrable function or
    \item $f$ is a linear combination of an absolutely integrable function and a Fourier transform (or inverse Fourier transform) of an absolutely integrable function.
  \end{enumerate}
\end{defn}

If $f$ is a classically transformable but not absolutely integrable function, then it can be shown that formulas (\ref{direct}) and (\ref{inv}) can still be used to define $\mathfrak{F}[f]$ and $\mathfrak{F}^{-1}[f]$ provided that the integrals are taken in the sense of Cauchy principal values.

We suppose that solutions of equations from Section \ref{complete} are classically transformable. It is a necessary condition to use integral transform methods.

\subsection{Discrete Fourier transform}

We are interested in discrete Fourier transforms in order to calculate the inverse integral Fourier transform numerically. This topic is discussed in \cite{recipes}, but the author does not agree with their way of calculating integral (\ref{inv}). Here, we should add some additional remarks. In \cite{recipes}, on page 503, one can find the formula:
$$
  H(f_n) = \int_{-\infty}^{\infty} h(t)e^{2\pi i f_n t}dt \approx
  \sum_{k=0}^{N-1} h_k e^{2\pi i f_n t_k} \Delta = \Delta
  \sum_{k=0}^{N-1} h_k e^{2\pi ikn/N}
$$
where $f_n := \frac{n}{N\Delta}$ and $t_k := k\Delta$. We cannot understand why the interval $(-\infty,0)$ is not taken into consideration. That is why, in the present work, we show how one should perform this discretization. This formula is used in the programs given in Appendix \ref{app2}.

Before starting the Fourier integral discretization, we recall the formulae of discrete Fourier transform used in very well-known scientific computing libraries \cite{fftw}, \cite{wwwfftw}. Incidentally, this formula is used in MatLab to perform FFT computations.

Direct transform:
\begin{equation*}
  X_k = \sum_{j=1}^N x_j\omega_N^{(j-1)(k-1)}.
\end{equation*}

Inverse transform:
\begin{equation}\label{discr-direct}
  x_j = \frac1N\sum_{k=1}^N X_k\omega_N^{-(j-1)(k-1)}.
\end{equation}
where $\omega_N=e^{\frac{-2\pi i}{N}}$.

Now we calculate numerically this integral:
\begin{equation}\label{discr-inverse}
  u(x) = \frac1{2\pi}\int\limits_{-\infty}^{+\infty}\hat{u} (\omega)
  e^{i\omega x}d\omega
\end{equation}

To perform our discretization, we need several notations: $\delta$ denotes the discretization step in the Fourier space domain, and $k_j$ is a wave number. Therefore
$$
  \omega_j:= j\delta = 2\pi k_j,
$$
$$
  x_n := \frac{n}{N\Delta}
$$
where $\Delta := \frac{\delta}{2\pi}$.
\begin{multline*}
  u(x_n) = \frac1{2\pi}\int\limits_{-\infty}^{+\infty}\hat{u}
  (\omega)e^{i\omega x_n}d\omega =\\= \int\limits_{0}^{+\infty}
  \hat{u}(2\pi k) e^{2\pi i k x_n} dk - \int\limits_{0}^{+\infty}\hat{u}(-2\pi
  k) e^{-2\pi i k x_n} dk \approx \\ \approx \int\limits_{0}^{N\delta}
  \hat{u}(2\pi k) e^{2\pi i k x_n} dk - \int\limits_{0}^{N\delta}\hat{u}(-2\pi
  k) e^{-2\pi i k x_n} dk \approx \\ \approx
  \Delta\left(
    \sum_{j=0}^{N-1}\hat{u}(2\pi k_j)e^{2\pi i k_j x_n} -
    \sum_{j=0}^{N-1}\hat{u}(-2\pi k_j)e^{-2\pi i k_j x_n}
  \right) =\\=
  \Delta\left(
    \sum_{j=0}^{N-1}\hat{u}_je^{2\pi i j n/N} -
    \sum_{j=0}^{N-1}\hat{u}_{-j}e^{-2\pi i j n/N}
  \right) = \\ =
  \Delta\left(
    \sum_{j=1}^{N}\hat{u}_{j-1}\omega_N^{-(j-1)n} -
    \sum_{j=1}^{N}\hat{u}_{-(j-1)}\omega_N^{(j-1)n}
  \right).
\end{multline*}

If we recall the definitions of direct and inverse discrete Fourier transforms, we can calculate the sums in the last equality
\begin{equation*}
 \boxed{\{u(x_n)\}_{n=-N/2+1}^{N/2} =
 \Delta\bigl(N \ifft(\{\hat{u}_{j+N}\}_{j=0}^{N-1}) -
  \fft(\{\hat{u}_{N-j}\}_{j=0}^{N-1})\bigr)}
\end{equation*}

\subsection{Filon's quadrature formulae}

There is an alternative way to compute Fourier-type integrals with a special class of quadrature formulae. These formulae are named Filon's formulae because Filon was the first to propose this idea. Unfortunately, we do not know the original paper. We discovered this method from \cite{bakh}.

These formulae are particularly useful for highly oscillating functions. The main idea of the method is not to interpolate the integrated function but to interpolate its amplitude with a polynomial of low order. In this work, we show how to do it in the case of linear interpolation. The reader can obtain better results by using quadratic interpolation. One more advantage of this method is that one can better approximate integrals by changing the order of interpolation but with the same number of discretization points. With FFT-based methods, one can obtain more accurate results only by adding discretization points. For some problems, the difference can be important. But the computational complexity of Filon's quadrature is $O(N^2)$ operations, while the method described in the previous subsection performs it with $O(N\ln N)$ operations.

We calculate numerically the integral
\begin{equation*}
  I = \int\limits_a^b f(x) e^{i\omega x}\;dx \equiv \int\limits_a^b
  F(x)\;dx.
\end{equation*}
We use the subintervals
$$
  a=x_0<x_1<\ldots<x_N=b.
$$
On the $k^{th}$ subinterval we replace $f(x)$ by its interpolation polynomial with $q+1$ nodes:
\begin{equation*}
  I_N=\sum_{k=1}^N I_k = \sum_{k=1}^N\int_{x_{k-1}}^{x_k}P_q(x)e^{i\omega
  x}\;dx.
\end{equation*}
Filon obtained his formula with $q=2$ (quadratic interpolation). In the present work, we deal with linear interpolation:
$$
  P_1(x)=f(x_{k-1}) + (x-x_{k-1})f(x_{k-1},x_k) = y_{k-1}+
  \frac{y_k-y_{k-1}}{x_k-x_{k-1}}(x-x_{k-1})
$$
where $f(x_{k-1},x_k)$ denotes the first divided difference of the function $f$ at the points $x_{k-1}, x_k$\footnote{A notion of divided differences is very common in Soviet numerical analysis literature. The definition of the first divided difference is $f(x_{k-1},x_k)\equiv\frac{y_k-y_{k-1}}{x_k-x_{k-1}}$. For second-order divided differences, we have a recurrent relation: $f(x_{k-1},x_k,x_{k+1})\equiv\frac{f(x_{k+1},x_k)-f(x_k,x_{k-1})}{x_{k+1}-x_{k-1}}$. And the definition of divided differences of $k^{th}$ order is: $f(x_i,x_{i+1},\ldots,x_{i+k}) = \frac{f(x_{i+1},\ldots,x_k)-f(x_i,\ldots,x_{k-1})}{x_{i+k}-x_i}$. These differences are useful to approximate derivatives.}.

Then we have:
\begin{multline*}
  I_k=\int_{x_{k-1}}^{x_k}P_1(x)e^{i\omega x}\;dx=\\=\int_{x_{k-1}}^{x_k}
  \left(y_{k-1}+\frac{y_k-y_{k-1}}{x_k-x_{k-1}}(x-x_{k-1})\right)e^{i\omega x}\;dx=\\
  =\left.y_{k-1}\frac{e^{i\omega x}}{i\omega}\right|_{x_{k-1}}^{x_k} +
  \frac{y_k-y_{k-1}}{x_k-x_{k-1}}\left\{
  \left.(x-x_{k-1})\frac{e^{i\omega x}}{i\omega}\right|_{x_{k-1}}^{x_k}-
  \int_{x_{k-1}}^{x_k}\frac{e^{i\omega x}}{i\omega}\;dx\right\} = \\
  = \frac{y_{k-1}}{i\omega}(e^{i\omega x_k}-e^{i\omega x_{k-1}}) +
  (y_k-y_{k-1})\frac{e^{i\omega x_k}}{i\omega} - \left.\frac{y_k-y_{k-1}}{x_k-x_{k-1}}
  \frac{e^{i\omega x}}{(i\omega)^2}\right|_{x_{k-1}}^{x_k} = \\
  = -\frac{F_{k-1}}{i\omega} + \frac{F_k}{i\omega} + \frac{y_k-y_{k-1}}{\omega^2h_k}
  (e^{i\omega x_k}-e^{i\omega x_{k-1}})=\\
  =\frac{F_k-F_{k-1}}{i\omega} + 2i\sin\left(\frac{\omega
  h_k}{2}\right)
  \left(\frac{y_k-y_{k-1}}{\omega^2h_k}\right)e^{i\omega x_{k-1/2}},
\end{multline*}
where $F_k\equiv y_k e^{i\omega x_k}$, $h_k \equiv x_k - x_{k-1}$.

Summation over $k$ gives
$$
\boxed{
  I_N=\sum_{k=1}^N I_k=\frac{F_N-F_0}{i\omega} + \frac{2i}{\omega^2}
  \sum_{k=1}^N\frac{\sin\omega h_k/2}{h_k}(y_k-y_{k-1})e^{i\omega x_{k-1/2}}
  }
$$

To conclude this subsection, the author would like to mention here that the above formula is useful only in the case where $\omega(b-a)\gg1$. To explain this, we can look at the last formula and see that it requires a lot of operations per discretization step (more than for Simpson's formula). In the case where $\omega(b-a)\ll1$, the function $F(x)$ is no longer highly oscillating, and we suggest using another quadrature formulae.

\section{Homogeneous ice layer}\label{hom}

\subsection{Water pressure determination}

First, we begin with the linear acoustic equation. We repeat the problem that we are going to solve:
\begin{equation*}
  \frac1{\gamma^2}\pd{^2p}{t^2} = \Delta p, \qquad
  (x,y)\in\mathbb{R}\times(-H,0),
\end{equation*}
\begin{equation*}
  \rho_g\pd{^2v}{t^2} = \pd{p}{y}, \qquad y = 0,
\end{equation*}
\begin{equation*}
  \pd{p}{y} = 0, \qquad y = -H.
\end{equation*}
In this work, we look for travelling wave solutions of our equations. In mathematical notation, this means
\begin{equation*}
  p(x,y,t) := p(x-c_0t,y)
\end{equation*}
where $c_0$ is the moving load velocity. We substitute this particular choice of the solution in (\ref{linacoustic}) and obtain:
\begin{equation}\label{ellipt}
\chi\pd{^2p}{x^2} + \pd{^2p}{y^2} = 0
\end{equation}
where $\chi$ denotes the dimensionless quantity:
\begin{equation*}
  \chi := 1 - \frac{c_0^2}{\gamma^2}
\end{equation*}
It is interesting to note that equation (\ref{ellipt}) is
\begin{itemize}
    \item elliptic if $|c_0|<|\gamma|$
    \item hyperbolic, otherwise.
\end{itemize}
In this work, we consider only the first case since it is the most physically relevant. Recall that the speed of sound in water is $\gamma\approx 1500m/s$.

Let us multiply equation (\ref{ellipt}) by $e^{-i\omega x}$ and integrate on $\mathbb{R}$:
\begin{equation*}
  \frac{d^2\hat{p}}{dy^2} - \chi\omega^2\hat{p}= 0
\end{equation*}
We have assumed that we are in the elliptic case. In other words $0<\chi\le1$. A general solution of this equation can be written as
\begin{equation*}
  \hat{p}(\omega,y) = C_1(\omega)\cosh(\sqrt\chi\omega y) + C_2(\omega)
  \sinh (\sqrt\chi\omega y)
\end{equation*}
We take the Fourier transform of the boundary condition (\ref{dpdy}) on $\Gamma_1$:
\begin{equation*}
  \pd{p}{y} = 0 \Rightarrow \frac{d\hat{p}}{dy} = 0, \qquad y = -H
\end{equation*}
Next, we use this condition in order to establish a relationship between the unknown functions $C_1(\omega)$ and $C_2(\omega)$:
\begin{equation*}
  \frac{d\hat{p}}{dy}= C_1(\omega)\sqrt{\chi}\omega
  \sinh(\sqrt\chi\omega y) + C_2(\omega)\sqrt\chi\omega
  \cosh(\sqrt\chi\omega y)
\end{equation*}
Evaluating it at $y=-H$ yields
\begin{equation*}
  C_2(\omega)= \tanh(\sqrt\chi\omega H) C_1(\omega).
\end{equation*}
Now, there is only one independent function:
\begin{equation*}
  \hat{p}(\omega,y) = C_1(\omega)(
    \cosh(\sqrt\chi\omega y) + \tanh(\sqrt\chi\omega H)
    \sinh(\sqrt\chi\omega y))
\end{equation*}
The second condition (\ref{lam3}) on $\Gamma_2$ is
\begin{equation*}
  y=0: \pd{p}{y} = \rho_g\pd{^2v}{t^2}.
\end{equation*}
We take the Fourier transform of this equation and substitute the travelling wave solution $v(x,y,t) := v(x-c_0t,y)$:
\begin{equation*}
  \frac{d\hat{p}}{dy} = -\rho_g c_0^2\omega^2 \hat{v}(\omega,0)
\end{equation*}
Finally we can find the unknown function $C_1(\omega)$:
\begin{equation*}
  \left.\frac{d\hat{p}}{dy}\right|_{y=0} = C_1(\omega)\sqrt\chi
  \omega\tanh(\sqrt\chi\omega H) = -\rho_g c_0^2\omega^2 \hat{v}
  (\omega,0)
\end{equation*}
\begin{equation*}
  C_1(\omega) = -\frac{\rho_g\omega c_0^2}{\sqrt\chi}
  \coth(\sqrt\chi\omega H)\hat{v}(\omega,0)
\end{equation*}
The solution for $\hat p$ is
\begin{multline}\label{pressure}
  \hat{p}(\omega,y) = -\frac{\rho_g\omega c_0^2}{\sqrt\chi}
  \coth(\sqrt\chi\omega H)\hat{v}(\omega,0)(
    \cosh(\sqrt\chi\omega y) +\\+ \tanh(\sqrt\chi\omega H)
    \sinh(\sqrt\chi\omega y))
\end{multline}

\subsection{Solution of Lam\'e's equations for an homogeneous layer}

Recall that Lam\'{e}'s equations are given by (\ref{lam1}), (\ref{lam2}). In order to obtain the dimensionless equations below we need to divide equation (\ref{lam1}) by $\lambda+2\mu$ and (\ref{lam2}) by $\mu$:
\begin{equation}\label{lam12}
  \pd{^2u}{x^2} + \frac{\mu}{\lambda+2\mu}\pd{^2u}{y^2} +
  \frac{\lambda+\mu}{\lambda+2\mu}\pd{^2v}{x\partial y} =
  \frac{\rho_g}{\lambda+2\mu}\pd{^2u}{t^2},
\end{equation}
\begin{equation}\label{lam22}
  \pd{^2v}{x^2} + \frac{\lambda+2\mu}{\mu}\pd{^2v}{y^2} +
  \frac{\lambda+\mu}{\mu}\pd{^2u}{x\partial y} =
  \frac{\rho_g}{\mu}\pd{^2v}{t^2}.
\end{equation}

Let us introduce the velocities:
\begin{equation*}
  c_1 = \sqrt{\frac{\lambda+2\mu}{\rho_g}} =
  \sqrt{\frac{E(1-\nu)}{\rho_g(1+\nu)(1-2\nu)}} \approx 3900 m/s
\end{equation*}
and
\begin{equation*}
  c_2 = \sqrt{\frac{\mu}{\rho_g}} =
  \sqrt{\frac{E}{2\rho_g(1+\nu)}} \approx 1900 m/s.
\end{equation*}
These constants represent the propagation velocities of elastic waves in two orthogonal directions $x$ and $y$.

Again, we look for travelling-wave solutions:
\begin{equation*}
  u(x,y,t) := u(x-c_0t,y),
\end{equation*}
\begin{equation*}
  v(x,y,t) := v(x-c_0t,y)
\end{equation*}
where $c_0$ is the moving load velocity.

After substituting this particular form in equations (\ref{lam12}) and (\ref{lam22}), and using the relations between different elastic constants (\ref{lambda}) and (\ref{mu}), one obtains:
\begin{equation*}
  \left(1-\frac{c_0^2}{c_1^2}\right)\pd{^2u}{x^2} +
  \frac{1-2\nu}{2(1-\nu)}\pd{^2u}{y^2} +
  \frac1{2(1-\nu)}\pd{^2v}{x\partial y} = 0
\end{equation*}
\begin{equation*}
  \left(1-\frac{c_0^2}{c_2^2}\right)\pd{^2v}{x^2} +
  \frac{2(1-\nu)}{1-2\nu}\pd{^2v}{y^2} +
  \frac1{1-2\nu}\pd{^2u}{x\partial y} = 0
\end{equation*}
We introduce the notation
$$
  \chi_i := 1 - \frac{c_0^2}{c_i^2}, \qquad i=1,2.
$$

In Fourier space, one has a system of ODE
\begin{equation*}
  \od{^2\hat u}{y^2} - \frac{i\omega}{1-2\nu}\od{\hat v}{y} -
  \frac{2(1-\nu)}{1-2\nu}\chi_1\omega^2\hat u = 0
\end{equation*}
\begin{equation*}
  \od{^2\hat v}{y^2} - \frac{i\omega}{2(1-\nu)}\od{\hat u}{y} -
  \frac{1-2\nu}{2(1-\nu)}\chi_2\omega^2\hat u = 0
\end{equation*}
One can rewrite this system as a first-order system by introducing the new functions $\hat{w}:=\od{\hat{u}}{y}$ and
$\hat{z}:=\od{\hat{v}}{y}$:
\begin{equation*}
 \left\{
 \begin{array}{ll}
   \od{\hat{u}}{y} =& \hat{w} \\
   \od{\hat{w}}{y} =& \frac{2(1-\nu)}{1-2\nu}\chi_1\omega^2 \hat{u} +
   \frac{i\omega}{1-2\nu}\hat{z} \\
   \od{\hat{v}}{y} =& \hat{z} \\
   \od{\hat{z}}{y} =& \frac{i\omega}{2(1-\nu)}\hat{w} +
   \frac{1-2\nu}{2(1-\nu)}\chi_2\omega^2 \hat{v}
 \end{array}
 \right.
\end{equation*}
or in matrix form
$$
  \od{\hat{X}}{y} = A\hat{X}
$$
where
\begin{equation*}
A =
\begin{pmatrix}
  0 & 1 & 0 & 0 \\
  \frac{2(1-\nu)}{1-2\nu}\chi_1\omega^2 & 0 & 0 & \frac{i\omega}{1-2\nu}\\
  0 & 0 & 0 & 1 \\
  0 & \frac{i\omega}{2(1-\nu)} & \frac{1-2\nu}{2(1-\nu)}\chi_2\omega^2 & 0 \\
\end{pmatrix},
\end{equation*}
\begin{equation*}
  \hat X = (\hat u, \hat w, \hat v, \hat z)^t
\end{equation*}

The general solution of this system of ODE depends on four
functions:
\begin{multline}\label{sasu}
  \hat u(\omega,y) = C_1(\omega)\cosh(k_1\omega y) + C_2(\omega)
  \sinh(k_1\omega y) +\\+ C_3(\omega)\sinh(k_2\omega y) + C_4(\omega)
  \cosh(k_2\omega y),
\end{multline}
\begin{multline}\label{sasv}
  \hat v(\omega,y) = \frac{i}{k_1^2}\bigl(
    k_1\sinh(k_1\omega y)C_1(\omega) + k_1\cosh(k_1\omega y)
    C_2(\omega) +\\+ F\cosh(k_2\omega y) C_3(\omega) + F\sinh(k_2\omega
    y) C_4(\omega)\bigr).
\end{multline}
The notation
\begin{equation*}
  k_1 := \sqrt{1-\frac{2c_0^2\rho_g(1+\nu)}{E}}
\end{equation*}
\begin{equation*}
  k_2 := \sqrt{1 - \frac{c_0^2\rho_g}{E(1-\nu)}(1-\nu-2\nu^2)}
\end{equation*}
\begin{equation*}
  F :=k_2k_1^2
\end{equation*}
has been introduced.

We would like to make a remark. From equality (\ref{sasv}) it is clear that the ice deflection tends to infinity when $k_1\to 0$. The block velocity corresponding to $k_1=0$ is
\begin{equation*}
  c_0 = \sqrt{\frac{E}{2\rho_g(1+\nu)}} = c_2
\end{equation*}
Thus, we found analytically a resonant velocity.

Recall that all boundary conditions are written in terms of the stress tensor. That is why we need to determine the two components of this tensor $\sigma_y$, $\tau_{xy}$ in order to build below a system of linear equations with unknown functions $C_1(\omega), C_2(\omega), C_3(\omega), C_4(\omega)$. The connection between the displacements and the stress tensor is given by Hooke's law
\begin{equation*}
  \sigma_y = \lambda\pd{u}{x} + (\lambda+2\mu)\pd{v}{y},
\end{equation*}
\begin{equation*}
  \tau_{xy} = \mu\left(
    \pd{u}{y} + \pd{v}{x}
  \right)
\end{equation*}
We take Fourier transforms of these two equations
\begin{equation}\label{hook1}
  \hat{\sigma}_y = -i\lambda\omega\hat{u} + (\lambda+2\mu)
  \od{\hat v}{y}
\end{equation}
\begin{equation}\label{hook2}
  \hat\tau_{xy} = \mu\left(
    \od{\hat u}{y} - i\omega\hat v
  \right)
\end{equation}
and substitute the solutions (\ref{sasu}), (\ref{sasv}) in
(\ref{hook1}), (\ref{hook2}):
\begin{multline*}
  \hat\sigma_y = i E\omega\bigl(
    B_1\cosh(k_1\omega y)C_1(\omega) + B_1\sinh(k_1\omega y)
    C_2(\omega) +\\+ B_2\sinh(k_2\omega y)C_3(\omega) +
    B_2\cosh(k_2\omega y)C_4(\omega)\bigr)
\end{multline*}
where
\begin{equation*}
  B_1 := \frac1{1+\nu},\qquad B_2 := \frac{(1-\nu)Fk_2 - \nu
  k_1^2}{k_1^2(1+\nu)(1-2\nu)} =
  \frac{(1-\nu)k_2^2-\nu}{(1+\nu)(1-2\nu)}.
\end{equation*}
\begin{multline*}
  \hat\tau_{xy} = \mu\omega(
    A_1\sinh(k_1\omega y)C_1(\omega) + A_1\cosh(k_1\omega
    y)C_2(\omega) +\\+ A_2\cosh(k_2\omega y)C_3(\omega) +
    A_2\sinh(k_2\omega y)C_4(\omega))
\end{multline*}
where
\begin{equation*}
  A_1\equiv A := k_1+\frac1{k_1},\qquad A_2 := k_2+\frac{F}{k_1^2} =
  2k_2.
\end{equation*}

We rewrite the expressions for $\hat u$, $\hat v$, $\hat \sigma_y$, $\hat \tau_{xy}$ in vector form:
\begin{equation}\label{imc}
  \begin{pmatrix}
    \hat u(\omega, y) \\
    \hat v(\omega, y) \\
    \hat \sigma_y(\omega, y) \\
    \hat \tau_{xy}(\omega, y) \\
  \end{pmatrix} = I\cdot M(\omega,y)
  \begin{pmatrix}
    C_1(\omega) \\
    C_2(\omega) \\
    C_3(\omega) \\
    C_4(\omega) \\
  \end{pmatrix}
\end{equation}
where
$$
  I=\diag{I_{11},I_{22},I_{33},I_{44}} = \diag{1,i/k_1^2,i E\omega,
  \mu\omega}
$$
and $M(\omega,y)$ is a $4\times4$ real matrix with components:
\begin{equation*}
  m_{11}(\omega,y) = \cosh(k_1\omega y),\qquad m_{12}(\omega,y) =
  \sinh(k_1\omega y),
\end{equation*}
\begin{equation*}
  m_{13}(\omega,y) = \sinh(k_2\omega y),\qquad m_{14}(\omega,y) =
  \cosh(k_2\omega y),
\end{equation*}
\begin{equation*}
  m_{21}(\omega,y) = k_1\sinh(k_1\omega y), \qquad
  m_{22}(\omega,y) = k_1\cosh(k_1\omega y),
\end{equation*}
\begin{equation*}
  m_{23}(\omega,y) = F\cosh(k_2\omega y),\qquad
  m_{24}(\omega,y) = F\sinh(k_2\omega y),
\end{equation*}
\begin{equation*}
  m_{31}(\omega,y) = B_1\cosh(k_1\omega y),\qquad
  m_{32}(\omega,y) = B_1\sinh(k_1\omega y),
\end{equation*}
\begin{equation*}
  m_{33}(\omega,y) = B_2\sinh(k_2\omega y),\qquad
  m_{34}(\omega,y) = B_2\cosh(k_2\omega y),
\end{equation*}
\begin{equation*}
  m_{41}(\omega,y) = A_1\sinh(k_1\omega y),\qquad
  m_{42}(\omega,y) = A_1\cosh(k_1\omega y),
\end{equation*}
\begin{equation*}
  m_{43}(\omega,y) = A_2\cosh(k_2\omega y),\qquad
  m_{44}(\omega,y) = A_2\sinh(k_2\omega y).
\end{equation*}

In order to solve this problem, we have to find the four unknown functions $C_1(\omega), C_2(\omega), C_3(\omega), C_4(\omega)$. We will obtain these unknowns by using the boundary conditions.

On $\Gamma_3$:
\begin{equation*}
  \sigma_y(x,h) = l(x),
\end{equation*}
or, in Fourier space,
\begin{equation*}
  \hat\sigma_y(\omega,h) =\hat l(\omega).
\end{equation*}
From this we obtain the first equation:
\begin{equation*}
  \hat\sigma_y(\omega,h) = I_{33}\sum_{j=1}^4 m_{3j}(\omega,h)
  C_j(\omega) = \hat l(\omega),
\end{equation*}
The same equation in expanded form reads
\begin{multline*}
  B_1\cosh(k_1\omega h)C_1(\omega) + B_1\sinh(k_1\omega h)
  C_2(\omega) + \\ + B_2\sinh(k_2\omega h) C_3(\omega) + B_2
  \cosh(k_2\omega h) C_4(\omega) = -i\frac{\hat l(\omega)}{E\omega}
\end{multline*}

In this work, we take for numerical computations a constant load concentrated on the segment $[-a,a]$:
\begin{equation*}
  l(x) = \left\{%
\begin{array}{ll}
    P, &  |x|<a\\
    0, & \hbox{otherwise.} \\
\end{array}%
\right.
\end{equation*}
The Fourier transform of this function is
\begin{equation*}
  \hat l(\omega) = P\frac{\sin(\omega a)}{\omega}.
\end{equation*}

Then, the first equation takes the form:
\begin{multline}\label{eq1}
  B_1\cosh(k_1\omega h)C_1(\omega) + B_1\sinh(k_1\omega h)
  C_2(\omega) + \\ + B_2\sinh(k_2\omega h) C_3(\omega) + B_2
  \cosh(k_2\omega h) C_4(\omega) = -i\frac{P\sin(\omega a)}{E\omega^2}
\end{multline}

Similarly, the second condition on $\Gamma_3$ is
$$
  \tau_{xy}(x,h)=g(x)
$$
or, in Fourier space,
\begin{equation*}
  \hat\tau_{xy} (\omega,h) = \hat
  g(\omega)
\end{equation*}
In the present work, we take $g(x) = 0$ for the numerical computations (in other words, we neglect the friction between the moving load and the ice) in order to be able to compare our results with those of other researchers. Recall that the more general case $g(x)\neq0$ can be treated here as well.

In the general case, the short hand version of the second equation is
\begin{equation*}
  \hat\tau_{xy}(\omega,h) = I_{44}\sum_{j=1}^4 m_{4j}(\omega,h)
  C_j(\omega) = \hat g(\omega),
\end{equation*}
or, in expanded form,
\begin{multline}\label{eq2}
  A_1\sinh(k_1\omega h) C_1(\omega) + A_1\cosh(k_1\omega h)
  C_2(\omega) +\\+ A_2\cosh(k_2\omega h) C_3(\omega) +
  A_2\sinh(k_2\omega h) C_4(\omega) = \hat g(\omega).
\end{multline}

The next equation is obtained from the condition on $\Gamma_2$ that there is no friction between water and ice. In fact, the generalization is not difficult. One has
$$
  \tau_{xy}(x,0) = 0 \Rightarrow \hat\tau_{xy}(\omega,0) = 0,
$$
which yields
\begin{equation*}
  I_{44}\sum_{j=1}^4 m_{4j}(\omega,0) C_j(\omega) = 0,
\end{equation*}
or, in expanded form,
\begin{equation}\label{eq3}
  A_1C_2(\omega) + A_2C_3(\omega) = 0.
\end{equation}

The last equation is slightly more difficult to obtain. On $\Gamma_2$, we have the condition
$$
  \sigma_y(x,0) = -p(x,0) + \rho_w g v(x,0),
$$
or, in Fourier space,
\begin{equation*}
  \hat\sigma_y(\omega,0) = -\hat p(\omega,0) + \rho_g g \hat
  v(\omega,0).
\end{equation*}
Above we obtained the formula (\ref{pressure}):
\begin{multline*}
  \hat p(\omega,y) = -\frac{\omega\rho_g c_0^2}{\sqrt\chi} \coth(\sqrt\chi\omega
  H)\hat v(\omega,0) (\cosh(\sqrt\chi\omega y) +\\+ \tanh(\sqrt\chi\omega
  H)\sinh(\sqrt\chi\omega y))
\end{multline*}
For $y=0$, it becomes
\begin{equation*}
  \hat p(\omega,0) = -\frac{\omega\rho_g c_0^2}{\sqrt\chi}
  \coth(\sqrt\chi\omega H)\hat v(\omega,0)
\end{equation*}
The last equation follows:
\begin{equation*}
  \hat\sigma_y(\omega,0) = \left(\frac{\rho_g\omega c_0^2}{\sqrt\chi}
  \coth(\sqrt\chi\omega H) + \rho_w g\right)\hat v(\omega,0)
\end{equation*}
We introduce the short-hand notation
\begin{equation*}
  s(\omega) = \left(\frac{\rho_g\omega c^2}{\sqrt\chi}
  \coth(\sqrt\chi\omega H) + \rho_w g\right).
\end{equation*}
Therefore, the last equation is
\begin{equation*}
  \sum_{j=1}^4(m_{3j}(\omega,0) -
  I_{22}/I_{33}s(\omega)m_{2j}(\omega,0))C_j(\omega) = 0.
\end{equation*}
or, if we write the expressions for coefficients $m_{ij}$,
\begin{equation}\label{eq4}
  B_1C_1(\omega) - \frac{s(\omega)}{E\omega k_1}C_2(\omega) -
  \frac{F s(\omega)}{E\omega k_1^2} C_3(\omega) + B_2 C_4(\omega)
  =0.
\end{equation}

We have four unknown functions and four equations (\ref{eq1}), (\ref{eq2}), (\ref{eq3}), (\ref{eq4}). Thus, solving this linear system allows us to reconstruct the solution in Fourier space. The last step is to take the inverse Fourier transform. We do it with an FFT algorithm.

The author would like to mention here that the matrix of this system is ill-conditioned for $\omega$ in the neighbourhood of zero and for large $\omega$. This problem generates several purely numerical difficulties. In our program, we distinguish two cases, $\omega \ll 1$ and $\omega\gg1$. In the first case, we multiply the last equation by $\omega$ in order to avoid the division on $\omega$ and by $\sinh(\sqrt\chi\omega H)$ because $s(\omega)$ is singular at zero\footnote{In fact $s(\omega)$ contains $\coth(\sqrt\chi\omega H)$ which tends to $\infty$ when $\omega\to 0$}. For large $\omega$, we divide the first and the second equations by $\cosh(k_1\omega h)$ in order to avoid the operations with very large numbers. As an illustration, we give here the limit of the linear system matrix when $\omega\to 0$:
\begin{equation*}
  \lim_{\omega\to 0} A(\omega) = \begin{pmatrix}
    B_1 & 0 & 0 & B_2 \\
    0 & A_1 & A_2 & 0 \\
    0 & A_1 & A_2 & 0 \\
    0 & 0 & 0 & 0 \\
  \end{pmatrix}.
\end{equation*}

Appendix \ref{app3} shows the results of the numerical computations for different values of ice thickness and moving load velocity (Figures \ref{c01:fig} - \ref{c250:fig}). The program is given in Appendix \ref{app2}. The values of physical parameters can be found in this table:

\begin{center}
\begin{tabular}{||c|c||}
  \hline  \hline
  Moving block length & $2a = 2 m$ \\
  \hline
  Ice thickness & $h = 1.5 m$ \\
  \hline
  Water depth & $H = 5.0 m$ \\
  \hline
  Poisson ratio & $\nu = 0.33$ \\
  \hline
  Young's modulus & $E = 9.5 Gpa$ \\
  \hline
  The standard gravitational acceleration & $g = 9.80665 m/s^2$\\
  \hline
  Block velocity & $c_0 = 15.0 m/s$ \\
  \hline
  Sound velocity in the water & $\gamma = 1500 m/s$ \\
  \hline
  Ice density & $\rho_g = 926 kg/m^3$ \\
  \hline
  Water density & $\rho_w = 1027 kg/m^3$ \\
  \hline
  Block load & $F = 17 500 Pa$ \\
  \hline
  \hline
\end{tabular}
\end{center}

\subsection{Gravitation influence on a homogeneous layer deformation}

In this section, we are going to investigate the influence of the gravitational force on a homogeneous layer deformation.

Let us consider the governing equations. In the ice layer, we have Lam\'e's equations with a gravitational term
\begin{equation}\label{lamgrav1}
  (\lambda+2\mu)\pd{^2u}{x^2} + \mu\pd{^2u}{y^2} +
  (\lambda+\mu)\pd{^2v}{x\partial y} = \rho_g \pd{^2u}{t^2}
\end{equation}
\begin{equation}\label{lamgrav2}
  \mu\pd{^2v}{x^2} + (\lambda+2\mu)\pd{^2v}{y^2} +
  (\lambda+\mu)\pd{^2u}{x\partial y} = \rho_g\pd{^2v}{t^2}
  + \rho_g g
\end{equation}
In the water domain, fortunately, we have the same equation
\begin{equation*}
    \frac{1}{\gamma^2}\pd{^2p}{t^2} = \Delta p.
\end{equation*}
In fact, this is due to the fact that we write the equation (\ref{linacoustic}) for the perturbations of the steady flow, and this steady flow already contains the gravitational term. The boundary conditions are the same, too.

Recall that we could not apply the integral transforms method directly to the equations (\ref{lamgrav1}) (\ref{lamgrav2}) because they contain the term $\rho_g g$, which is not $L_2(-\infty,+\infty)$ integrable.

In order to avoid this technical difficulty, we apply the method of solution superposition. In other words, we use a very strong property of over problem  - the linearity.

We represent the solution of the equations (\ref{lamgrav1}), (\ref{lamgrav2}) in the following form
\begin{equation*}
    u(x,y) = u_1(x,y) + u_2(x,y),
\end{equation*}
\begin{equation*}
    v(x,y) = v_1(x,y) + v_2(x,y).
\end{equation*}

The displacements field $(u_1,v_1)$ satisfies the homogeneous equations but inhomogeneous boundary conditions
\begin{equation*}
  (\lambda+2\mu)\pd{^2u_1}{x^2} + \mu\pd{^2u_1}{y^2} +
  (\lambda+\mu)\pd{^2v_1}{x\partial y} = \rho_g \pd{^2u_1}{t^2},
\end{equation*}
\begin{equation*}
\mu\pd{^2v_1}{x^2} + (\lambda+2\mu)\pd{^2v_1}{y^2} +
  (\lambda+\mu)\pd{^2u_1}{x\partial y} = \rho_g\pd{^2v_1}{t^2}.
\end{equation*}
The boundary conditions are
\begin{equation*}
  \sigma_y^1(x,h) = l(x),
\end{equation*}
\begin{equation*}
  \tau_{xy}^1(x,h) = f(x),
\end{equation*}
\begin{equation*}
  \sigma_y^1(x,0) = -p(x,0) + \rho_w g v_1(x,0),
\end{equation*}
\begin{equation*}
  \tau_{xy}^1(x,0) = 0.
\end{equation*}

It is clear that we have already solved this problem in the previous section.

The displacements field $(u_2,v_2)$ satisfies inhomogeneous equations and homogeneous boundary conditions
\begin{equation*}
  (\lambda+2\mu)\pd{^2u_2}{x^2} + \mu\pd{^2u_2}{y^2} +
  (\lambda+\mu)\pd{^2v_2}{x\partial y} = \rho_g \pd{^2u_2}{t^2},
\end{equation*}
\begin{equation*}
\mu\pd{^2v_2}{x^2} + (\lambda+2\mu)\pd{^2v_2}{y^2} +
  (\lambda+\mu)\pd{^2u_2}{x\partial y} = \rho_g\pd{^2v_2}{t^2} + \rho_g g.
\end{equation*}
The boundary conditions are
\begin{equation}\label{cond1}
  \sigma_y^2(x,h) = 0,
\end{equation}
\begin{equation}\label{cond2}
  \tau_{xy}^2(x,h) = 0,
\end{equation}
\begin{equation}\label{cond3}
  \sigma_y^2(x,0) = 0,
\end{equation}
\begin{equation}\label{cond4}
  \tau_{xy}^2(x,0) = 0.
\end{equation}

In this section, we are going to determine $(u_2,v_2)$ from these equations in order to estimate the gravity influence on the layer deformation under the moving load. It is clear now that $(u_1+u_2,v_1+v_2)$ satisfies initial problem (\ref{lamgrav1}) (\ref{lamgrav2}). Now, we would like to make several physical assumptions that will considerably simplify our problem.

First of all, $u_2\equiv 0$ since the gravitation compresses only our elastic layer. Then, it is not difficult to see that $v_2(x,y,t) \equiv v_2(y,t)$. We can explain it by the symmetry of our problem. The fact that the gravitational field is constant yields that we have a steady solution $v_2 = v_2(y)$.

If we take into account all assumptions made above, we will have only an ordinary differential equation
\begin{equation*}
  (\lambda+2\mu) \od{^2v_2}{y^2} = \rho_g g,
\end{equation*}
or, if we recall that $c_1 = \sqrt{\frac{\lambda+2\mu}{\rho_g}}$, we obtain
\begin{equation*}
  \od{^2v_2}{y^2} = \frac{g}{c_1^2}.
\end{equation*}
Its solution is
\begin{equation*}
  v_2(y) = \frac{g}{2c_1^2} y^2 + Ay + B.
\end{equation*}

So, we have four boundary conditions (\ref{cond1}), (\ref{cond2}), (\ref{cond3}), (\ref{cond4}) and two unknown constants $A$ and $B$. Fortunately, there is no contradiction here because the boundary conditions (\ref{cond2}) (\ref{cond4}) are satisfied identically
\begin{equation*}
  \tau_{xy}^2(x,y) = \mu\Bigl(
    \pd{u_2}{y} + \pd{v_2}{x}
  \Bigr) \equiv 0.
\end{equation*}

We use the condition (\ref{cond1}) to determine $A$:
\begin{equation*}
  \sigma_y^2(x,y) = \lambda\pd{u_2}{x} + (\lambda+2\mu)\pd{v_2}{y} =
  (\lambda+2\mu)\Bigl(\frac{g}{c_1^2}y + A\Bigr).
\end{equation*}
For $y=h$ it becomes
\begin{equation*}
  \sigma_y^2(x,h) =
  (\lambda+2\mu)\Bigl(\frac{g}{c_1^2}h + A\Bigr)
\end{equation*}
which yields
\begin{equation*}
  A = -\frac{g}{c_1^2} h.
\end{equation*}

Unfortunately, we cannot satisfy at the same time the condition (\ref{cond3}), and we have to determine one more unknown constant $B$. That is why we replace the condition (\ref{cond3}) with another
\begin{equation}\label{conditionbidon}
  v_2(0) = 0.
\end{equation}
The physical meaning of this condition is that the ice-water interface is motionless. From (\ref{conditionbidon}) one can easily find
\begin{equation*}
    B = 0.
\end{equation*}

Thus, we have the solution
\begin{eqnarray*}
  v_2(y) &=& \frac{g}{2c_1^2}y(y-2h), \qquad y\in [0,h], \\
  u_2(y) &=& 0.
\end{eqnarray*}

We would like to analyze the obtained result. In order to estimate the gravitation influence on the layer deformation, we need to estimate the function $v_2(y)$
\begin{equation*}
  |v_2(y)| \leq |v_2(h)| = \frac{g}{2c_1^2} h^2 \approx
  3.125\times10^{-7}.
\end{equation*}

This estimation is true because the function $v_2(y)$ is obviously monotonous.

We conclude that the gravity effects are important since the solution $(u_2,v_2)$ is of the same order as the solution $(u_1, v_1)$ obtained in the previous section.

\section{Multilayer generalization and Lamzyuk-Privarnikov's functions method}

In this section, we consider almost the same problem, but a multilayer pack replaces the ice layer. The layers in this pack differ by their thickness and mechanical properties (density, Young's modulus, Poisson ratio). But they are constant in each layer. Within each layer, we attach a local coordinate system $x'O'y'$. This problem can be considered as an approximation of real ice layers with properties depending on the depth. Figure \ref{pict2} provides an illustration of this situation.

\begin{figure}[htbp]
\begin{center}
\unitlength 1mm
\begin{picture}(120.00,115.00)(0,0)

\linethickness{0.70mm}
\put(0.00,85.00){\line(1,0){115.00}}

\linethickness{0.70mm}
\put(0.00,35.00){\line(1,0){115.00}}

\linethickness{0.25mm}
\put(40.00,35.00){\line(1,0){85.00}}
\put(125.00,35.00){\vector(1,0){0.12}}

\put(125.00,30.00){\makebox(0,0)[cc]{$x$}}

\linethickness{0.25mm}
\put(55.00,35.00){\line(0,1){70.00}}
\put(55.00,105.00){\vector(0,1){0.12}}

\linethickness{0.25mm}
\put(65.00,90.00){\line(1,0){10.00}}
\put(75.00,90.00){\vector(1,0){0.12}}

\put(80.00,95.00){\makebox(0,0)[cc]{$c_0$}}

\linethickness{0.70mm}
\put(0.00,10.00){\line(1,0){115.00}}

\linethickness{0.20mm}
\multiput(0.00,5.00)(0.12,0.12){42}{\line(1,0){0.12}}

\linethickness{0.20mm}
\multiput(10.00,5.00)(0.12,0.12){42}{\line(1,0){0.12}}

\linethickness{0.20mm}
\multiput(15.00,5.00)(0.12,0.12){42}{\line(1,0){0.12}}

\linethickness{0.20mm}
\multiput(5.00,5.00)(0.12,0.12){42}{\line(1,0){0.12}}

\linethickness{0.20mm}
\multiput(20.00,5.00)(0.12,0.12){42}{\line(1,0){0.12}}

\linethickness{0.20mm}
\multiput(25.00,5.00)(0.12,0.12){42}{\line(1,0){0.12}}

\linethickness{0.20mm}
\multiput(30.00,5.00)(0.12,0.12){42}{\line(1,0){0.12}}

\linethickness{0.20mm}
\multiput(35.00,5.00)(0.12,0.12){42}{\line(1,0){0.12}}

\linethickness{0.20mm}
\multiput(40.00,5.00)(0.12,0.12){42}{\line(1,0){0.12}}

\linethickness{0.20mm}
\multiput(45.00,5.00)(0.12,0.12){42}{\line(1,0){0.12}}

\linethickness{0.20mm}
\multiput(50.00,5.00)(0.12,0.12){42}{\line(1,0){0.12}}

\linethickness{0.20mm}
\multiput(55.00,5.00)(0.12,0.12){42}{\line(1,0){0.12}}

\linethickness{0.20mm}
\multiput(60.00,5.00)(0.12,0.12){42}{\line(1,0){0.12}}

\linethickness{0.20mm}
\multiput(65.00,5.00)(0.12,0.12){42}{\line(1,0){0.12}}

\linethickness{0.20mm}
\multiput(70.00,5.00)(0.12,0.12){42}{\line(1,0){0.12}}

\linethickness{0.20mm}
\multiput(75.00,5.00)(0.12,0.12){42}{\line(1,0){0.12}}

\linethickness{0.20mm}
\multiput(80.00,5.00)(0.12,0.12){42}{\line(1,0){0.12}}

\linethickness{0.20mm}
\multiput(85.00,5.00)(0.12,0.12){42}{\line(1,0){0.12}}

\linethickness{0.20mm}
\multiput(90.00,5.00)(0.12,0.12){42}{\line(1,0){0.12}}

\linethickness{0.20mm}
\multiput(95.00,5.00)(0.12,0.12){42}{\line(1,0){0.12}}

\linethickness{0.20mm}
\multiput(100.00,5.00)(0.12,0.12){42}{\line(1,0){0.12}}

\linethickness{0.20mm}
\multiput(105.00,5.00)(0.12,0.12){42}{\line(1,0){0.12}}

\linethickness{0.20mm}
\multiput(110.00,5.00)(0.12,0.12){42}{\line(1,0){0.12}}

\linethickness{0.20mm}
\put(5.00,25.00){\line(1,0){10.00}}

\linethickness{0.20mm}
\put(20.00,30.00){\line(1,0){10.00}}

\linethickness{0.20mm}
\put(10.00,15.00){\line(1,0){10.00}}

\linethickness{0.20mm}
\put(55.00,20.00){\line(1,0){10.00}}

\linethickness{0.20mm}
\put(25.00,25.00){\line(1,0){10.00}}

\linethickness{0.20mm}
\put(10.00,20.00){\line(1,0){10.00}}

\linethickness{0.20mm}
\put(55.00,25.00){\line(1,0){10.00}}

\linethickness{0.20mm}
\put(40.00,20.00){\line(1,0){10.00}}

\linethickness{0.20mm}
\put(70.00,30.00){\line(1,0){10.00}}

\linethickness{0.20mm}
\put(70.00,15.00){\line(1,0){10.00}}

\linethickness{0.20mm}
\put(90.00,30.00){\line(1,0){10.00}}

\linethickness{0.20mm}
\put(105.00,15.00){\line(1,0){10.00}}

\linethickness{0.20mm}
\put(90.00,20.00){\line(1,0){10.00}}

\linethickness{0.20mm}
\put(0.00,75.00){\line(1,0){115.00}}

\linethickness{0.20mm}
\put(0.00,65.00){\line(1,0){115.00}}

\linethickness{0.20mm}
\put(0.00,55.00){\line(1,0){115.00}}

\linethickness{0.20mm}
\put(0.00,45.00){\line(1,0){115.00}}

\linethickness{0.20mm}
\put(95.00,55.00){\line(1,0){25.00}}
\put(120.00,55.00){\vector(1,0){0.12}}

\put(125.00,55.00){\makebox(0,0)[cc]{$x'$}}

\put(130.00,65.00){\makebox(0,0)[cc]{}}

\linethickness{0.20mm}
\put(55.00,55.00){\line(0,1){10.00}}
\put(55.00,65.00){\vector(0,1){0.12}}

\put(50.00,60.00){\makebox(0,0)[cc]{$y'$}}

\put(55.00,30.00){\makebox(0,0)[cc]{$O$}}

\put(50.00,50.00){\makebox(0,0)[cc]{$O'$}}

\linethickness{1.00mm}
\put(55.00,55.00){\circle{0.00}}

\put(80.00,80.00){\makebox(0,0)[cc]{$h_1,\rho_1,\nu_1,E_1$}}

\linethickness{0.20mm}
\put(45.00,95.00){\line(0,1){5.00}}
\put(45.00,95.00){\vector(0,-1){0.12}}

\linethickness{0.20mm}
\put(50.00,95.00){\line(0,1){10.00}}
\put(50.00,95.00){\vector(0,-1){0.12}}

\linethickness{0.20mm}
\put(60.00,95.00){\line(0,1){10.00}}
\put(60.00,95.00){\vector(0,-1){0.12}}

\linethickness{0.20mm}
\put(65.00,95.00){\line(0,1){5.00}}
\put(65.00,95.00){\vector(0,-1){0.12}}

\put(80.00,70.00){\makebox(0,0)[cc]{$h_2,\rho_2,\nu_2,E_2$}}

\put(85.00,60.00){\makebox(0,0)[cc]{$\cdots$}}

\put(90.00,50.00){\makebox(0,0)[cc]{$h_{n-1},\rho_{n-1},\nu_{n-1},E_{n-1}$}}

\put(80.00,40.00){\makebox(0,0)[cc]{$h_n,\rho_n,\nu_n,E_n$}}

\linethickness{0.70mm}
\put(45.00,85.00){\line(1,0){20.00}}
\put(45.00,85.00){\line(0,1){10.00}}
\put(65.00,85.00){\line(0,1){10.00}}
\put(45.00,95.00){\line(1,0){20.00}}

\put(55.00,110.00){\makebox(0,0)[cc]{$y$}}

\put(50.00,115.00){\makebox(0,0)[cc]{}}

\put(15.00,80.00){\makebox(0,0)[cc]{$1$}}

\put(15.00,70.00){\makebox(0,0)[cc]{$2$}}

\put(15.00,60.00){\makebox(0,0)[cc]{$\cdots$}}

\put(15.00,50.00){\makebox(0,0)[cc]{$n-1$}}

\put(15.00,40.00){\makebox(0,0)[cc]{$n$}}

\end{picture}
\end{center}
  \caption{Moving load on the multilayer plate}\label{pict2}
\end{figure}

Such multilayer approximations are also used to model road asphalt cover. In practice, the engineers usually take three layers. Very seldom five. We will use the results of this section below to obtain the solution for an inhomogeneous layer problem.

Let $\overrightarrow{w}(\omega,y) = ^t(\hat{u}(\omega,y), \hat{v}(\omega,y), \hat{\sigma_y}(\omega,y), \hat{\tau_{xy}}(\omega,y))$. In the previous section, we obtained the expression (\ref{imc}):
$$
  \overrightarrow{w}(\omega,y) = I M(\omega,y) \overrightarrow{C}(\omega)
$$
where the matrices $M$ and $I$ were defined earlier.

We introduce a new vector $\overrightarrow{\alpha}(\omega)$ which has a physical meaning:
\begin{equation*}
  \overrightarrow{w}(\omega,0) = I \overrightarrow{\alpha}(\omega).
\end{equation*}

The matrix $M(\omega,0)$ is not singular. It can be easily seen from its expression:
\begin{equation*}
M(\omega,0) = \begin{pmatrix}
  1 & 0 & 0 & 1 \\
  0 & k_1 & k_2k_1^2 & 0 \\
  B_1 & 0 & 0 & B_2 \\
  0 & A_1 & 2k_2 & 0 \\
\end{pmatrix}
\end{equation*}
After simplification, one obtains
$$
  \det M(\omega,0) = \frac{k_1k_2(1-k_1^2)(1-k_2^2)}{1-2\nu} \neq 0.
$$
One can easily find the connection between the vector
$\overrightarrow{\alpha}$ and $\overrightarrow{C}$:
$$
  \overrightarrow{C}(\omega) = M^{-1}(\omega,0)
  \overrightarrow{\alpha}(\omega).
$$
This formula indicates that we can reduce our problem to that of finding the vector $\overrightarrow{\alpha}$, which is more convenient and natural.

Suppose we have found $\overrightarrow{\alpha}$. Then, we can find the vector $\overrightarrow{w}(\omega,y)$ using this relation:
\begin{equation*}
  \overrightarrow{w}(\omega,y) = I M(\omega,y)
  \overrightarrow{C}(\omega) = I H(\omega, y)
  \overrightarrow{\alpha}(\omega)
\end{equation*}
with $H(\omega,y):= M(\omega,y)M^{-1}(\omega,0) = (h_{ij}(\omega,y))_{1\leq i,j\leq 4}$. It is easy to find explicitly the elements of this matrix\footnote{Below $A$ means $A_1$}:
$$
  h_{11}(\omega,y) = \frac{B_1\cosh(k_2\omega y)- B_2\cosh(k_1\omega
  y)}{B_1-B_2},
$$
$$
  h_{12}(\omega,y) = \frac{2k_2\sinh(k_1\omega y) -
  A\sinh(k_2\omega y)}{k_1k_2(2-Ak_1)},
$$
$$
  h_{13}(\omega,y) = \frac{\cosh(k_1\omega y) - \cosh(k_2\omega
  y)}{B_1-B_2},
$$
$$
  h_{14}(\omega,y) = \frac{\sinh(k_2\omega y) -
  k_1k_2\sinh(k_1\omega y)}{k_2(2-Ak_1)},
$$
$$
  h_{21}(\omega,y) = \frac{B_1k_1^2k_2\sinh(k_2\omega y) -
  B_2k_1\sinh(k_1\omega y)}{B_1-B_2},
$$
$$
  h_{22}(\omega,y) = \frac{2\cosh(k_1\omega y) -Ak_1 \cosh(k_2\omega
  y)}{2-Ak_1},
$$
$$
  h_{23}(\omega,y) = k_1\frac{\sinh(k_1\omega y) -
  k_1k_2\sinh(k_2\omega y)}{B_1-B_2},
$$
$$
  h_{24}(\omega,y) = k_1^2\frac{\cosh(k_2\omega y) -
  \cosh(k_1\omega y)}{2-Ak_1},
$$
$$
  h_{31}(\omega,y) = B_1B_2\frac{\cosh(k_2\omega y) -
  \cosh(k_1\omega y)}{B_1-B_2},
$$
$$
  h_{32}(\omega,y) = \frac{2B_1k_2\sinh(k_1\omega y) -
  AB_2\sinh(k_2\omega y)}{k_1k_2(2-Ak_1)},
$$
$$
  h_{33}(\omega,y) = \frac{B_1\cosh(k_1\omega y) -
  B_2\cosh(k_2\omega y)}{B_1-B_2},
$$
$$
  h_{34}(\omega,y) = \frac{B_2\sinh(k_2\omega y) -
  B_1k_1k_2\sinh(k_1\omega y)}{k_2(2-Ak_1)},
$$
$$
  h_{41}(\omega,y) = \frac{2k_2B_1\sinh(k_2\omega y) -
  AB_2\sinh(k_1\omega y)}{B_1-B_2},
$$
$$
  h_{42}(\omega,y) = 2A\frac{\cosh(k_1\omega y) -
  \cosh(k_2\omega y)}{k_1(2-Ak_1)},
$$
$$
  h_{43}(\omega,y) = \frac{A\sinh(k_1\omega y) -
  2k_2\sinh(k_2\omega y)}{B_1-B_2},
$$
$$
  h_{44}(\omega,y) = \frac{2\cosh(k_2\omega y) -
  Ak_1\cosh(k_1\omega y)}{2-Ak_1}.
$$

As we established earlier, the solution to our problem is completely determined by the four functions $\overrightarrow{\alpha}(\omega) = ^t( \alpha(\omega),\beta(\omega),\gamma(\omega),\delta(\omega))$. In the multilayer pack with $n$ layers, we associate to each layer number $j$ its vector $\overrightarrow{\alpha}_j(\omega)$. So the solution is determined by $n$ vectors $\{\overrightarrow{\alpha}_j (\omega)\}_{j=1}^n$.

We assume that the layers do not detach during deformation. Thus, at the interface between two layers, we have the natural conditions of continuity:
\begin{eqnarray*}
  u^{(j)}(x,-h_j) &=& u^{(j+1)}(x,0), \\
  v^{(j)}(x,-h_j) &=& v^{(j+1)}(x,0), \\
  \sigma_y^{(j)}(x,-h_j) &=& \sigma_y^{(j+1)}(x,0) \\
  \tau_{xy}^{(j)}(x,-h_j) &=& \tau_{xy}^{(j+1)}(x,0)
\end{eqnarray*}
Here $h_j$ denotes the thickness $j^{th}$ layer. These conditions are written in a local coordinate system associated with each layer. Now we take Fourier transforms of these equalities and write them in vector form:
\begin{equation*}
  \overrightarrow{w}_j(-h_j) = \overrightarrow{w}_{j+1}(0).
\end{equation*}
We prefer to rewrite this identity in terms of $\overrightarrow{\alpha}$:
\begin{equation*}
  I_{j+1}H_{j+1}(\omega,0)\overrightarrow{\alpha}_{j+1}(\omega) =
  I_j H_j(\omega, -h_j) \overrightarrow{\alpha}_j(\omega).
\end{equation*}
Note that $H(\omega,0) = M(\omega,0)M^{-1} (\omega,0) = Id$. Therefore
\begin{equation}\label{recrelalpha}
  \boxed{
  \overrightarrow{\alpha}_{j+1}(\omega) = I_{j+1}^{-1}I_j
  H_j(\omega,-h_j)\overrightarrow{\alpha}_j(\omega)}
\end{equation}

If we knew the vector $\overrightarrow{\alpha}_1(\omega)$, we could find all the others recurrence (\ref{recrelalpha}). The problem is that the boundary conditions give us only two components of the vector $\overrightarrow{\alpha}_1$ and two components of $\overrightarrow{\alpha}_n$. Thus, we have some kind of discrete boundary problem.

With the recurrence relation (\ref{recrelalpha}) we can find the connection between $\overrightarrow{\alpha}_1(\omega)$ and $\overrightarrow{\alpha}_n(\omega)$:
\begin{multline*}
  \overrightarrow{\alpha}_n(\omega) = I_n^{-1}I_{n-1}H_{n-1}
  (\omega,-h_{n-1}) \overrightarrow{\alpha}_{n-1}(\omega) =\\=
  I_n^{-1}I_{n-1}H_{n-1}(\omega,-h_{n-1})\left(
  I_{n-1}^{-1}I_{n-2}H_{n-2}(\omega,-h_{n-2})
  \overrightarrow{\alpha}_{n-2}(\omega)\right) = \ldots\\ =
  \prod_{k=1}^{n-1} I_{n-k+1}^{-1}I_{n-k}H_{n-k}(\omega,-h_{n-k})
  \overrightarrow{\alpha}_1(\omega).
\end{multline*}
Let $$D = (d_{ij})_{1\le i,j\le 4} := \prod_{k=1}^{n-1} I_{n-k+1}^{-1}I_{n-k}H_{n-k}(\omega,-h_{n-k}) \in Mat_{4\times4}(\mathbb{C}).$$ We have
\begin{equation*}
  \overrightarrow{\alpha}_n(\omega) =
  D(\omega,h_1,h_2,\ldots,h_{n-1})\overrightarrow{\alpha}_1(\omega).
\end{equation*}

In Fourier space at the interface $\Gamma_2$, we have the boundary conditions
\begin{eqnarray*}
  \hat\tau_{xy}^{(n)} &=& 0, \\
  \hat\sigma_y^{(n)} &=& -\hat p(\omega,0) + \rho_g g \hat v(\omega,0) =
  s(\omega)\hat v(\omega,0) = s(\omega)\hat v^{(n)}.
\end{eqnarray*}
By definition $\overrightarrow{w}(\omega,0) = \overrightarrow{w}_n(\omega) = I_n\overrightarrow{\alpha}_n(\omega)$. So, we can obtain the first equation:
\begin{multline*}
  \hat\tau_{xy}^{(n)} = \mu_n\omega\delta_n(\omega) = \mu_n\omega
  \bigl( d_{41}(\omega,h_1,h_2,\ldots,h_{n-1})\alpha_1(\omega) +\\+
  d_{42}(\omega,h_1,h_2,\ldots,h_{n-1})\beta_1(\omega) +
  d_{43}(\omega,h_1,h_2,\ldots,h_{n-1})\gamma_1(\omega) +\\+
  d_{44}(\omega,h_1,h_2,\ldots,h_{n-1})\delta_1(\omega)\bigr) = 0,
\end{multline*}
or in equivalent form:
\begin{multline}\label{d1}
  d_{41}(\omega,h_1,h_2,\ldots,h_{n-1})\alpha_1(\omega) +
  d_{42}(\omega,h_1,h_2,\ldots,h_{n-1})\beta_1(\omega) +\\+
  d_{43}(\omega,h_1,h_2,\ldots,h_{n-1})\gamma_1(\omega) +
  d_{44}(\omega,h_1,h_2,\ldots,h_{n-1})\delta_1(\omega) = 0.
\end{multline}

Similarly we have the condition $\hat\sigma_{y}^{(n)} = s(\omega) \hat v^{(n)}$ that gives us one more equation:
\begin{multline*}
  iE_n\omega\bigl(
  d_{31}(\omega,h_1,h_2,\ldots,h_{n-1})\alpha_1(\omega) +
  d_{32}(\omega,h_1,h_2,\ldots,h_{n-1})\beta_1(\omega) +\\+
  d_{33}(\omega,h_1,h_2,\ldots,h_{n-1})\gamma_1(\omega) +
  d_{34}(\omega,h_1,h_2,\ldots,h_{n-1})\delta_1(\omega)\bigr) =\\=
  s(\omega)\frac{i}{k_{1n}^2}\bigl(
  d_{21}(\omega,h_1,h_2,\ldots,h_{n-1})\alpha_1(\omega) +\\+
  d_{22}(\omega,h_1,h_2,\ldots,h_{n-1})\beta_1(\omega) +
  d_{23}(\omega,h_1,h_2,\ldots,h_{n-1})\gamma_1(\omega) +\\+
  d_{24}(\omega,h_1,h_2,\ldots,h_{n-1})\delta_1(\omega)\bigr)
\end{multline*}
The second equation can be rewritten as
\begin{multline}\label{d2}
 \Bigl(E_n\omega d_{31} - \frac{s(\omega)}{k_{1n}^2} d_{21}\Bigr)
 \alpha_1(\omega) +
 \Bigl(E_n\omega d_{32} - \frac{s(\omega)}{k_{1n}^2} d_{22}\Bigr)
 \beta_1(\omega) +\\+
 \Bigl(E_n\omega d_{33} - \frac{s(\omega)}{k_{1n}^2} d_{23}\Bigr)
 \gamma_1(\omega) +
 \Bigl(E_n\omega d_{34} - \frac{s(\omega)}{k_{1n}^2} d_{24}\Bigr)
 \delta_1(\omega) = 0
\end{multline}

If we put together equations (\ref{d1}) and (\ref{d2}), we have a system of two linear equations with four unknowns $\alpha_1$, $\beta_1$, $\gamma_1$ and $\delta_1$. In order to obtain the full problem, we also use the two boundary conditions on the interface $\Gamma_3$:
\begin{eqnarray*}
  \hat\sigma_y(\omega,h) &=& \hat l(\omega) =
  iE\omega\gamma_1(\omega) \\
  \hat\tau_{xy}(\omega,h) &=& \hat g(\omega) =
  \mu\omega\delta_1(\omega).
\end{eqnarray*}
In other words we know $\gamma_1(\omega)$ and $\delta_1(\omega)$. After putting these values in the system (\ref{d1}), (\ref{d2}) we will be able to determine the two unknown functions $\alpha_1(\omega)$ and $\beta_1(\omega)$. So, we will have the full vector $\overrightarrow{\alpha}_1(\omega)$ and with the recurrence relation (\ref{recrelalpha}) we can find the deformations of each layer in our multilayer pack.

It would be great if we could calculate the elements $d_{ij}$, but in practice, it is almost impossible\footnote{We have to calculate the products of matrices. Analytically, it is awkward. Numerically, it is not interesting.} and we will use a less obvious way.

We repeat one more time that we have the two equations (\ref{d1}), (\ref{d2}) with unknowns $\alpha_1(\omega)$ and $\beta_1(\omega)$. Suppose we have solved these equations:
\begin{eqnarray*}
  \alpha_1(\omega) &=& -A_n(\omega,h_1,\ldots,h_{n-1})
  \gamma_1(\omega) - C_n(\omega,h_1,\ldots,h_{n-1}) \delta_1(\omega)
  \\
  \beta_1(\omega) &=& -B_n(\omega,h_1,\ldots,h_{n-1})
  \gamma_1(\omega) - D_n(\omega,h_1,\ldots,h_{n-1}) \delta_1(\omega)
\end{eqnarray*}
The functions $A_n$, $B_n$, $C_n$, and $D_n$ are named Lamzyuk-Privarnikov functions because this method was proposed by A.K. Privarnikov and developed later by V. D. Lamzyuk.\footnote{Many thanks to my teacher V.D.~Lamzyuk who learned me this technic and not only this.} Our main problem is to obtain these four functions. We construct recurrence relations for them. Let us assume that we have constructed these functions for a multilayer pack with $(n-1)$ layers. The numbering of layers starts from $2$ to $n$:

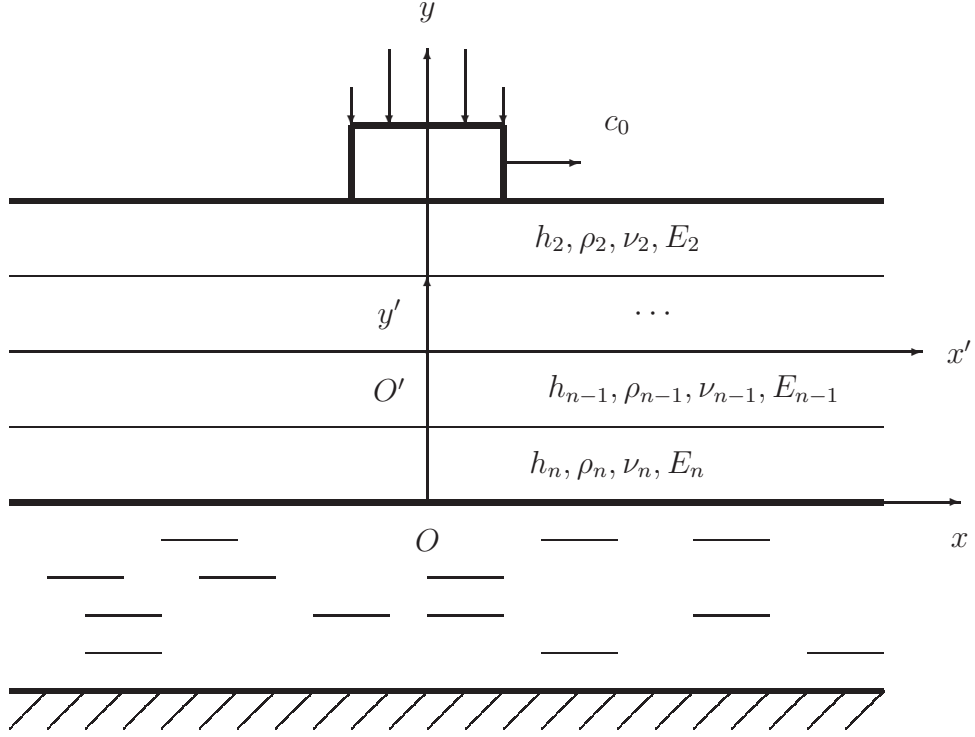
\begin{figure}[htbp]
\begin{center}
\unitlength 1mm
\begin{picture}(120.00,95.00)(0,0)

\linethickness{0.70mm}
\put(0.00,70.00){\line(1,0){115.00}}

\linethickness{0.70mm}
\put(0.00,30.00){\line(1,0){115.00}}

\linethickness{0.25mm}
\put(40.00,30.00){\line(1,0){85.00}}
\put(125.00,30.00){\vector(1,0){0.12}}

\put(125.00,25.00){\makebox(0,0)[cc]{$x$}}

\linethickness{0.25mm}
\put(55.00,30.00){\line(0,1){60.00}}
\put(55.00,90.00){\vector(0,1){0.12}}

\linethickness{0.25mm}
\put(65.00,75.00){\line(1,0){10.00}}
\put(75.00,75.00){\vector(1,0){0.12}}

\put(80.00,80.00){\makebox(0,0)[cc]{$c_0$}}

\linethickness{0.70mm}
\put(0.00,5.00){\line(1,0){115.00}}

\linethickness{0.20mm}
\multiput(0.00,0.00)(0.12,0.12){42}{\line(1,0){0.12}}

\linethickness{0.20mm}
\multiput(10.00,0.00)(0.12,0.12){42}{\line(1,0){0.12}}

\linethickness{0.20mm}
\multiput(15.00,0.00)(0.12,0.12){42}{\line(1,0){0.12}}

\linethickness{0.20mm}
\multiput(5.00,0.00)(0.12,0.12){42}{\line(1,0){0.12}}

\linethickness{0.20mm}
\multiput(20.00,0.00)(0.12,0.12){42}{\line(1,0){0.12}}

\linethickness{0.20mm}
\multiput(25.00,0.00)(0.12,0.12){42}{\line(1,0){0.12}}

\linethickness{0.20mm}
\multiput(30.00,0.00)(0.12,0.12){42}{\line(1,0){0.12}}

\linethickness{0.20mm}
\multiput(35.00,0.00)(0.12,0.12){42}{\line(1,0){0.12}}

\linethickness{0.20mm}
\multiput(40.00,0.00)(0.12,0.12){42}{\line(1,0){0.12}}

\linethickness{0.20mm}
\multiput(45.00,0.00)(0.12,0.12){42}{\line(1,0){0.12}}

\linethickness{0.20mm}
\multiput(50.00,0.00)(0.12,0.12){42}{\line(1,0){0.12}}

\linethickness{0.20mm}
\multiput(55.00,0.00)(0.12,0.12){42}{\line(1,0){0.12}}

\linethickness{0.20mm}
\multiput(60.00,0.00)(0.12,0.12){42}{\line(1,0){0.12}}

\linethickness{0.20mm}
\multiput(65.00,0.00)(0.12,0.12){42}{\line(1,0){0.12}}

\linethickness{0.20mm}
\multiput(70.00,0.00)(0.12,0.12){42}{\line(1,0){0.12}}

\linethickness{0.20mm}
\multiput(75.00,0.00)(0.12,0.12){42}{\line(1,0){0.12}}

\linethickness{0.20mm}
\multiput(80.00,0.00)(0.12,0.12){42}{\line(1,0){0.12}}

\linethickness{0.20mm}
\multiput(85.00,0.00)(0.12,0.12){42}{\line(1,0){0.12}}

\linethickness{0.20mm}
\multiput(90.00,0.00)(0.12,0.12){42}{\line(1,0){0.12}}

\linethickness{0.20mm}
\multiput(95.00,0.00)(0.12,0.12){42}{\line(1,0){0.12}}

\linethickness{0.20mm}
\multiput(100.00,0.00)(0.12,0.12){42}{\line(1,0){0.12}}

\linethickness{0.20mm}
\multiput(105.00,0.00)(0.12,0.12){42}{\line(1,0){0.12}}

\linethickness{0.20mm}
\multiput(110.00,0.00)(0.12,0.12){42}{\line(1,0){0.12}}

\linethickness{0.20mm}
\put(5.00,20.00){\line(1,0){10.00}}

\linethickness{0.20mm}
\put(20.00,25.00){\line(1,0){10.00}}

\linethickness{0.20mm}
\put(10.00,10.00){\line(1,0){10.00}}

\linethickness{0.20mm}
\put(55.00,15.00){\line(1,0){10.00}}

\linethickness{0.20mm}
\put(25.00,20.00){\line(1,0){10.00}}

\linethickness{0.20mm}
\put(10.00,15.00){\line(1,0){10.00}}

\linethickness{0.20mm}
\put(55.00,20.00){\line(1,0){10.00}}

\linethickness{0.20mm}
\put(40.00,15.00){\line(1,0){10.00}}

\linethickness{0.20mm}
\put(70.00,25.00){\line(1,0){10.00}}

\linethickness{0.20mm}
\put(70.00,10.00){\line(1,0){10.00}}

\linethickness{0.20mm}
\put(90.00,25.00){\line(1,0){10.00}}

\linethickness{0.20mm}
\put(105.00,10.00){\line(1,0){10.00}}

\linethickness{0.20mm}
\put(90.00,15.00){\line(1,0){10.00}}

\linethickness{0.20mm}
\put(0.00,60.00){\line(1,0){115.00}}

\linethickness{0.20mm}
\put(0.00,50.00){\line(1,0){115.00}}

\linethickness{0.20mm}
\put(0.00,40.00){\line(1,0){115.00}}

\linethickness{0.20mm}
\put(95.00,50.00){\line(1,0){25.00}}
\put(120.00,50.00){\vector(1,0){0.12}}

\put(125.00,50.00){\makebox(0,0)[cc]{$x'$}}

\put(130.00,60.00){\makebox(0,0)[cc]{}}

\linethickness{0.20mm}
\put(55.00,50.00){\line(0,1){10.00}}
\put(55.00,60.00){\vector(0,1){0.12}}

\put(50.00,55.00){\makebox(0,0)[cc]{$y'$}}

\put(55.00,25.00){\makebox(0,0)[cc]{$O$}}

\put(50.00,45.00){\makebox(0,0)[cc]{$O'$}}

\linethickness{1.00mm}
\put(55.00,50.00){\circle{0.00}}

\linethickness{0.20mm}
\put(45.00,80.00){\line(0,1){5.00}}
\put(45.00,80.00){\vector(0,-1){0.12}}

\linethickness{0.20mm}
\put(50.00,80.00){\line(0,1){10.00}}
\put(50.00,80.00){\vector(0,-1){0.12}}

\linethickness{0.20mm}
\put(60.00,80.00){\line(0,1){10.00}}
\put(60.00,80.00){\vector(0,-1){0.12}}

\linethickness{0.20mm}
\put(65.00,80.00){\line(0,1){5.00}}
\put(65.00,80.00){\vector(0,-1){0.12}}

\put(80.00,65.00){\makebox(0,0)[cc]{$h_2,\rho_2,\nu_2,E_2$}}

\put(85.00,55.00){\makebox(0,0)[cc]{$\cdots$}}

\put(90.00,45.00){\makebox(0,0)[cc]{$h_{n-1},\rho_{n-1},\nu_{n-1},E_{n-1}$}}

\put(80.00,35.00){\makebox(0,0)[cc]{$h_n,\rho_n,\nu_n,E_n$}}

\linethickness{0.70mm}
\put(45.00,70.00){\line(1,0){20.00}}
\put(45.00,70.00){\line(0,1){10.00}}
\put(65.00,70.00){\line(0,1){10.00}}
\put(45.00,80.00){\line(1,0){20.00}}

\put(55.00,95.00){\makebox(0,0)[cc]{$y$}}

\put(50.00,95.00){\makebox(0,0)[cc]{}}

\end{picture}
\end{center}
  \caption{Moving load on the multilayer plate. Illustration for
  special layers numeration}\label{pict3}
\end{figure}

By definition of the Lamzyuk-Privarnikov functions, we have:
\begin{eqnarray*}
  \alpha_2(\omega) &=& -A_{n-1}(\omega,h_2,\ldots,h_{n-1})
  \gamma_2(\omega) - C_{n-1}(\omega,h_2,\ldots,h_{n-1}) \delta_2(\omega)
  \\
  \beta_2(\omega) &=& -B_{n-1}(\omega,h_2,\ldots,h_{n-1})
  \gamma_2(\omega) - D_{n-1}(\omega,h_2,\ldots,h_{n-1}) \delta_2(\omega)
\end{eqnarray*}
Next, we put one more layer on top of the upper boundary of the previous multilayer pack (Figure \ref{pict3}). This new layer is labelled $1$. So, we obtain the situation illustrated in Figure \ref{pict2}.

Earlier, we established the recurrence relation (\ref{recrelalpha}) between $\overrightarrow{\alpha}_j$ and $\overrightarrow{\alpha}_{j+1}$. Now we use it with $j=1$:
$$
  \overrightarrow{\alpha}_2(\omega) = I_2^{-1}I_1 H_1(-h_1)
  \overrightarrow{\alpha}_1(\omega) = B(\omega,h_1)
  \overrightarrow{\alpha}_1(\omega),
$$
where we introduced the notation $B(\omega,h) = (b_{ij})_{1\le i,j\le 4} := I_2^{-1}I_1H_1(-h)$. Recall the definition of the diagonal matrix $I_1=\diag{1,\frac{i}{k_{11}^2},iE_1\omega,\mu_1\omega}$ and $I_2^{-1} = \diag{1,\frac{k_{12}^2}{i},\frac1{iE_2\omega},\frac1{\mu_2\omega}}$. It is not difficult to calculate explicitly the coefficients $(b_{ij})$:
$$
  b_{11}(\omega,h) = \frac{B_{11}\cosh(k_{21}\omega h) -
  B_{21}\cosh(k_{11}\omega h)}{B_{11}-B_{21}},
$$
$$
  b_{12}(\omega,h) = \frac{A_1\sinh(k_{21}\omega h) -
  2k_{21}\sinh(k_{11}\omega h)}{k_{11}k_{21}(2-A_1k_{11})},
$$
$$
  b_{13}(\omega,h) = \frac{\cosh(k_{11}\omega h) -
  \cosh(k_{21}\omega h)}{B_{11}-B_{21}},
$$
$$
  b_{14}(\omega,h) = \frac{k_{11}k_{21}\sinh(k_{11}\omega h)
  -\sinh(k_{21}\omega h)}{k_{21}(2-A_1k_{11})},
$$
$$
  b_{21}(\omega,h) = \frac{k_{12}^2}{k_{11}}
  \frac{B_{21}\sinh(k_{11}\omega h) -
  B_{11}k_{11}k_{21}\sinh(k_{21}\omega h)}{B_{11} - B_{21}},
$$
$$
  b_{22}(\omega,h) = \frac{k_{12}^2}{k_{11}^2}
  \frac{2\cosh(k_{11}\omega h)-A_1k_{11}\cosh(k_{21}\omega h)}
  {2-A_1k_{11}},
$$
$$
  b_{23}(\omega,h) = \frac{k_{12}^2}{k_{11}}
  \frac{k_{11}k_{21}\sinh(k_{21}\omega h) -
  \sinh(k_{11}\omega h)}{B_{11}-B_{21}},
$$
$$
  b_{24}(\omega,h) = k_{12}^2 \frac{\cosh(k_{21}\omega h) -
  \cosh(k_{11}\omega h)}{2-A_1k_{11}},
$$
$$
  b_{31}(\omega,h) = \frac{E_1}{E_2} B_{11}B_{21}
  \frac{\cosh(k_{21}\omega h) - \cosh(k_{11}\omega h)}
  {B_{11}-B_{21}},
$$
$$
  b_{32}(\omega,h) = \frac{E_1}{E_2}
  \frac{A_1B_{21}\sinh(k_{21}\omega h) -
  2B_{11}k_{21}\sinh(k_{11}\omega h)}
  {k_{11}k_{21}(2-A_1k_{11})},
$$
$$
  b_{33}(\omega,h) = \frac{E_1}{E_2}
  \frac{B_{11}\cosh(k_{11}\omega h) -
  B_{21}\cosh(k_{21}\omega h)}
  {B_{11} - B_{21}},
$$
$$
  b_{34}(\omega,h) = \frac{E_1}{E_2}
  \frac{B_{11}k_{11}k_{21}\sinh(k_{11}\omega h) -
  B_{21}\sinh(k_{21}\omega h)}
  {k_{21}(2-A_1k_{11})},
$$
$$
  b_{41}(\omega,h) = \frac{\mu_1}{\mu_2}
  \frac{A_1B_{21}\sinh(k_{11}\omega h) -
  2B_{11}k_{21}\sinh(k_{21}\omega h)}
  {B_{11}-B_{21}},
$$
$$
  b_{42}(\omega,h) = \frac{\mu_1}{\mu_2} 2A_1
  \frac{\cosh(k_{11}\omega h) - \cosh(k_{21}\omega h)}
  {k_{11}(2-A_1k_{11})},
$$
$$
  b_{43}(\omega,h) = \frac{\mu_1}{\mu_2}
  \frac{2k_{21}\sinh(k_{21}\omega h) -
  A_1\sinh(k_{11}\omega h)}
  {B_{11}-B_{21}},
$$
$$
  b_{44}(\omega,h) = \frac{\mu_1}{\mu_2}
  \frac{2\cosh(k_{21}\omega h) -
  A_1k_{11}\cosh(k_{11}\omega h)}
  {2-A_1k_{11}}.
$$

The matrix $B$ provides us with a linear relation between $\overrightarrow{\alpha}_2(\omega)$ and $\overrightarrow{\alpha}_1(\omega)$. The extended formulae are
\begin{eqnarray*}
 \alpha_2(\omega) &=& b_{11}(\omega,h)\alpha_1(\omega) +
 b_{12}(\omega,h)\beta_1(\omega) +
 b_{13}(\omega,h)\gamma_1(\omega) +
 b_{14}(\omega,h)\delta_1(\omega), \\
 \beta_2(\omega) &=& b_{21}(\omega,h)\alpha_1(\omega) +
 b_{22}(\omega,h)\beta_1(\omega) +
 b_{23}(\omega,h)\gamma_1(\omega) +
 b_{24}(\omega,h)\delta_1(\omega), \\
 \gamma_2(\omega) &=& b_{31}(\omega,h)\alpha_1(\omega) +
 b_{32}(\omega,h)\beta_1(\omega) +
 b_{33}(\omega,h)\gamma_1(\omega) +
 b_{34}(\omega,h)\delta_1(\omega), \\
 \delta_2(\omega) &=& b_{41}(\omega,h)\alpha_1(\omega) +
 b_{42}(\omega,h)\beta_1(\omega) +
 b_{43}(\omega,h)\gamma_1(\omega) +
 b_{44}(\omega,h)\delta_1(\omega).
\end{eqnarray*}
In the definition of Lamzyuk-Privarnikov functions for $(n-1)$ layers, we have
\begin{eqnarray*}
  \alpha_2(\omega) &=& -A_{n-1}(\omega,h_2,\ldots,h_{n-1})
  \gamma_2(\omega) - C_{n-1}(\omega,h_2,\ldots,h_{n-1})
  \delta_2(\omega),
  \\
  \beta_2(\omega) &=& -B_{n-1}(\omega,h_2,\ldots,h_{n-1})
  \gamma_2(\omega) - D_{n-1}(\omega,h_2,\ldots,h_{n-1})
  \delta_2(\omega).
\end{eqnarray*}
One can easily obtain two identities:
\begin{multline*}
  (b_{11} + A_{n-1}b_{31} + C_{n-1}b_{41})\alpha_1(\omega) +
  (b_{12} + A_{n-1}b_{32} + C_{n-1}b_{42})\beta_1(\omega) +\\+
  (b_{13} + A_{n-1}b_{33} + C_{n-1}b_{43})\gamma_1(\omega) +
  (b_{14} + A_{n-1}b_{34} + C_{n-1}b_{44})\delta_1(\omega) = 0,
\end{multline*}
\begin{multline*}
  (b_{21} + B_{n-1}b_{31} + D_{n-1}b_{41})\alpha_1(\omega) +
  (b_{22} + B_{n-1}b_{32} + D_{n-1}b_{42})\beta_1(\omega) +\\+
  (b_{23} + B_{n-1}b_{33} + D_{n-1}b_{43})\gamma_1(\omega) +
  (b_{24} + B_{n-1}b_{34} + D_{n-1}b_{44})\delta_1(\omega) = 0.
\end{multline*}
If we remember that in these identities, the unknowns are $\alpha_1(\omega)$ and $\beta_1(\omega)$ we can easily obtain a linear system of two equations:
\begin{multline*}
  (b_{11} + A_{n-1}b_{31} + C_{n-1}b_{41})\alpha_1(\omega) +
  (b_{12} + A_{n-1}b_{32} + C_{n-1}b_{42})\beta_1(\omega) =\\=
  -(b_{13} + A_{n-1}b_{33} + C_{n-1}b_{43})\gamma_1(\omega) -
  (b_{14} + A_{n-1}b_{34} + C_{n-1}b_{44})\delta_1(\omega),
\end{multline*}
\begin{multline*}
  (b_{21} + B_{n-1}b_{31} + D_{n-1}b_{41})\alpha_1(\omega) +
  (b_{22} + B_{n-1}b_{32} + D_{n-1}b_{42})\beta_1(\omega) =\\=
  -(b_{23} + B_{n-1}b_{33} + D_{n-1}b_{43})\gamma_1(\omega) -
  (b_{24} + B_{n-1}b_{34} + D_{n-1}b_{44})\delta_1(\omega).
\end{multline*}
The solution of this system is
\begin{eqnarray}\label{sol1}
  \alpha_1(\omega) &=& -\frac{Y_n(\omega)\gamma_1(\omega)+
  Q_n(\omega)\delta_1(\omega)}{R_n(\omega)},\\
  \beta_1(\omega) &=& -\frac{X_n(\omega)\gamma_1(\omega)+
  Z_n(\omega)\delta_1(\omega)}{R_n(\omega)},\label{sol2}
\end{eqnarray}
where the functions $Y_n(\omega)$, $Q_n(\omega)$, $X_n(\omega)$, $Z_n(\omega)$, $R_n(\omega)$ are determined by Cramer's rule:
\begin{equation*}
  R_n(\omega) =
  \begin{vmatrix}
    b_{11} + A_{n-1}b_{31} + C_{n-1}b_{41} &
    b_{12} + A_{n-1}b_{32} + C_{n-1}b_{42}\\
    b_{21} + B_{n-1}b_{31} + D_{n-1}b_{41} &
    b_{22} + B_{n-1}b_{32} + D_{n-1}b_{42}\\
  \end{vmatrix},
\end{equation*}
\begin{equation*}
  Y_n(\omega) =
  \begin{vmatrix}
    b_{13} + A_{n-1}b_{33} + C_{n-1}b_{43} &
    b_{12} + A_{n-1}b_{32} + C_{n-1}b_{42}\\
    b_{23} + B_{n-1}b_{33} + D_{n-1}b_{43} &
    b_{22} + B_{n-1}b_{32} + D_{n-1}b_{42}\\
  \end{vmatrix},
\end{equation*}
\begin{equation*}
  Q_n(\omega) =
  \begin{vmatrix}
    b_{14} + A_{n-1}b_{34} + C_{n-1}b_{44} &
    b_{12} + A_{n-1}b_{32} + C_{n-1}b_{42}\\
    b_{24} + B_{n-1}b_{34} + D_{n-1}b_{44} &
    b_{22} + B_{n-1}b_{32} + D_{n-1}b_{42}\\
  \end{vmatrix},
\end{equation*}
\begin{equation*}
  X_n(\omega) =
  \begin{vmatrix}
    b_{11} + A_{n-1}b_{31} + C_{n-1}b_{41} &
    b_{13} + A_{n-1}b_{33} + C_{n-1}b_{43}\\
    b_{21} + B_{n-1}b_{31} + D_{n-1}b_{41} &
    b_{23} + B_{n-1}b_{33} + D_{n-1}b_{43}\\
  \end{vmatrix},
\end{equation*}
\begin{equation*}
  Z_n(\omega) =
  \begin{vmatrix}
    b_{11} + A_{n-1}b_{31} + C_{n-1}b_{41} &
    b_{14} + A_{n-1}b_{34} + C_{n-1}b_{44}\\
    b_{21} + B_{n-1}b_{31} + D_{n-1}b_{41} &
    b_{24} + B_{n-1}b_{34} + D_{n-1}b_{44}\\
  \end{vmatrix}.
\end{equation*}

If the reader compares the identities (\ref{sol1}), (\ref{sol2}) with the definition of Lamzyuk-Privarnikov functions
\begin{eqnarray*}
  \alpha_1(\omega) &=& -A_n(\omega,h_1,\ldots,h_{n-1})
  \gamma_1(\omega) - C_n(\omega,h_1,\ldots,h_{n-1}) \delta_1(\omega)
  \\
  \beta_1(\omega) &=& -B_n(\omega,h_1,\ldots,h_{n-1})
  \gamma_1(\omega) - D_n(\omega,h_1,\ldots,h_{n-1}) \delta_1(\omega)
\end{eqnarray*}
it is easy to conclude that
\begin{eqnarray*}
  A_n(\omega) = \frac{Y_n(\omega)}{R_n(\omega)}, \qquad
  B_n(\omega) &=& \frac{X_n(\omega)}{R_n(\omega)}, \\
  C_n(\omega) = \frac{Q_n(\omega)}{R_n(\omega)}, \qquad
  D_n(\omega) &=& \frac{Z_n(\omega)}{R_n(\omega)}.
\end{eqnarray*}

We have calculated analytical expressions of the functions $Y_n(\omega)$, $Q_n(\omega)$, $X_n(\omega)$, $Z_n(\omega)$, $R_n(\omega)$. Here they are:
\begin{multline*}
  Y_n(\omega) = g_{11}(\omega) + g_{12}(\omega)A_{n-1}(\omega) +
  g_{13}(\omega)B_{n-1}(\omega) +\\+ g_{14}(\omega)C_{n-1}(\omega) +
  g_{15}(\omega)D_{n-1}(\omega) +
  g_{16}(\omega)\Delta_{n-1}(\omega),
\end{multline*}
\begin{multline*}
  X_n(\omega) = g_{21}(\omega) + g_{22}(\omega)A_{n-1}(\omega) +
  g_{23}(\omega)B_{n-1}(\omega) +\\+ g_{24}(\omega)C_{n-1}(\omega) +
  g_{25}(\omega)D_{n-1}(\omega) +
  g_{26}(\omega)\Delta_{n-1}(\omega),
\end{multline*}
\begin{multline*}
  Q_n(\omega) = g_{31}(\omega) + g_{32}(\omega)A_{n-1}(\omega) +
  g_{33}(\omega)B_{n-1}(\omega) +\\+ g_{34}(\omega)C_{n-1}(\omega) +
  g_{35}(\omega)D_{n-1}(\omega) +
  g_{36}(\omega)\Delta_{n-1}(\omega),
\end{multline*}
\begin{multline*}
  Z_n(\omega) = g_{41}(\omega) + g_{42}(\omega)A_{n-1}(\omega) +
  g_{43}(\omega)B_{n-1}(\omega) +\\+ g_{44}(\omega)C_{n-1}(\omega) +
  g_{45}(\omega)D_{n-1}(\omega) +
  g_{46}(\omega)\Delta_{n-1}(\omega),
\end{multline*}
\begin{multline*}
  R_n(\omega) = g_{51}(\omega) + g_{52}(\omega)A_{n-1}(\omega) +
  g_{53}(\omega)B_{n-1}(\omega) +\\+ g_{54}(\omega)C_{n-1}(\omega) +
  g_{55}(\omega)D_{n-1}(\omega) +
  g_{56}(\omega)\Delta_{n-1}(\omega),
\end{multline*}
where
\begin{equation*}
  \Delta_{n-1}(\omega) = \begin{vmatrix}
    A_{n-1}(\omega) & B_{n-1}(\omega) \\
    C_{n-1}(\omega) & D_{n-1}(\omega) \\
  \end{vmatrix} = A_{n-1}(\omega)D_{n-1}(\omega) -
  B_{n-1}(\omega)C_{n-1}(\omega).
\end{equation*}

The expressions of the coefficients $g_{ij}(\omega)$, $1\le i\le 5$, $1\le j\le 6$ are:

\begin{multline*}
  g_{11}(\omega) = b_{13}b_{22}-b_{12}b_{23} =
  \frac{k_{12}^2}{k_1k_2(2-Ak_1)(B_1-B_2)}\Bigl(
  k_2(2+Ak_1)\times \\ \times (1-\cosh(k_1\omega h)\cosh(k_2\omega h))
  +(A+2k_1k_2^2)\sinh(k_1\omega h)\sinh(k_2\omega h)
  \Bigr),
\end{multline*}
\begin{multline*}
  g_{12}(\omega) = b_{33}b_{22}-b_{23}b_{32} =
  \frac{E_1k_{12}^2}{E_2k_1k_2(2-Ak_1)(B_1-B_2)}\times
  \\ \Bigl(
  (2B_1+AB_2k_1)k_2 - k_2(AB_1k_1+2B_2)\cosh(k_1\omega
  h)\cosh(k_2\omega h) +\\+ (AB_2+2B_1k_1k_2^2)
  \sinh(k_1\omega h)\sinh(k_2\omega h),
  \Bigr)
\end{multline*}
\begin{multline*}
  g_{13}(\omega) = b_{13}b_{32} - b_{12}b_{33} =
  \frac{E_1}{E_2k_2(2-Ak_1)}\times\\ \Bigl(
  2k_2\sinh(k_1\omega h)\cosh(k_2\omega h) -
  A\cosh(k_1\omega h)\sinh(k_2\omega h)\Bigr)
\end{multline*}
\begin{multline*}
  g_{14}(\omega) = b_{43}b_{22} - b_{42}b_{23} =
  \frac{\mu_1k_{12}^2}{\mu_2k_1(B_1-B_2)}\times \\
  \Bigl(
  2k_2\cosh(k_1\omega h)\sinh(k_2\omega h) -
  A\sinh(k_1\omega h)\cosh(k_2\omega h)\Bigr),
\end{multline*}
\begin{multline*}
  g_{15}(\omega) = b_{13}b_{42} - b_{12}b_{43} =
  \frac{\mu_1}{\mu_2k_2(2-Ak_1)(B_1-B_2)}\times \\
  \Bigl(4Ak_2(1-\cosh(k_1\omega h)\cosh(k_2\omega h)) +\\+
  (A^2+4k_2^2)\sinh(k_1\omega h)\sinh(k_2\omega h)\Bigr),
\end{multline*}
\begin{multline*}
  g_{16}(\omega) = b_{33}b_{42} - b_{32}b_{43} =
  \Bigl(
  2Ak_2(B_1+B_2)\times \\
  (1-\cosh(k_1\omega h)\cosh(k_2\omega h)) +
  (A^2B_2+4k_2^2B_1)\sinh(k_1\omega h)\sinh(k_2\omega h)\Bigr),
\end{multline*}


\begin{multline*}
  g_{21}(\omega) = b_{11}b_{23} - b_{13}b_{21} =
  \frac{k_{12}^2}{B_1-B_2}\Bigl(
  k_1k_2\cosh(k_1\omega h)\sinh(k_2\omega h) -\\-
  \sinh(k_1\omega h)\cosh(k_2\omega h)
  \Bigr),
\end{multline*}

\begin{multline*}
  g_{22}(\omega) = b_{31}b_{23} - b_{33}b_{21} =
  \frac{E_1k_{12}^2}{E_2(B_1-B_2)}\times \\ \times
  \Bigl(
  B_1k_1k_2\cosh(k_1\omega h)\sinh(k_2\omega h) -
  B_2\sinh(k_1\omega h)\cosh(k_2\omega h)
  \Bigr),
\end{multline*}

\begin{equation*}
   g_{23}(\omega) = b_{11}b_{33} - b_{13}b_{31} =
  \frac{E_1k_1}{E_2}\cosh(k_1\omega h)\cosh(k_2\omega h),
\end{equation*}

\begin{equation*}
    g_{24}(\omega) = b_{41}b_{23} - b_{43}b_{21} =
    \frac{\mu_1k_{12}^2k_2(2-Ak_1)}{\mu_2(B_1-B_2)}
    \sinh(k_1\omega h)\sinh(k_2\omega h),
\end{equation*}
\begin{multline*}
  g_{25}(\omega) = b_{11}b_{43} - b_{13}b_{41} =
  \frac{\mu_1k_1}{\mu_2(B_1-B_2)}\Bigl(
  2k_2\cosh(k_1\omega h)\sinh(k_2\omega h) -\\-
  A\sinh(k_1\omega h)\cosh(k_2\omega h)\Bigr),
\end{multline*}
\begin{multline*}
  g_{26}(\omega) = b_{31}b_{43} - b_{33}b_{41} =
  \frac{E_1\mu_1k_1}{E_2\mu_2(B_1-B_2)}\times \\ \times
  \Bigl(
  2B_1k_2\cosh(k_1\omega h)\sinh(k_2\omega h) -
  AB_2\sinh(k_1\omega h)\cosh(k_2\omega h)
  \Bigr),
\end{multline*}
\begin{multline*}
  g_{31}(\omega) = b_{14}b_{22} - b_{12}b_{24} =
  \frac{k_{12}^2}{k_1k_2(2-Ak_1)}\times \\ \times
  \Bigl(
  k_1k_2\sinh(k_1\omega h)\cosh(k_2\omega h) -
  \cosh(k_1\omega h)\sinh(k_2\omega h)\Bigr),
\end{multline*}

\begin{multline*}
  g_{32}(\omega) = b_{34}b_{22} - b_{32}b_{24} =
  \frac{E_1k_{12}^2}{E_2k_1k_2(2-Ak_1)}\times \\ \times
  \Bigl(
  B_1k_1k_2\sinh(k_1\omega h)\cosh(k_2\omega h) -
  B_2\cosh(k_1\omega h)\sinh(k_2\omega h)
  \Bigr),
\end{multline*}

\begin{equation*}
  g_{33}(\omega) = b_{14}b_{32} - b_{12}b_{34} =
  \frac{E_1(B_1-B_2)}{E_2k_2(2-Ak_1)}
  \sinh(k_1\omega h)\sinh(k_2\omega h),
\end{equation*}

\begin{equation*}
  g_{34}(\omega) = b_{44}b_{22} - b_{42}b_{24} =
  \frac{\mu_1k_{12}^2}{\mu_2k_{11}}
  \cosh(k_1\omega h)\cosh(k_2\omega h),
\end{equation*}

\begin{multline*}
  g_{35}(\omega) = b_{14}b_{42} - b_{12}b_{44} =
  \frac{\mu_1}{\mu_2k_2(2-Ak_1)}\times \\ \times
  \Bigl(
  2k_2\sinh(k_1\omega h)\cosh(k_2\omega h) -
  A\cosh(k_1\omega h)\sinh(k_2\omega h)
  \Bigr),
\end{multline*}

\begin{multline*}
  g_{36}(\omega) = b_{34}b_{42} - b_{32}b_{44} =
  \frac{E_1\mu_1}{E_2\mu_2k_2(2-Ak_1)}\times \\ \times
  \Bigl(
  2B_1k_2\sinh(k_1\omega h)\cosh(k_2\omega h) -
  AB_2\cosh(k_1\omega h)\sinh(k_2\omega h)
  \Bigr),
\end{multline*}

\begin{multline*}
  g_{41}(\omega) = b_{11}b_{24} - b_{14}b_{21} =
  \frac{k_{12}^2}{k_2(B_1-B_2)(2-Ak_1)}\times \\ \times
  \Bigl(
  k_1k_2(B_1 + B_2)(1-\cosh(k_1\omega h)\cosh(k_2\omega h)) +\\+
  (B_1k_1^2k_2^2 + B_2)\sinh(k_1\omega h)\sinh(k_2\omega h)
  \Bigr),
\end{multline*}

\begin{multline*}
  g_{42}(\omega) = b_{31}b_{24} - b_{34}b_{21} =
  \frac{E_1k_{12}^2}{E_2k_2(2-Ak_1)(B_1-B_2)}\times \\ \times
  \Bigl(
  2B_1B_2k_1k_2(1-\cosh(k_1\omega h)\cosh(k_2\omega h)) +\\+
  (B_1^2k_1^2k_2^2 + B_2^2)\sinh(k_1\omega h)\sinh(k_2\omega h)
  \Bigr),
\end{multline*}

\begin{multline*}
  g_{43}(\omega) = b_{11}b_{34} - b_{14}b_{31} =
  \frac{E_1k_1}{E_2k_2(2-Ak_1)}\times \\ \times
  \Bigl(
  B_1k_1k_2\sinh(k_1\omega h)\cosh(k_2\omega h) -
  B_2\cosh(k_1\omega h)\sinh(k_2\omega h)\Bigr),
\end{multline*}

\begin{multline*}
  g_{44}(\omega) = b_{41}b_{24} - b_{44}b_{21} =
  \frac{\mu_1k_{12}^2}{\mu_2(B_1-B_2)}\times \\ \times
  \Bigl(
  B_1k_1k_2\cosh(k_1\omega h)\sinh(k_2\omega h) -
  B_2\sinh(k_1\omega h)\cosh(k_2\omega h)
  \Bigr),
\end{multline*}

\begin{multline*}
  g_{45}(\omega) = b_{11}b_{44} - b_{14}b_{41} =
  \frac{\mu_1k_{11}}{\mu_2k_{21}(2-Ak_1)(B_1-B_2)}
  \times \\ \times
  \Bigl(
  (2B_1k_2+AB_2k_1k_2) - (2B_2k_2+AB_1k_1k_2)
  \cosh(k_1\omega h)\cosh(k_2\omega h) +\\+
  (2B_1k_1k_2^2 + AB_2)\sinh(k_1\omega h)\sinh(k_2\omega h)
  \Bigr),
\end{multline*}

\begin{multline*}
  g_{46}(\omega) = b_{31}b_{44} - b_{34}b_{41} =
  \frac{E_1\mu_1k_1}{\mu_2E_2(2-Ak_1)(B_1-B_2)}
  \times \\ \times
  \Bigl(
  B_1B_2k_2(2+Ak_1)(1-\cosh(k_1\omega h)\cosh(k_2\omega h)) +\\+
  (AB_2^2+2B_1^2k_1k_2^2)\sinh(k_1\omega h)\sinh(k_2\omega h)
  \Bigr),
\end{multline*}
\begin{multline*}
  g_{51}(\omega) = b_{11}b_{22} - b_{12}b_{21} =
  \frac{k_{12}^2}{k_1k_2(2-Ak_1)(B_1-B_2)}\times \\
  \Bigl(
  (2B_1k_2+AB_2k_1k_2)\cosh(k_1\omega h)\cosh(k_2\omega h) -\\-
  (2k_2B_2 + AB_1k_1k_2) -
  (AB_2 + 2k_1k_2^2B_1)\sinh(k_1\omega h)\sinh(k_2\omega h)
  \Bigr),
\end{multline*}
\begin{multline*}
  g_{52}(\omega) = b_{31}b_{22} - b_{32}b_{21} =
  \frac{E_1k_{12}^2}{E_2k_1k_2(2-Ak_1)(B_1-B_2)}\times \\
  \Bigl(B_1B_2k_2(2+Ak_1)[\cosh(k_1\omega h)\cosh(k_2\omega h)-1] -\\-
  (AB_{21}^2 + 2B_1^2k_1k_2^2)\sinh(k_1\omega h)\sinh(k_2\omega h)
  \Bigr),
\end{multline*}
\begin{multline*}
  g_{53}(\omega) = b_{11}b_{32} - b_{12}b_{31} =
  \frac{E_1}{E_2k_2(2-Ak_1)}\times \\ \Bigl(
  AB_2\cosh(k_1\omega h)\sinh(k_2\omega h) -
  2B_1k_2\sinh(k_1\omega h)\cosh(k_2\omega h)\Bigr),
\end{multline*}
\begin{multline*}
  g_{54}(\omega) = b_{41}b_{22} - b_{42}b_{21} =
  \frac{\mu_1k_{12}^2}{\mu_2k_1(B_1-B_2)}\times \\ \Bigr(
  AB_2\sinh(k_1\omega h)\cosh(k_2\omega h) -
  2B_1k_2\cosh(k_1\omega h)
  \sinh(k_2\omega h)\Bigl),
\end{multline*}
\begin{multline*}
  g_{55}(\omega) = b_{11}b_{42} - b_{12}b_{41} =
  \frac{\mu_1}{\mu_2k_2(2-Ak_1)(B_1-B_2)}\Bigl(
  2Ak_2(B_1+B_2)\times \\ \times
  [\cosh(k_1\omega h)\cosh(k_2\omega h)-1] -\\-
  (A^2B_2+4k_2^2B_1)\sinh(k_1\omega h)\sinh(k_2\omega h)\Bigr),
\end{multline*}
\begin{multline*}
  g_{56}(\omega) = b_{31}b_{42} - b_{32}b_{41} =
  \frac{E_1\mu_1}{E_2\mu_2k_2(2-Ak_1)(B_1-B_2)}
  \Bigl(4AB_1B_2k_2\times \\ \times
  (\cosh(k_1\omega h)\cosh(k_2\omega h)-1) -\\-
  (A^2B_2^2+4B_1^2k_2^2)\sinh(k_1\omega h)\sinh(k_2\omega h)\Bigr),
\end{multline*}

Here, we constructed the Lamzyuk-Privarnikov functions for a multilayer pack. These functions depend only on the mechanical properties and geometrical characteristics of the pack and do not depend on the boundary conditions. It means that once constructed, these functions can be used to solve different problems. That is why we find this method amazing and very useful.

\section{Inhomogeneous layer limit}

\begin{figure}[htbp]
\begin{center}
\unitlength 1mm
\begin{picture}(90.50,40.00)(0,0)

\linethickness{0.40mm}
\put(5.00,28.00){\line(1,0){69.50}}

\linethickness{0.40mm}
\put(5.00,20.00){\line(1,0){69.50}}

\linethickness{0.20mm}
\put(74.50,20.00){\line(1,0){10.50}}
\put(85.00,20.00){\vector(1,0){0.12}}

\linethickness{0.15mm}
\put(42.00,20.00){\line(0,1){21.00}}
\put(42.00,41.00){\vector(0,1){0.12}}

\linethickness{0.40mm}
\put(6.00,4.00){\line(1,0){69.00}}

\linethickness{0.20mm}
\multiput(6.00,0.50)(0.16,0.12){29}{\line(1,0){0.16}}

\linethickness{0.20mm}
\multiput(12.00,0.50)(0.16,0.12){29}{\line(1,0){0.16}}

\linethickness{0.20mm}
\multiput(9.00,0.50)(0.16,0.12){29}{\line(1,0){0.16}}

\linethickness{0.20mm}
\multiput(15.00,0.50)(0.16,0.12){29}{\line(1,0){0.16}}

\linethickness{0.20mm}
\multiput(21.00,0.50)(0.16,0.12){29}{\line(1,0){0.16}}

\linethickness{0.20mm}
\multiput(18.00,0.50)(0.16,0.12){29}{\line(1,0){0.16}}

\linethickness{0.20mm}
\multiput(24.00,0.50)(0.16,0.12){29}{\line(1,0){0.16}}

\linethickness{0.20mm}
\multiput(30.00,0.50)(0.16,0.12){29}{\line(1,0){0.16}}

\linethickness{0.20mm}
\multiput(27.00,0.50)(0.16,0.12){29}{\line(1,0){0.16}}

\linethickness{0.20mm}
\multiput(33.00,0.50)(0.16,0.12){29}{\line(1,0){0.16}}

\linethickness{0.20mm}
\multiput(39.00,0.50)(0.16,0.12){29}{\line(1,0){0.16}}

\linethickness{0.20mm}
\multiput(36.00,0.50)(0.16,0.12){29}{\line(1,0){0.16}}

\linethickness{0.20mm}
\multiput(42.00,0.50)(0.16,0.12){29}{\line(1,0){0.16}}

\linethickness{0.20mm}
\multiput(48.00,0.50)(0.16,0.12){29}{\line(1,0){0.16}}

\linethickness{0.20mm}
\multiput(45.00,0.50)(0.16,0.12){29}{\line(1,0){0.16}}

\linethickness{0.20mm}
\multiput(51.00,0.50)(0.16,0.12){29}{\line(1,0){0.16}}

\linethickness{0.20mm}
\multiput(57.00,0.50)(0.16,0.12){29}{\line(1,0){0.16}}

\linethickness{0.20mm}
\multiput(54.00,0.50)(0.16,0.12){29}{\line(1,0){0.16}}

\linethickness{0.20mm}
\multiput(60.00,0.50)(0.16,0.12){29}{\line(1,0){0.16}}

\linethickness{0.20mm}
\multiput(66.00,0.50)(0.16,0.12){29}{\line(1,0){0.16}}

\linethickness{0.20mm}
\multiput(63.00,0.50)(0.16,0.12){29}{\line(1,0){0.16}}

\linethickness{0.20mm}
\multiput(69.00,0.50)(0.16,0.12){29}{\line(1,0){0.16}}

\linethickness{0.40mm}
\put(34.50,28.50){\line(1,0){16.00}}
\put(34.50,28.50){\line(0,1){8.00}}
\put(50.50,28.50){\line(0,1){8.00}}
\put(34.50,36.50){\line(1,0){16.00}}

\put(82.00,16.00){\makebox(0,0)[cc]{$x$}}

\put(123.50,64.00){\makebox(0,0)[cc]{}}

\put(40.00,41.00){\makebox(0,0)[cc]{$y$}}

\linethickness{0.20mm}
\put(50.50,32.00){\line(1,0){8.00}}
\put(58.50,32.00){\vector(1,0){0.12}}

\put(56.00,35.00){\makebox(0,0)[cc]{$c_0$}}

\put(57.50,5.00){\makebox(0,0)[cc]{}}

\linethickness{0.20mm}
\put(62.00,4.00){\line(0,1){16.00}}
\put(62.00,20.00){\vector(0,1){0.12}}
\put(62.00,4.00){\vector(0,-1){0.12}}

\linethickness{0.20mm}
\put(62.00,20.00){\line(0,1){8.00}}
\put(62.00,28.00){\vector(0,1){0.12}}
\put(62.00,20.00){\vector(0,-1){0.12}}

\put(65.00,13.00){\makebox(0,0)[cc]{$H$}}

\put(63.50,13.00){\makebox(0,0)[cc]{}}

\put(65.00,24.00){\makebox(0,0)[cc]{$h$}}

\linethickness{0.15mm}
\put(12.00,19.00){\line(1,0){4.00}}

\linethickness{0.15mm}
\put(32.00,14.00){\line(1,0){4.00}}

\linethickness{0.15mm}
\put(47.00,14.00){\line(1,0){4.00}}

\linethickness{0.15mm}
\put(17.00,10.50){\line(1,0){4.00}}

\linethickness{0.15mm}
\put(22.50,16.50){\line(1,0){4.00}}

\linethickness{0.15mm}
\put(6.50,7.50){\line(1,0){4.00}}

\linethickness{0.15mm}
\put(6.50,16.50){\line(1,0){4.00}}

\linethickness{0.15mm}
\put(26.50,6.50){\line(1,0){4.00}}

\linethickness{0.15mm}
\put(27.50,12.50){\line(1,0){4.00}}

\linethickness{0.15mm}
\put(35.00,8.00){\line(1,0){4.00}}

\linethickness{0.15mm}
\put(38.00,12.00){\line(1,0){4.00}}

\linethickness{0.15mm}
\put(16.00,7.00){\line(1,0){4.00}}

\linethickness{0.15mm}
\put(44.50,6.50){\line(1,0){4.00}}

\linethickness{0.15mm}
\put(53.00,16.50){\line(1,0){4.00}}

\linethickness{0.15mm}
\put(53.00,7.00){\line(1,0){4.00}}

\linethickness{0.15mm}
\put(64.50,7.50){\line(1,0){4.00}}

\linethickness{0.15mm}
\put(67.50,16.50){\line(1,0){4.00}}

\linethickness{0.15mm}
\put(69.00,13.00){\line(1,0){4.00}}

\linethickness{0.15mm}
\put(70.00,6.50){\line(1,0){4.00}}

\linethickness{0.15mm}
\put(39.50,14.00){\line(1,0){4.00}}

\linethickness{0.15mm}
\put(45.50,9.00){\line(1,0){4.00}}

\linethickness{0.15mm}
\put(34.00,19.00){\line(0,1){2.00}}

\linethickness{0.15mm}
\put(50.00,19.00){\line(0,1){2.00}}

\put(32.00,17.00){\makebox(0,0)[cc]{$-a$}}

\put(50.00,17.00){\makebox(0,0)[cc]{$a$}}

\put(42.00,17.00){\makebox(0,0)[cc]{$o$}}

\put(41.00,16.00){\makebox(0,0)[cc]{}}

\put(81.00,4.00){\makebox(0,0)[cc]{$\Gamma_1$}}

\put(76.00,4.00){\makebox(0,0)[cc]{}}

\put(81.00,24.00){\makebox(0,0)[cc]{$\Gamma_2$}}

\put(81.00,29.00){\makebox(0,0)[cc]{$\Gamma_3$}}

\put(11.00,34.00){\makebox(0,0)[cc]{$T=T_1$}}

\put(11.00,14.00){\makebox(0,0)[cc]{$T=T_0$}}

\linethickness{0.15mm}
\put(52.00,9.00){\line(1,0){4.00}}

\linethickness{0.15mm}
\put(6.00,9.00){\line(1,0){4.00}}

\linethickness{0.15mm}
\put(22.00,14.00){\line(1,0){4.00}}

\linethickness{0.15mm}
\put(56.00,14.00){\line(1,0){4.00}}

\put(20.00,25.00){\makebox(0,0)[cc]{$\rho(y),E(y),\nu(y)$}}

\end{picture}
\end{center}
  \caption{Moving load on the inhomogeneous layer.}\label{pict4}
\end{figure}
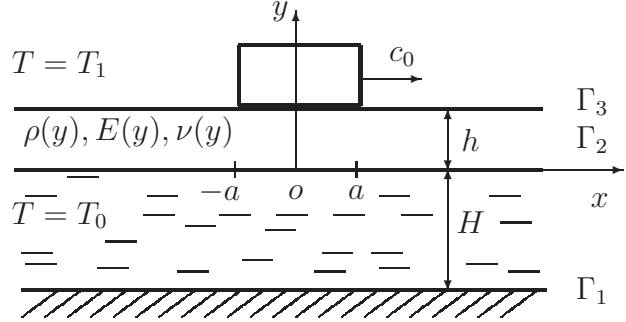

In this section, we take the limit as the number of layers tends to infinity, and the maximum thickness tends to zero with the constraint
$$
  \sum_{i=1}^n h_i = h.
$$

Let us introduce continuous notation. So far, we have considered the piecewise constant function $\overrightarrow{\alpha}(\omega,y)$ which is defined by
\begin{equation*}
  \overrightarrow{\alpha}(\omega,y) =
  \overrightarrow{\alpha}_k(\omega), \qquad
  \sum_{i=0}^{n-k-1} h_{n-i} \le y < \sum_{i=0}^{n-k} h_{n-i},
  k=1,\ldots,n-1.
\end{equation*}
To simplify the notation, we assume that all layers have the same thickness. In other words, we have
$$
  \alpha_k(\omega)\equiv\alpha(\omega,y),\qquad \alpha_{k+1}(\omega)\equiv\alpha(\omega,y-h)
$$
$$
  \beta_k(\omega)\equiv\beta(\omega,y),\qquad \beta_{k+1}(\omega)\equiv\beta(\omega,y-h)
$$
$$
  \delta_k(\omega)\equiv\delta(\omega,y),\qquad \delta_{k+1}(\omega)\equiv\delta(\omega,y-h)
$$
$$
  \gamma_k(\omega)\equiv\gamma(\omega,y),\qquad \gamma_{k+1}(\omega)\equiv\gamma(\omega,y-h)
$$

In the previous section, we obtained the recurrence relation (\ref{recrelalpha}) for the vector $\overrightarrow{\alpha}_{k}(\omega)$. This relation gives the connection between two vectors $\overrightarrow{\alpha}(\omega)$ for consecutive layers in the multilayer pack. We have also established that knowing the vector $\overrightarrow{\alpha}$ is equivalent to knowing the solution to our problem since
\begin{equation}\label{gensol}
  \begin{pmatrix}
    u(x,y) \\
    v(x,y) \\
    \sigma_y(x,y) \\
    \tau_{xy}(x,y) \\
  \end{pmatrix} = \frac1{2\pi}\int\limits_{-\infty}^{+\infty}
  IH(\omega,y)\overrightarrow{\alpha}(\omega,y)
  e^{-i\omega x}d\omega.
\end{equation}

Let
$$
  B_j(\omega,h_k)=(b_{ij})_{1\le i,j\le 4} :=
  I_{k+1}^{-1}I_kH_k(\omega,-h_k).
$$
We can rewrite the recurrence relation (\ref{recrelalpha}) in the form
$$
  \begin{Vmatrix}
    \alpha_{k+1}(\omega) \\
    \beta_{k+1}(\omega) \\
    \gamma_{k+1}(\omega) \\
    \delta_{k+1}(\omega)
  \end{Vmatrix} =
  \begin{Vmatrix}
    b_{11}(\omega,h) & b_{12}(\omega,h) & b_{13}(\omega,h) & b_{14}(\omega,h) \\
    b_{21}(\omega,h) & b_{22}(\omega,h) & b_{23}(\omega,h) & b_{24}(\omega,h) \\
    b_{31}(\omega,h) & b_{32}(\omega,h) & b_{33}(\omega,h) & b_{34}(\omega,h) \\
    b_{41}(\omega,h) & b_{42}(\omega,h) & b_{43}(\omega,h) & b_{44}(\omega,h) \\
  \end{Vmatrix} \cdot
  \begin{Vmatrix}
    \alpha_{k}(\omega) \\
    \beta_{k}(\omega) \\
    \gamma_{k}(\omega) \\
    \delta_{k}(\omega)
  \end{Vmatrix}
$$
or in continous notations:
\begin{equation*}
  \begin{Vmatrix}
    \alpha(\omega,y-h) \\
    \beta(\omega,y-h) \\
    \gamma(\omega,y-h) \\
    \delta(\omega,y-h)
  \end{Vmatrix} =
  \begin{Vmatrix}
    b_{11}(\omega,h) & b_{12}(\omega,h) & b_{13}(\omega,h) & b_{14}(\omega,h) \\
    b_{21}(\omega,h) & b_{22}(\omega,h) & b_{23}(\omega,h) & b_{24}(\omega,h) \\
    b_{31}(\omega,h) & b_{32}(\omega,h) & b_{33}(\omega,h) & b_{34}(\omega,h) \\
    b_{41}(\omega,h) & b_{42}(\omega,h) & b_{43}(\omega,h) & b_{44}(\omega,h) \\
  \end{Vmatrix} \cdot
  \begin{Vmatrix}
    \alpha(\omega,y) \\
    \beta(\omega,y) \\
    \gamma(\omega,y) \\
    \delta(\omega,y)
  \end{Vmatrix}
\end{equation*}

In order to obtain a system of differential equations, we subtract from both parts the vector $\overrightarrow{\alpha}(\omega,y)$ and divide then by $-h$:
$$
  \frac{\overrightarrow{\alpha}(\omega,y-h) -
  \overrightarrow{\alpha}(\omega,y)}{-h} =
  \frac{B_k(\omega,h) - Id}{-h}
  \overrightarrow{\alpha}(\omega,y)
$$
or in expanded form:
$$
  \begin{Vmatrix}
    \frac{\alpha(\omega,y-h)-\alpha(\omega,y)}{-h} \\
    \frac{\beta(\omega,y-h)-\beta(\omega,y)}{-h} \\
    \frac{\gamma(\omega,y-h)-\gamma(\omega,y)}{-h} \\
    \frac{\delta(\omega,y-h)-\delta(\omega,y)}{-h}
  \end{Vmatrix} =
  \begin{Vmatrix}
    \frac{b_{11}(\omega,y)-1}{-h} & \frac{b_{12}(\omega,y)}{-h} & \frac{b_{13}(\omega,y)}{-h} &
    \frac{b_{14}(\omega,y)}{-h} \\

    \frac{b_{21}(\omega,y)}{-h} & \frac{b_{22}(\omega,y)-1}{-h} & \frac{b_{23}(\omega,y)}{-h} &
    \frac{b_{24}(\omega,y)}{-h} \\

    \frac{b_{31}(\omega,y)}{-h} & \frac{b_{32}(\omega,y)}{-h} & \frac{b_{33}(\omega,y)-1}{-h} &
    \frac{b_{34}(\omega,y)}{-h} \\

    \frac{b_{41}(\omega,y)}{-h} & \frac{b_{42}(\omega,y)}{-h} & \frac{b_{43}(\omega,y)}{-h} &
    \frac{b_{44}(\omega,y)-1}{-h} \\
  \end{Vmatrix}\cdot
  \begin{Vmatrix}
    \alpha(\omega,y) \\
    \beta(\omega,y) \\
    \gamma(\omega,y) \\
    \delta(\omega,y)
  \end{Vmatrix}
$$

Let $h\rightarrow 0$. One obtains a system of linear differential equations:
\begin{equation*}
  \begin{Vmatrix}
    \frac{d\alpha(\omega,y)}{dy} \\
    \frac{d\beta(\omega,y)}{dy} \\
    \frac{d\gamma(\omega,y)}{dy} \\
    \frac{d\delta(\omega,y)}{dy}
  \end{Vmatrix} =
  \begin{Vmatrix}
    \lim\limits_{h\to 0}\frac{b_{11}(\omega,y)-1}{-h} & \lim\limits_{h\to 0}\frac{b_{12}(\omega,y)}{-h} &
    \lim\limits_{h\to 0}\frac{b_{13}(\omega,y)}{-h} & \lim\limits_{h\to 0}\frac{b_{14}(\omega,y)}{-h} \\

    \lim\limits_{h\to 0}\frac{b_{21}(\omega,y)}{-h} & \lim\limits_{h\to 0}\frac{b_{22}(\omega,y)-1}{-h} &
    \lim\limits_{h\to 0}\frac{b_{23}(\omega,y)}{-h} & \lim\limits_{h\to 0}\frac{b_{24}(\omega,y)}{-h} \\

    \lim\limits_{h\to 0}\frac{b_{31}(\omega,y)}{-h} & \lim\limits_{h\to 0}\frac{b_{32}(\omega,y)}{-h} &
    \lim\limits_{h\to 0}\frac{b_{33}(\omega,y)-1}{-h} & \lim\limits_{h\to 0}\frac{b_{34}(\omega,y)}{-h} \\

    \lim\limits_{h\to 0}\frac{b_{41}(\omega,y)}{-h} & \lim\limits_{h\to 0}\frac{b_{42}(\omega,y)}{-h} &
    \lim\limits_{h\to 0}\frac{b_{43}(\omega,y)}{-h} & \lim\limits_{h\to 0}\frac{b_{44}(\omega,y)-1}{-h} \\
  \end{Vmatrix}\cdot
  \begin{Vmatrix}
    \alpha(\omega,y) \\
    \beta(\omega,y) \\
    \gamma(\omega,y) \\
    \delta(\omega,y)
  \end{Vmatrix}
\end{equation*}
or, in matrix form,
$$
  \frac{d\overrightarrow{\alpha}}{dy} = A \overrightarrow{\alpha}(\omega,y).
$$
Now, we have to calculate the matrix $A$.
$$
  \lim_{h\to 0}\frac{b_{11}-1}{-h} =
  \lim_{h\to 0}\frac{b_{13}}{-h} = 0,
$$
$$
  \lim_{h\to 0}\frac{b_{12}}{-h} = \frac{2k_1-A}{k_1(2-Ak_1)}\omega,
$$
$$
  \lim_{h\to 0}\frac{b_{14}}{-h} = \frac{1-k_1^2}{2-Ak_1}\omega.
$$
So, we obtain the first differential equation:
$$
  \frac{d\alpha}{dy} = \frac{2k_1-A}{k_1(2-Ak_1)}\omega\beta(\omega,y) +
  \frac{1-k_1^2}{2-Ak_1}\omega \delta(\omega,y)
$$
or after simplification
\begin{equation}\label{alpha1}
  \frac{d\alpha}{dy} = -\frac{\omega}{k_1^2}\beta(\omega,y) +
  \omega \delta(\omega,y)
\end{equation}
We consider the second equation:
$$
  \lim_{h\to 0}\frac{b_{21}}{-h} = \omega k_1^2
  \frac{B_1k_2^2-B_2}{B_1-B_2},
$$
$$
  \lim_{h\to 0}\frac{b_{22}-1}{-h} = 2\frac{d}{dy}\ln k_1(y),
$$
$$
  \lim_{h\to 0}\frac{b_{23}}{-h} = \omega k_1^2
  \frac{1-k_1^2}{B_1-B_2},
$$
$$
  \lim_{h\to 0}\frac{b_{24}}{-h} = 0.
$$
Therefore
\begin{multline}\label{alpha2}
  \od{\beta}{y} = \omega k_1^2
  \frac{B_1k_2^2-B_2}{B_1-B_2}\alpha(\omega,y) +\\+
  2\frac{d}{dy}\ln k_1(y)\beta(\omega,y) +
  \omega k_1^2\frac{1-k_1^2}{B_1-B_2}\gamma(\omega,y)
\end{multline}
Third equation:
$$
  \lim_{h\to 0}\frac{b_{31}}{-h} = 0,
$$
$$
  \lim_{h\to 0}\frac{b_{32}}{-h} = \omega
  \frac{2B_1k_1-AB_2}{k_1(2-Ak_1)},
$$
$$
  \lim_{h\to 0}\frac{b_{33}-1}{-h} =
  -\od{}{y}\ln E(y),
$$
$$
  \lim_{h\to 0}\frac{b_{34}}{-h} =
  \omega\frac{B_2-B_1k_1^2}{2-Ak_1}.
$$
\begin{multline}\label{alpha3}
  \od{\gamma}{y} = \omega\frac{2B_1k_1-AB_2}{k_1(2-Ak_1)}
  \beta(\omega,y) - \od{}{y}\ln E(y)\gamma(\omega,y) +\\+
  \omega\frac{B_2-B_1k_1^2}{2-Ak_1}\delta(\omega,y)
\end{multline}
Fourth equation:
$$
  \lim_{h\to 0}\frac{b_{41}}{-h} =
  \omega\frac{2B_1k_2^2-AB_2k_1}{B_1-B_2},
$$
$$
  \lim_{h\to 0}\frac{b_{42}}{-h} = 0,
$$
$$
  \lim_{h\to 0}\frac{b_{43}}{-h} = \omega
  \frac{Ak_1-2k_2^2}{B_1-B_2},
$$
$$
  \lim_{h\to 0}\frac{b_{44}-1}{-h} =
  -\od{}{y}\ln \mu(\omega,y).
$$
\begin{multline}\label{alpha4}
  \od{\delta}{y} = \omega\frac{2B_1k_2^2-AB_2k_1}{B_1-B_2}
  \alpha(\omega,y) +
  \omega\frac{Ak_1-2k_2^2}{B_1-B_2}\gamma(\omega,y) -\\
  -\od{}{y}\ln \mu(\omega,y)\delta(\omega,y)
\end{multline}

Putting together the new equations (\ref{alpha1}), (\ref{alpha2}), (\ref{alpha3}), (\ref{alpha4}) leads to the following system of ordinary differential equations:
\begin{eqnarray*}
  \od{\alpha}{y} &=& -\frac{\omega}{k_1^2}\beta(\omega,y) +
  \omega \delta(\omega,y), \\
  \od{\beta}{y} &=& \omega k_1^2
  \frac{B_1k_2^2-B_2}{B_1-B_2}\alpha(\omega,y) +
  2\frac{d}{dy}\ln k_1(y)\beta(\omega,y) +
  \omega k_1^2\frac{1-k_1^2}{B_1-B_2}\gamma(\omega,y), \\
  \od{\gamma}{y} &=& \omega\frac{2B_1k_1-AB_2}{k_1(2-Ak_1)}
  \beta(\omega,y) - \od{}{y}\ln E(y)\gamma(\omega,y) +
  \omega\frac{B_2-B_1k_1^2}{2-Ak_1}\delta(\omega,y), \\
  \od{\delta}{y} &=& \omega\frac{2B_1k_2^2-AB_2k_1}{B_1-B_2}
  \alpha(\omega,y) +
  \omega\frac{Ak_1-2k_2^2}{B_1-B_2}\gamma(\omega,y)
  -\od{}{y}\ln \mu(\omega,y)\delta(\omega,y).
\end{eqnarray*}

\subsection{Lamzyuk-Privarnikov functions in the case of
inhomogeneous layer}

Now, we have to take the limit in the recurrence relations for Lamzyuk-Privarnikov's functions. It will be a little bit more complicated because these relations are not linear as in the previous case.

First of all, we start by introducing continuous notation for Lamzyuk-Privarnikov functions:
$$
  A_k(\omega)\equiv A(\omega,y),\qquad A_{k+1}(\omega)\equiv
  A(\omega,y+h),
$$
$$
  B_k(\omega)\equiv B(\omega,y),\qquad B_{k+1}(\omega)\equiv
  B(\omega,y+h),
$$
$$
  C_k(\omega)\equiv C(\omega,y),\qquad C_{k+1}(\omega)\equiv
  C(\omega,y+h),
$$
$$
  D_k(\omega)\equiv D(\omega,y),\qquad D_{k+1}(\omega)\equiv
  D(\omega,y+h).
$$
Also, earlier, we established recurrence relations for these functions:
\begin{eqnarray*}
  A(\omega,y+h) = \frac{Y_n(\omega)}{R_n(\omega)}, \qquad
  B(\omega,y+h) &=& \frac{X_n(\omega)}{R_n(\omega)}, \\
  C(\omega,y+h) = \frac{Q_n(\omega)}{R_n(\omega)}, \qquad
  D(\omega,y+h) &=& \frac{Z_n(\omega)}{R_n(\omega)}.
\end{eqnarray*}
From both parts of these identities we subtract $A(\omega,y)$, $B(\omega,y)$, $C(\omega,y)$, $D(\omega,y)$ respectively and divide them by $h$:
\begin{eqnarray*}
  \frac{A(\omega,y+h)-A(\omega,y)}{h} &=&
  \frac{Y_n(\omega)-A_n(\omega)R_n(\omega)}{hR_n(\omega)}, \\
  \frac{B(\omega,y+h)-B(\omega,y)}{h} &=&
  \frac{X_n(\omega)-B_n(\omega)R_n(\omega)}{hR_n(\omega)}, \\
  \frac{C(\omega,y+h)-C(\omega,y)}{h} &=&
  \frac{Q_n(\omega)-C_n(\omega)R_n(\omega)}{hR_n(\omega)}, \\
  \frac{D(\omega,y+h)-D(\omega,y)}{h} &=&
  \frac{Z_n(\omega)-D_n(\omega)R_n(\omega)}{hR_n(\omega)}.
\end{eqnarray*}
Taking the limit as $h\to 0$ yields differential equations:
\begin{eqnarray*}
  \od{A}{y} &=& \lim_{h\to 0}
  \frac{Y_n(\omega)-A_n(\omega)R_n(\omega)}{hR_n(\omega)}, \\
  \od{B}{y} &=& \lim_{h\to 0}
  \frac{X_n(\omega)-B_n(\omega)R_n(\omega)}{hR_n(\omega)}, \\
  \od{C}{y} &=& \lim_{h\to 0}
  \frac{Q_n(\omega)-C_n(\omega)R_n(\omega)}{hR_n(\omega)}, \\
  \od{D}{y} &=& \lim_{h\to 0}
  \frac{Z_n(\omega)-D_n(\omega)R_n(\omega)}{hR_n(\omega)}.
\end{eqnarray*}

Let us calculate these limits. We start with the first equation. In expanded form, this limit is:
\begin{multline*}
  \lim_{h\to 0}\frac{Y_n(\omega,h)-A_n(\omega,y)R_n(\omega,h)}{R_n(\omega,h)}=
  \lim_{h\to 0}
  ((g_{11}(\omega)+g_{12}(\omega)A_n(\omega)+\\+
  g_{13}(\omega)B_n(\omega)+ g_{14}(\omega)C_n(\omega)
  +g_{15}(\omega)D_n(\omega)+g_{16}(\omega)\Delta_n(\omega))-\\-A_n(\omega)(
  g_{51}(\omega)+g_{52}(\omega)A_n(\omega)+g_{53}(\omega)
  B_n(\omega)+g_{54}(\omega)C_n(\omega)]+\\+g_{55}(\omega)D_n(\omega)+g_{56}(\omega)
  \Delta_n(\omega)))/((g_{51}(\omega)+g_{52}(\omega)A_n(\omega)+g_{53}(\omega)
  B_n(\omega)+\\+g_{54}C_n(\omega)+g_{55}(\omega)D_n(\omega)+g_{56}(\omega)\Delta_n(\omega))h).
\end{multline*}

The limit of the denominator is:
\begin{equation*}
  \lim_{h\to 0} g_{51} = k_1(y),
\end{equation*}
\begin{equation*}
  \lim_{h\to 0} g_{52} = \lim_{h\to 0} g_{53} =
  \lim_{h\to 0} g_{54} = \lim_{h\to 0} g_{55} =
  \lim_{h\to 0} g_{56} = 0.
\end{equation*}
Absolute term of the equation:
\begin{equation*}
  \lim_{h\to 0}\frac{g_{11}}{h} = 0
\end{equation*}
Factor of $A$:
$$
  \lim_{h\to 0} \frac{g_{12}-g_{51}}{h} = k_1(y)\od{}{y}\ln E(y)
$$
Factor of $B$:
$$
  \lim_{h\to 0} \frac{g_{13}}{h} = -\frac{\omega}{k_1(y)}
$$
Factor of C:
$$
  \lim_{h\to 0} \frac{g_{14}}{h} =
  -\frac{k_1k_2\omega(2k_2-Ak_1)}{B_1-B_2}
$$
Factor of D:
$$
  \lim_{h\to 0} \frac{g_{15}}{h} = 0
$$
Factor of $BC$:
$$
  \lim_{h\to 0} \frac{-g_{16}}{h} = 0
$$
Factor of $AD$:
$$
  \lim_{h\to 0} \frac{g_{16}-g_{55}}{h} = 0
$$
Factor of $A(AD-BC)$:
$$
  \lim_{h\to 0} \frac{g_{56}}{h} = 0
$$
Factor of $A^2$:
$$
  \lim_{h\to 0} \frac{-g_{52}}{h} = 0
$$
Factor of $AB$:
$$
  \lim_{h\to 0} \frac{-g_{53}}{h} = -\omega
  \frac{AB_2-2B_1k_1}{2-Ak_1}
$$
Factor of $AC$:
$$
  \lim_{h\to 0} \frac{-g_{54}}{h} = -\omega
  \frac{k_1(AB_2k_1-2B_1k_2^2)}{B_1-B_2}.
$$
So, we obtain the first equation:
\begin{multline}\label{Aeq}
  \od{A}{y} = \od{}{y}\ln E(y) A(y,\omega) -
  \frac{\omega}{k_1^2(y)}B(y,\omega) -
  \omega\frac{k_2(2k_2-Ak_1)}{B_1-B_2}C(y,\omega) -\\-
  \omega\frac{AB_2-2B_1k_1}{k_1(1-k_1^2)}A(y,\omega)B(y,\omega) -
  \omega\frac{AB_2k_1-2B_1k_2^2}{B_1-B_2}A(y,\omega)C(y,\omega).
\end{multline}

Now, we perform the calculations to obtain the second equation. We have to compute the limit:
\begin{multline*}
  \lim_{h\to 0}\frac{X_n(\omega,h)-B_n(\omega,y)R_n(\omega,h)}{R_n(\omega,h)}=
  \lim_{h\to 0}
  ((g_{21}(\omega)+g_{22}(\omega)A_n(\omega)+\\+
  g_{23}(\omega)B_n(\omega)+ g_{24}(\omega)C_n(\omega)
  +g_{25}(\omega)D_n(\omega)+g_{26}(\omega)\Delta_n(\omega))-\\-B_n(\omega)(
  g_{51}(\omega)+g_{52}(\omega)A_n(\omega)+g_{53}(\omega)
  B_n(\omega)+g_{54}(\omega)C_n(\omega)]+\\+g_{55}(\omega)D_n(\omega)+g_{56}(\omega)
  \Delta_n(\omega)))/((g_{51}(\omega)+g_{52}(\omega)A_n(\omega)+g_{53}(\omega)
  B_n(\omega)+\\+g_{54}C_n(\omega)+g_{55}(\omega)D_n(\omega)+g_{56}(\omega)\Delta_n(\omega))h).
\end{multline*}

Absolute term of the equation:
$$
  \lim_{h\to 0} \frac{g_{21}}{h} = -\omega
  \frac{k_1^3(1-k_2^2)}{B_1-B_2}
$$
Factor of $A$:
$$
  \lim_{h\to 0} \frac{g_{22}}{h} = -\omega
  \frac{k_1^3(B_2-B_1k_2^2)}{B_1-B_2}
$$
Factor of $B$:
$$
  \lim_{h\to 0} \frac{g_{23}-g_{51}}{h} =
  2\od{k_1}{y} + k_1(y)\od{k_1}{y}\ln E(y)
$$
Factor of $C$:
$$
  \lim_{h\to 0} \frac{g_{24}}{h} = 0
$$
Factor of $D$:
$$
  \lim_{h\to 0} \frac{g_{25}}{h} = -\omega
  \frac{k_1(Ak_1-2k_2^2)}{B_1-B_2}
$$
Factor of $AD$:
$$
  \lim_{h\to 0} \frac{g_{26}}{h} = -\omega
  \frac{k_1(AB_2k_1-2B_1k_2^2)}{B_1-B_2}
$$
Factor of $BC$:
$$
  \lim_{h\to 0} \frac{-g_{26}-g_{54}}{h} = 0
$$
Factor of $AB$:
$$
  \lim_{h\to 0} \frac{-g_{52}}{h} = 0
$$
Factor of $B^2$:
$$
  \lim_{h\to 0} \frac{-g_{53}}{h} = \omega
  \frac{2B_1k_1-AB_2}{2-Ak_1}
$$
Factor of $BD$:
$$
  \lim_{h\to 0} \frac{-g_{55}}{h} = 0
$$
Factor of $B(AD-BC)$:
$$
  \lim_{h\to 0} \frac{-g_{56}}{h} = 0.
$$
The second equation is
\begin{multline}\label{Beq}
  \od{B}{y} = -\omega\frac{k_1^2(1-k_2^2)}{B_1-B_2} -
  -\omega\frac{k_1^2(B_2-B_1k_2^2)}{B_1-B_2}A(y,\omega) +
  \od{}{y}\ln(k_1^2E)B -\\-
  \omega\frac{Ak_1-2k_2^2}{B_1-B_2}D(y,\omega)
  -\omega\frac{AB_2k_1-2B_1k_2^2}{B_1-B_2}A(y,\omega)D(y,\omega)
  -\\-\omega\frac{AB_2-2B_1k_1}{k_1(1-k_1^2)}B^2(y,\omega).
\end{multline}

We consider the third equation:
\begin{multline*}
  \lim_{h\to 0}\frac{Q_n(\omega,h)-C_n(\omega,y)R_n(\omega,h)}{R_n(\omega,h)}=
  \lim_{h\to 0}
  ((g_{31}(\omega)+g_{32}(\omega)A_n(\omega)+\\+
  g_{33}(\omega)B_n(\omega)+ g_{34}(\omega)C_n(\omega)
  +g_{35}(\omega)D_n(\omega)+g_{36}(\omega)\Delta_n(\omega))-\\-C_n(\omega)(
  g_{51}(\omega)+g_{52}(\omega)A_n(\omega)+g_{53}(\omega)
  B_n(\omega)+g_{54}(\omega)C_n(\omega)]+\\+g_{55}(\omega)D_n(\omega)+g_{56}(\omega)
  \Delta_n(\omega)))/((g_{51}(\omega)+g_{52}(\omega)A_n(\omega)+g_{53}(\omega)
  B_n(\omega)+\\+g_{54}C_n(\omega)+g_{55}(\omega)D_n(\omega)+g_{56}(\omega)\Delta_n(\omega))h).
\end{multline*}

Absolute term of the equation:
$$
  \lim_{h\to 0} \frac{g_{31}}{h} = -\omega k_1
$$
Factor of $A$:
$$
  \lim_{h\to 0} \frac{g_{32}}{h} = -\omega
  \frac{k_1(B_2-B_1k_1^2)}{2-Ak_1}
$$
Factor of $B$:
$$
  \lim_{h\to 0} \frac{g_{33}}{h} = 0
$$
Factor of $C$:
$$
  \lim_{h\to 0} \frac{g_{34}-g_{51}}{h} = k_1\od{}{y}\ln\mu(y)
$$
Factor of $D$:
$$
  \lim_{h\to 0} \frac{g_{35}}{h} = -\frac{\omega}{k_1}
$$
Factor of $AD$:
$$
  \lim_{h\to 0} \frac{g_{36}}{h} = -\omega
  \frac{AB_2-2B_1k_1}{2-Ak_1}
$$
Factor of $BC$:
$$
  \lim_{h\to 0} \frac{-g_{36}-g_{53}}{h} = 0
$$
Factor of $AC$:
$$
  \lim_{h\to 0} \frac{-g_{52}}{h} = 0
$$
Factor of $C^2$:
$$
  \lim_{h\to 0} \frac{-g_{54}}{h} =
  -\omega\frac{k_1(AB_2k_1-2B_1k_2^2)}{B_1-B_2}
$$
Factor of $CD$:
$$
  \lim_{h\to 0} \frac{-g_{55}}{h} = 0
$$
Factor of $C(AD-BC)$:
$$
  \lim_{h\to 0} \frac{-g_{56}}{h} = 0.
$$
Here is the third equation:
\begin{multline}\label{Ceq}
  \od{C}{y} = -\omega -\omega\frac{B_2-B_1k_1^2}{1-k_1^2}A(y,\omega)
  +\od{}{y}\ln\mu(y)C(y,\omega) - \frac{\omega}{k_1^2}D(y,\omega)
  -\\-\omega\frac{AB_2-2B_1k_1}{k_1(1-k_1^2)}A(y,\omega)D(y,\omega) -
  \omega\frac{AB_2k_1-2B_1k_2^2}{B_1-B_2}C^2(y,\omega).
\end{multline}
Now we consider the last one:
\begin{multline*}
  \lim_{h\to 0}\frac{Z_n(\omega,h)-D_n(\omega,y)R_n(\omega,h)}{R_n(\omega,h)}=
  \lim_{h\to 0}
  ((g_{41}(\omega)+g_{42}(\omega)A_n(\omega)+\\+
  g_{43}(\omega)B_n(\omega)+ g_{44}(\omega)C_n(\omega)
  +g_{45}(\omega)D_n(\omega)+g_{46}(\omega)\Delta_n(\omega))-\\-D_n(\omega)(
  g_{51}(\omega)+g_{52}(\omega)A_n(\omega)+g_{53}(\omega)
  B_n(\omega)+g_{54}(\omega)C_n(\omega)]+\\+g_{55}(\omega)D_n(\omega)+g_{56}(\omega)
  \Delta_n(\omega)))/((g_{51}(\omega)+g_{52}(\omega)A_n(\omega)+g_{53}(\omega)
  B_n(\omega)+\\+g_{54}C_n(\omega)+g_{55}(\omega)D_n(\omega)+g_{56}(\omega)\Delta_n(\omega))h).
\end{multline*}

Absolute term of the equation:
$$
  \lim_{h\to 0} \frac{g_{41}}{h} = 0
$$
Factor of $A$:
$$
  \lim_{h\to 0} \frac{g_{42}}{h} = 0
$$
Factor of $B$:
$$
  \lim_{h\to 0} \frac{g_{43}}{h} = -\omega
  \frac{k_1(B_2-B_1k_1^2)}{1-k_1^2}
$$
Factor of $C$:
$$
  \lim_{h\to 0} \frac{g_{44}}{h} =
  -\omega\frac{k_1^3(B_2-B_1k_2^2)}{B_1-B_2}
$$
Factor of $D$:
$$
  \lim_{h\to 0} \frac{g_{45}-g_{51}}{h} =
  2\od{k_1}{y} + k_1\od{}{y}\ln \mu(y)
$$
Factor of $BC$:
$$
  \lim_{h\to 0} \frac{-g_{46}}{h} = 0
$$
Factor of $AD$:
$$
  \lim_{h\to 0} \frac{g_{46}-g_{52}}{h} = 0
$$
Factor of $BD$:
$$
  \lim_{h\to 0} \frac{-g_{53}}{h} = \omega
  \frac{AB_2-2B_1k_1}{1-k_1^2}
$$
Factor of $DC$:
$$
  \lim_{h\to 0} \frac{-g_{54}}{h} = -\omega
  \frac{k_1(AB_2k_1-2k_2^2B_1)}{B_1-B_2}
$$
Factor of $D^2$:
$$
  \lim_{h\to 0} \frac{-g_{55}}{h} = 0
$$
Factor of $D(AD-BC)$:
$$
  \lim_{h\to 0} \frac{g_{56}}{h} = 0.
$$
The fourth equation reads:
\begin{multline}\label{Deq}
  \od{D}{y} = -\omega\frac{B_2-B_1k_1^2}{1-k_1^2}B(y,\omega)
  -\omega\frac{k_1^2(B_2-B_1k_2^2)}{B_1-B_2}C(y,\omega) +\\+
  \od{}{y}\ln(k_1\mu)D(y,\omega) -
  \omega\frac{AB_2-2B_1k_1}{k_1(1-k_1^2)}B(y,\omega)D(y,\omega) +\\+
  \omega\frac{AB_2k_1-2k_2^2B_1}{B_1-B_2}D(y,\omega)C(y,\omega).
\end{multline}

After putting together equations (\ref{Aeq}), (\ref{Beq}), (\ref{Ceq}), (\ref{Deq}) we obtain a system of ordinary differential equations of Riccati type:
\begin{eqnarray*}
  \od{A}{y} &=& \od{}{y}\ln E(y) A(y,\omega) -
  \frac{\omega}{k_1^2(y)}B(y,\omega) -
  \omega\frac{k_2(2k_2-Ak_1)}{B_1-B_2}C(y,\omega) -\\
  & & \omega\frac{AB_2-2B_1k_1}{k_1(1-k_1^2)}A(y,\omega)B(y,\omega) -
  \omega\frac{AB_2k_1-2B_1k_2^2}{B_1-B_2}A(y,\omega)C(y,\omega), \\
  \od{B}{y} &=& -\omega\frac{k_1^2(1-k_2^2)}{B_1-B_2}
  -\omega\frac{k_1^2(B_2-B_1k_2^2)}{B_1-B_2}A(y,\omega) +
  \od{}{y}\ln(k_1^2E)B(y,\omega) -\\
  & & -\omega\frac{Ak_1-2k_2^2}{B_1-B_2}D(y,\omega)
  -\omega\frac{AB_2k_1-2B_1k_2^2}{B_1-B_2}A(y,\omega)D(y,\omega) \\
  & &-\omega\frac{AB_2-2B_1k_1}{k_1(1-k_1^2)}B^2(y,\omega), \\
  \od{C}{y} &=& -\omega -\omega\frac{B_2-B_1k_1^2}{1-k_1^2}A(y,\omega)
  +\od{}{y}\ln\mu(y)C(y,\omega) - \frac{\omega}{k_1^2}D(y,\omega)
  -\\ & &-\omega\frac{AB_2-2B_1k_1}{k_1(1-k_1^2)}A(y,\omega)D(y,\omega) -
  \omega\frac{AB_2k_1-2B_1k_2^2}{B_1-B_2}C^2(y,\omega), \\
  \od{D}{y} &=& -\omega\frac{B_2-B_1k_1^2}{1-k_1^2}B(y,\omega)
  -\omega\frac{k_1^2(B_2-B_1k_2^2)}{B_1-B_2}C(y,\omega) +
  \od{}{y}\ln(k_1\mu)D(y,\omega) -\\
  & & -\omega\frac{AB_2-2B_1k_1}{k_1(1-k_1^2)}B(y,\omega)D(y,\omega) +
  \omega\frac{AB_2k_1-2k_2^2B_1}{B_1-B_2}D(y,\omega)C(y,\omega).
\end{eqnarray*}

This system of ordinary differential equations plays exactly the same role as recurrence relations for Lamzyuk-Privarnikov functions in discrete cases. Similarly, the equations of this system depend only on the mechanical properties of the layer. Thus, the solution of this system can be used for a number of problems. The initial conditions for this system of ODE are discussed below.

We would like to mention an open problem here. It is a well-known fact that solutions of Riccati-type differential equations can achieve infinity for finite values of the argument. Numerical experiences show that this situation can be realized, but not always. It would be interesting to know what are conditions on mechanical properties that do not allow the solutions to tend to infinity when $y\in [0,h]$. And what is the interpretation of this behaviour in the context of the theory of elasticity boundary value problems?

\subsection{Asymptotic analysis of the resulting equations}

It is useful to know the asymptotic behaviour of the solutions of system (\ref{Aeq}), (\ref{Beq}), (\ref{Ceq}), (\ref{Deq}). More precisely, we have to calculate the limit:
$$
  B_{\infty} := \lim_{\omega\to\infty} B(h,\omega),
$$
because it appears in a singular integral equation.

In order to be able to calculate it, we write asymptotic expansion for $\omega\gg 1$:
\begin{eqnarray*}
  A(y,\omega) &=& A_0(y) + \frac1{\omega}A_1(y) + \ldots \\
  B(y,\omega) &=& B_0(y) + \frac1{\omega}B_1(y) + \ldots \\
  C(y,\omega) &=& C_0(y) + \frac1{\omega}C_1(y) + \ldots \\
  D(y,\omega) &=& D_0(y) + \frac1{\omega}D_1(y) + \ldots
\end{eqnarray*}
and divide the system of equations (\ref{Aeq}), (\ref{Beq}), (\ref{Ceq}), (\ref{Deq}) by $\omega$:
\begin{eqnarray*}
  \frac1{\omega}\od{A}{y} &=& \frac1{\omega}\od{}{y}\ln E(y) A(y,\omega) -
  \frac1{k_1^2(y)}B(y,\omega) -
  \frac{k_2(2k_2-Ak_1)}{B_1-B_2}C(y,\omega) -\\
  & & \frac{AB_2-2B_1k_1}{k_1(1-k_1^2)}A(y,\omega)B(y,\omega) -
  \frac{AB_2k_1-2B_1k_2^2}{B_1-B_2}A(y,\omega)C(y,\omega), \\
  \frac1{\omega}\od{B}{y} &=& -\frac{k_1^2(1-k_2^2)}{B_1-B_2}
  -\frac{k_1^2(B_2-B_1k_2^2)}{B_1-B_2}A(y,\omega) +
  \frac1{\omega}\od{}{y}\ln(k_1^2E)B(y,\omega) -\\
  & & -\frac{Ak_1-2k_2^2}{B_1-B_2}D(y,\omega)
  -\frac{AB_2k_1-2B_1k_2^2}{B_1-B_2}A(y,\omega)D(y,\omega) \\
  & &-\frac{AB_2-2B_1k_1}{k_1(1-k_1^2)}B^2(y,\omega), \\
  \frac1{\omega}\od{C}{y} &=& -1 -\frac{B_2-B_1k_1^2}{1-k_1^2}A(y,\omega)
  +\frac1{\omega}\od{}{y}\ln\mu(y)C(y,\omega) - \frac1{k_1^2}D(y,\omega)
  -\\ & &-\frac{AB_2-2B_1k_1}{k_1(1-k_1^2)}A(y,\omega)D(y,\omega) -
  \frac{AB_2k_1-2B_1k_2^2}{B_1-B_2}C^2(y,\omega), \\
  \frac1{\omega}\od{D}{y} &=& -\frac{B_2-B_1k_1^2}{1-k_1^2}B(y,\omega)
  -\frac{k_1^2(B_2-B_1k_2^2)}{B_1-B_2}C(y,\omega) +
  \frac1{\omega}\od{}{y}\ln(k_1\mu)D(y,\omega) -\\
  & & -\frac{AB_2-2B_1k_1}{k_1(1-k_1^2)}B(y,\omega)D(y,\omega) +
  \frac{AB_2k_1-2k_2^2B_1}{B_1-B_2}D(y,\omega)C(y,\omega).
\end{eqnarray*}

Now we substitute the asymptotic expansions in these equations, and we look only at the zeroth order in $\omega$. We easily obtain the following system of algebraic equations:
\begin{eqnarray*}
  -\frac1{k_1^2(y)}B_0 -
  \frac{k_2(2k_2-Ak_1)}{B_1-B_2}C_0 -
  \frac{AB_2-2B_1k_1}{k_1(1-k_1^2)}A_0B_0 - & & \\
  \frac{AB_2k_1-2B_1k_2^2}{B_1-B_2}A_0C_0 &=& 0, \\
  -\frac{k_1^2(1-k_2^2)}{B_1-B_2}-\frac{k_1^2(B_2-B_1k_2^2)}{B_1-B_2}A_0
  -\frac{Ak_1-2k_2^2}{B_1-B_2}D_0 - &&\\
  -\frac{AB_2k_1-2B_1k_2^2}{B_1-B_2}A_0D_0
  -\frac{AB_2-2B_1k_1}{k_1(1-k_1^2)}B_0^2 &=& 0, \\
  -1 -\frac{B_2-B_1k_1^2}{1-k_1^2}A_0 - \frac1{k_1^2}D_0
  -\frac{AB_2-2B_1k_1}{k_1(1-k_1^2)}A_0D_0 - &&\\
  \frac{AB_2k_1-2B_1k_2^2}{B_1-B_2}C_0^2 &=& 0, \\
  -\frac{B_2-B_1k_1^2}{1-k_1^2}B_0 -
  \frac{k_1^2(B_2-B_1k_2^2)}{B_1-B_2}C_0
  -\frac{AB_2-2B_1k_1}{k_1(1-k_1^2)}B_0D_0 + &&\\
  \frac{AB_2k_1-2k_2^2B_1}{B_1-B_2}D_0C_0 &=& 0.
\end{eqnarray*}
Note that this system does not have a unique solution\footnote{In fact, this system has ten solutions in the complex plane.} but only one of its solutions has a physical sense. Unfortunately, the author could not solve analytically this system. That is why we used Maple to solve numerically this system. The Maple code listing is cited in Appendix \ref{app5}.

\subsection{Resume}

In the previous subsections, we obtained two systems of ordinary differential equations. The first system
\begin{eqnarray*}
  \od{\alpha}{y} &=& -\frac{\omega}{k_1^2}\beta(\omega,y) +
  \omega \delta(\omega,y), \\
  \od{\beta}{y} &=& \omega k_1^2
  \frac{B_1k_2^2-B_2}{B_1-B_2}\alpha(\omega,y) +
  2\frac{d}{dy}\ln k_1(y)\beta(\omega,y) +
  \omega k_1^2\frac{1-k_1^2}{B_1-B_2}\gamma(\omega,y), \\
  \od{\gamma}{y} &=& \omega\frac{2B_1k_1-AB_2}{k_1(2-Ak_1)}
  \beta(\omega,y) - \od{}{y}\ln E(y)\gamma(\omega,y) +
  \omega\frac{B_2-B_1k_1^2}{2-Ak_1}\delta(\omega,y), \\
  \od{\delta}{y} &=& \omega\frac{2B_1k_2^2-AB_2k_1}{B_1-B_2}
  \alpha(\omega,y) +
  \omega\frac{Ak_1-2k_2^2}{B_1-B_2}\gamma(\omega,y)
  -\od{}{y}\ln \mu(\omega,y)\delta(\omega,y).
\end{eqnarray*}
gives the vector $\overrightarrow{\alpha}(\omega,y)$. Knowing this vector is equivalent to solving our problem, as we established earlier in (\ref{gensol}). For this system we have a boundary value problem because at the interface $\Gamma_3$ we know two components $\gamma(\omega,h)$, $\delta(\omega,h)$ and at the interface $\Gamma_2$ we know $\delta(\omega,0)$ and a linear relation between $\beta(\omega,0)$ and $\gamma(\omega,0)$.

In order to avoid solving the boundary value problem, we used the Lamzyuk-Privarnikov functions method. This method provides two conditions on the boundary $\Gamma_3$ that we did not have:
\begin{eqnarray}\label{contlp}
  \alpha(\omega,h) &=& -A(h,\omega)\gamma(\omega,h) -
  C(h,\omega)\delta(\omega,h), \\
  \beta(\omega,h) &=& -B(h,\omega)\gamma(\omega,h) -
  D(h,\omega)\delta(\omega,h)\label{contlp2}.
\end{eqnarray}
To determine these functions, we obtained the nonlinear system of ordinary differential equations (\ref{Aeq}), (\ref{Beq}), (\ref{Ceq}), (\ref{Deq}).

We have to establish initial conditions for this system. It is not very difficult. To obtain them, we consider the limit when the ice thickness $h\to 0$:
\begin{eqnarray*}
  \alpha(\omega,0) &=& -A(0,\omega)\gamma(\omega,0) -
  C(0,\omega)\delta(\omega,0), \\
  \beta(\omega,0) &=& -B(0,\omega)\gamma(\omega,0) -
  D(0,\omega)\delta(\omega,0).
\end{eqnarray*}
In the limit, we do not have any ice plates. Thus, its displacements ($\alpha$, $\beta$) are identically equal to zero for any applied forces ($\gamma$, $\delta$). From this simple consideration, we deduce initial conditions for system (\ref{Aeq}), (\ref{Beq}), (\ref{Ceq}), (\ref{Deq}):
\begin{equation*}
  A(0,\omega) = B(0,\omega) = C(0,\omega) = D(0,\omega)
  \equiv 0.
\end{equation*}

We are able to formulate the algorithm of our problem solution:
\begin{enumerate}
    \item\label{item1} First of all, we solve a Cauchy-type problem for the system (\ref{Aeq}), (\ref{Beq}), (\ref{Ceq}), (\ref{Deq}) and find the four functions $A(h,\omega)$, $B(h,\omega)$, $C(h,\omega)$, $D(h,\omega)$.
    \item With these functions determined in item (\ref{item1}) we find the two unknown components $\alpha$ and $\beta$ of the vector $\overrightarrow{\alpha}$ at the boundary $\Gamma_3$.
    \item Now we solve a Cauchy-type problem to determine the vector $\overrightarrow{\alpha}$.
    \item To obtain displacements and stresses, we have to perform an inverse Fourier transform.
\end{enumerate}

\section{Contact problem}

Actually, contact mechanics is a very developed field of the theory of elasticity. The best general references on this topic in the Russian literature are \cite{muskh1,muskh2}. A more mathematical treatment of the contact mechanics boundary value problems is \cite{gakhov}. In the Western literature, the best monograph on this subject is \cite{cont}.

\begin{figure}[htbp]
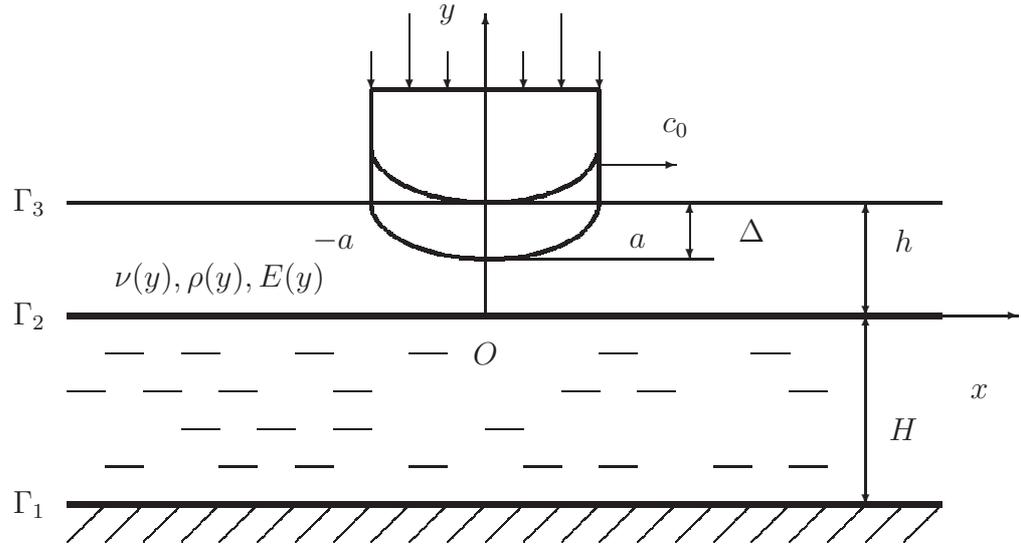

\begin{center}
\input contact1.tex
\end{center}
  \caption{Illustration for contact problem
  formulation.}\label{pict5}
\end{figure}

We make the following assumptions:
\begin{enumerate}
    \item The block is assumed to be rigid.
    \item The bottom of the block is given by the function $y=f(x)$, $x\in
[-a,a]$ with the following properties:
\begin{itemize}
    \item $f(x) \in C^{(1)}(-a,a)$,
    \item $f(x) = f(-x)$, $x\in [-a,a]$ in other words the block is
    symmetric.
\end{itemize}
    \item The block is immersed in the ice plate in such a way that there is full contact at the corner points $(-a,h)$ and $(a,h)$. This assumption gives us an explicit contact zone\footnote{In fact, we could consider the more general problem with an unknown contact zone, but in the frame of this work, we decided not to complicate it too much. Anyway, we should remark that at this point, a generalization is possible.}. This situation is illustrated in Figure~\ref{pict6}.
    \item Also, in order to simplify the singular integral equation, which we will obtain below, we suppose that there is no friction between the block and the ice plate.
\end{enumerate}

\begin{figure}[htbp]
\begin{center}
\unitlength 1mm
\begin{picture}(111.88,63.75)(0,0)

\linethickness{0.75mm}
\put(1.88,3.75){\line(1,0){100.00}}

\linethickness{0.15mm}
\multiput(1.88,-1.25)(0.12,0.12){42}{\line(1,0){0.12}}

\linethickness{0.15mm}
\multiput(6.88,-1.25)(0.12,0.12){42}{\line(1,0){0.12}}

\linethickness{0.15mm}
\multiput(11.88,-1.25)(0.12,0.12){42}{\line(1,0){0.12}}

\linethickness{0.15mm}
\multiput(16.88,-1.25)(0.12,0.12){42}{\line(1,0){0.12}}

\linethickness{0.15mm}
\multiput(21.88,-1.25)(0.12,0.12){42}{\line(1,0){0.12}}

\linethickness{0.15mm}
\multiput(26.88,-1.25)(0.12,0.12){42}{\line(1,0){0.12}}

\linethickness{0.15mm}
\multiput(31.88,-1.25)(0.12,0.12){42}{\line(1,0){0.12}}

\linethickness{0.15mm}
\multiput(36.88,-1.25)(0.12,0.12){42}{\line(1,0){0.12}}

\linethickness{0.15mm}
\multiput(41.88,-1.25)(0.12,0.12){42}{\line(1,0){0.12}}

\linethickness{0.15mm}
\multiput(46.88,-1.25)(0.12,0.12){42}{\line(1,0){0.12}}

\linethickness{0.15mm}
\multiput(51.88,-1.25)(0.12,0.12){42}{\line(1,0){0.12}}

\linethickness{0.15mm}
\multiput(56.88,-1.25)(0.12,0.12){42}{\line(1,0){0.12}}

\linethickness{0.15mm}
\multiput(66.88,-1.25)(0.12,0.12){42}{\line(1,0){0.12}}

\linethickness{0.15mm}
\multiput(61.88,-1.25)(0.12,0.12){42}{\line(1,0){0.12}}

\linethickness{0.15mm}
\multiput(71.88,-1.25)(0.12,0.12){42}{\line(1,0){0.12}}

\linethickness{0.15mm}
\multiput(76.88,-1.25)(0.12,0.12){42}{\line(1,0){0.12}}

\linethickness{0.15mm}
\multiput(81.88,-1.25)(0.12,0.12){42}{\line(1,0){0.12}}

\linethickness{0.15mm}
\multiput(86.88,-1.25)(0.12,0.12){42}{\line(1,0){0.12}}

\linethickness{0.15mm}
\multiput(91.88,-1.25)(0.12,0.12){42}{\line(1,0){0.12}}

\linethickness{0.15mm}
\multiput(96.88,-1.25)(0.12,0.12){42}{\line(1,0){0.12}}

\linethickness{0.65mm}
\put(1.88,23.75){\line(1,0){100.00}}

\linethickness{0.60mm}
\put(1.88,38.75){\line(1,0){35.00}}

\linethickness{0.65mm}
\put(66.88,38.75){\line(1,0){35.00}}

\linethickness{0.60mm}
\qbezier(36.88,38.75)(51.88,23.75)(66.88,38.75)

\linethickness{0.60mm}
\put(36.88,38.75){\line(0,1){10.00}}

\linethickness{0.60mm}
\put(36.88,48.75){\line(1,0){30.00}}

\linethickness{0.60mm}
\put(66.88,38.75){\line(0,1){10.00}}

\linethickness{0.15mm}
\put(96.88,38.75){\line(1,0){15.00}}
\put(111.88,38.75){\vector(1,0){0.12}}

\linethickness{0.15mm}
\put(51.88,38.75){\line(1,0){55.00}}

\linethickness{0.15mm}
\put(51.88,38.75){\line(0,1){25.00}}
\put(51.88,63.75){\vector(0,1){0.12}}

\put(111.88,33.75){\makebox(0,0)[cc]{$x$}}

\put(46.88,63.75){\makebox(0,0)[cc]{$y$}}

\linethickness{0.15mm}
\put(1.88,18.75){\line(1,0){5.00}}

\linethickness{0.15mm}
\put(16.88,18.75){\line(1,0){5.00}}

\linethickness{0.15mm}
\put(6.88,13.75){\line(1,0){5.00}}

\linethickness{0.15mm}
\put(31.88,18.75){\line(1,0){5.00}}

\linethickness{0.15mm}
\put(21.88,8.75){\line(1,0){5.00}}

\linethickness{0.15mm}
\put(46.88,18.75){\line(1,0){5.00}}

\linethickness{0.15mm}
\put(36.88,8.75){\line(1,0){5.00}}

\linethickness{0.15mm}
\put(1.88,8.75){\line(1,0){5.00}}

\linethickness{0.15mm}
\put(21.88,13.75){\line(1,0){5.00}}

\linethickness{0.15mm}
\put(66.88,18.75){\line(1,0){5.00}}

\linethickness{0.15mm}
\put(51.88,13.75){\line(1,0){5.00}}

\linethickness{0.15mm}
\put(56.88,8.75){\line(1,0){5.00}}

\linethickness{0.15mm}
\put(86.88,18.75){\line(1,0){5.00}}

\linethickness{0.15mm}
\put(76.88,8.75){\line(1,0){5.00}}

\linethickness{0.15mm}
\put(91.88,8.75){\line(1,0){5.00}}

\linethickness{0.15mm}
\put(66.88,13.75){\line(1,0){5.00}}

\linethickness{0.15mm}
\put(81.88,13.75){\line(1,0){5.00}}

\linethickness{0.15mm}
\put(41.88,48.75){\line(0,1){5.00}}
\put(41.88,48.75){\vector(0,-1){0.12}}

\linethickness{0.15mm}
\put(61.88,48.75){\line(0,1){5.00}}
\put(61.88,48.75){\vector(0,-1){0.12}}

\linethickness{0.15mm}
\put(46.88,48.75){\line(0,1){10.00}}
\put(46.88,48.75){\vector(0,-1){0.12}}

\linethickness{0.15mm}
\put(56.88,48.75){\line(0,1){10.00}}
\put(56.88,48.75){\vector(0,-1){0.12}}

\linethickness{0.15mm}
\put(66.88,43.75){\line(1,0){10.00}}
\put(76.88,43.75){\vector(1,0){0.12}}

\put(76.88,48.75){\makebox(0,0)[cc]{$c_0$}}

\put(31.88,33.75){\makebox(0,0)[cc]{$-a$}}

\put(66.88,33.75){\makebox(0,0)[cc]{$a$}}

\put(16.88,28.75){\makebox(0,0)[cc]{$\nu(y), \rho(y), E(y)$}}

\end{picture}
\end{center}
  \caption{Illustration for the immersion of the block in the
floating ice plate.}\label{pict6}
\end{figure}

Now, we write the mathematical formulation of the contact problem formulation:
\begin{eqnarray}\label{condcont1}
  \sigma_y(x,h) &=& 0, \qquad |x| > a, \\
  \tau_{xy}(x,h) &=& 0, \qquad |x| < \infty,\label{condcont2} \\
  v(x,h) &=& -\Delta + f(x), \qquad |x| < a.\label{condcont3}
\end{eqnarray}

Let us explain the meaning of these conditions. The first equation means that there are no forces applied to the upper ice boundary outside of the block in the region $[-a, a]$. The second means the absence of friction on the interface $\Gamma_3$. The last one gives us the vertical ice displacement in the interval $(-a, a)$ on $\Gamma_3$. By $\Delta>0$, we denote the depth at which the block was immersed in the ice (see Figure \ref{pict5}).

\subsection{Singular integral equation derivation}

Earlier, we obtained equations (\ref{contlp}) and (\ref{contlp2}) for the continuous case. We are interested here in the second identity (\ref{contlp2}):
\begin{equation*}
  \beta(\omega,h) = -B(h,\omega)\gamma(\omega,h) -
  D(h,\omega)\delta(\omega,h)
\end{equation*}
The vector $\overrightarrow{\alpha}(\omega,y)$ was defined as
\begin{eqnarray*}
  \hat u(\omega,h) &=& \alpha(\omega,h), \\
  \hat v(\omega,h) &=& \frac{i}{k_1^2(h)}\beta(\omega,h), \\
  \hat \sigma_y(\omega,h) &=& i E(h)\omega \gamma(\omega,h), \\
  \hat \tau_{xy}(\omega,h) &=& \mu(h)\omega \delta(\omega,h).
\end{eqnarray*}

The condition (\ref{condcont2}) gives us
$$
  \delta(\omega,h) = \frac{\hat \tau_{xy}(\omega,h)}
  {\mu(h)\omega} = 0.
$$
Then equation (\ref{contlp2}) becomes:
\begin{equation}\label{simplified}
  \beta(\omega,h) = -B(h,\omega)\gamma(\omega,h).
\end{equation}

Now we are interested in condition (\ref{condcont3}). In this condition, we do not know the constant $\Delta$. In order to eliminate this constant, we take the $x$-derivative:
\begin{equation}\label{dvdx}
  \od{v(x,h)}{x} = f'(x).
\end{equation}
A little calculation shows that
\begin{equation*}
  \left.\widehat{\od{v}{x}}\right|_{y=h} = (-i\omega)\hat
  v(\omega,h) = (-i\omega)\frac{i}{k_1^2(h)}\beta(\omega,h) =
  \frac{\omega}{k_1^2(h)}\beta(\omega,h).
\end{equation*}
Let us multiply the identity (\ref{simplified}) by
$\frac{\omega}{k_1^2}$:
\begin{equation*}
  \frac{\omega}{k_1^2}\beta(\omega,h) = -\frac{\omega}{k_1^2}
  B(h,\omega)\gamma(\omega,h).
\end{equation*}
Taking the inverse Fourier transform of this equality gives us:
\begin{equation*}
  \od{v(x,h)}{x} = -\frac1{2\pi
  k_1^2(h)}\int\limits_{-\infty}^\infty \omega
  B(h,\omega)\gamma(\omega,h)e^{-i\omega x}\; d\omega.
\end{equation*}
Using (\ref{dvdx}) we obtain:
\begin{equation}\label{formule2}
  -\frac{1}{2\pi k_1^2(h)} \int\limits_{-\infty}^{\infty}
  \omega B(h,\omega)\gamma(\omega,h)e^{-i\omega x} \; d\omega =
  f'(x), \qquad x\in (-a,a).
\end{equation}

In the contact problem, the contact pressure is defined by
\begin{equation*}
  q(x) := -\sigma_y(x,h), \qquad |x| < a.
\end{equation*}

At the beginning of this section, we assumed that the block is symmetric. This fact implies that the function $x\mapsto q(x)$ is even. This can be seen from a physical point of view.

By definition $iE(h)\omega\gamma(\omega,h) = \hat\sigma_y(\omega,h)$. So, we can make $q(x)$ transparent in our integral equation, since
\begin{multline*}
  \gamma(\omega,h) = \frac1{i E(h)\omega} \hat\sigma_y(\omega,h) =
  \frac1{i E(h)\omega}\int\limits_{-\infty}^{\infty} \sigma_y(x,h)
  e^{i\omega x} \; dx =\\=
  -\frac1{i E(h)\omega}\int\limits_{-a}^a q(x)e^{i\omega x} \; dx.
\end{multline*}
Here, we used the condition (\ref{condcont1}). Finally equation (\ref{formule2}) becomes
\begin{equation*}
  -\frac1{2\pi k_1^2(h)}\int\limits_{-\infty}^{\infty}
  \omega B(h,\omega) \left(
  -\frac1{i E(h)\omega}\int\limits_{-a}^a q(t)e^{i\omega t} \; dt
  \right) e^{-i\omega x}\; d\omega = f'(x),
\end{equation*}
or after simplification
\begin{equation}\label{int1}
  \frac1{2\pi i k_1^2E}\int\limits_{-\infty}^{+\infty}
  B(h,\omega)e^{-i\omega x} \left(
  \int\limits_{-a}^a q(t)e^{i\omega t} \; dt
  \right) \; d\omega = f'(x), x\in (-a,a).
\end{equation}

\begin{lemm}\label{lemme1}
  The function $\omega \mapsto B(h,\omega)$ is odd.
\end{lemm}

\begin{proof}
The proof follows from the fact that the system of ordinary differential equations (\ref{Aeq}), (\ref{Beq}), (\ref{Ceq}), (\ref{Deq}) remains invariant if we put
$$
  \omega \leftarrow -\omega,
$$
$$
  A(\omega,y) \leftarrow A(-\omega,y), \qquad
  B(\omega,y) \leftarrow -B(-\omega,y),
$$
$$
  C(\omega,y) \leftarrow -C(-\omega,y), \qquad
  D(\omega,y) \leftarrow D(-\omega,y).
$$
\end{proof}

Using Lemma \ref{lemme1} and the fact that the function $q(x)$ is even we can transform equation (\ref{int1}) in an equivalent form:
\begin{multline*}
  \frac1{2\pi i k_1^2 E}\int\limits_{0}^{+\infty} B_h(\omega)
  \int\limits_{-a}^a q(t) e^{i\omega(t-x)}\; dt \; d\omega +\\+
  \frac1{2\pi i k_1^2 E}\int\limits_{-\infty}^{0} B_h(\omega)
  \int\limits_{-a}^a q(t) e^{i\omega(t-x)}dtd\omega = f'(x),
\end{multline*}
where we introduced the notation $B_h(\omega) := B(h,\omega)$, or
\begin{multline*}
  \frac1{2\pi i k_1^2 E}\int\limits_{0}^{+\infty} B_h(\omega)
  \int\limits_{-a}^a q(t) e^{i\omega(t-x)}\; dt \; d\omega -\\-
  \frac1{2\pi i k_1^2 E}\int\limits_{0}^{+\infty} B_h(\omega)
  \int\limits_{-a}^a q(t) e^{i\omega(x-t)}\; dt \; d\omega = f'(x),
\end{multline*}
\begin{multline*}
  \frac1{2\pi i k_1^2 E}\int\limits_{0}^{+\infty} B_h(\omega)
  \Bigl(
  \int\limits_{-a}^a q(t) e^{-i\omega(x-t)}\; dt -\\-
  \int\limits_{-a}^a q(t) e^{i\omega(x-t)}\; dt
  \Bigr)\; d\omega = f'(x).
\end{multline*}
Since
$$
  e^{-i\omega(x-t)} - e^{i\omega(x-t)} = -2i\sin(\omega(x-t)),
$$
\begin{equation*}
  \frac{1}{2\pi i k_1^2E}
  \int\limits_0^{+\infty} B_h(\omega)
  \int\limits_{-a}^a q(t)\Bigl(
  -2i\sin(\omega(x-t))
  \Bigr)\; dt \; d\omega = f'(x),
\end{equation*}
\begin{equation*}
  \frac{1}{\pi k_1^2E}
  \int\limits_0^{+\infty} B_h(\omega)
  \int\limits_{-a}^a q(t)\sin(\omega(x-t))
  \; dt \; d\omega = -f'(x),
\end{equation*}
\begin{multline*}
  \frac1{\pi k_1^2 E}\int\limits_0^{+\infty} B_h(\omega)
  \Bigl(
  \int\limits_0^a q(t)\sin(\omega(x-t))\; dt +\\+
  \int\limits_{-a}^0 q(t)\sin(\omega(x-t))\; dt
  \Bigr) \; d\omega = -f'(x),
\end{multline*}
\begin{multline*}
  \frac1{\pi k_1^2 E}\int\limits_0^{+\infty} B_h(\omega)
  \Bigl(
  \int\limits_0^a q(t)\sin(\omega(x-t))\; dt +\\+
  \int\limits_{0}^a q(t)\sin(\omega(x+t))\; dt
  \Bigr) \; d\omega = -f'(x),
\end{multline*}
\begin{multline*}
  \frac1{\pi k_1^2 E}\int\limits_0^{+\infty} B_h(\omega)
  \int\limits_0^a q(t)\Bigl(\sin(\omega(x-t))\; dt +\\
  +\sin(\omega(x+t))
  \Bigr)\; dt \; d\omega = -f'(x),
\end{multline*}
\begin{multline}\label{last}
  \int\limits_0^{+\infty} B_h(\omega)
  \int\limits_0^a q(t)\Bigl(\sin(\omega(x-t))\; dt +\\
  +\sin(\omega(x+t))
  \Bigr)\; dt \; d\omega = -\pi k_1^2 E f'(x),
\end{multline}

Now we represent the function $B_h(\omega)$ in the form:
$$
  B_h(\omega) = B_{\infty} + (B_h(\omega)-B_\infty),
$$
where $B_\infty = \lim_{\omega\to +\infty} B(h,\omega)$. We put this representation in the last integral equation (\ref{last}):
\begin{multline*}
  \int\limits_0^{+\infty} B_\infty
  \int\limits_0^a q(t)\Bigl(\sin(\omega(x-t))\; dt+\sin(\omega(x+t))
  \Bigr)\; dt \; d\omega +\\+
  \int\limits_0^{+\infty} [B_h(\omega) - B_\infty]
  \int\limits_0^a q(t)\Bigl(\sin(\omega(x-t))\; dt+\sin(\omega(x+t))
  \Bigr)\; dt \; d\omega =\\= -\pi k_1^2 E f'(x), \qquad x\in (-a,a),
\end{multline*}
\begin{multline*}
  \int\limits_0^a q(t) \; dt
  \int\limits_0^{+\infty} B_\infty \sin(\omega(x+t))\; d\omega +
  \int\limits_0^a q(t) \; dt
  \int\limits_0^{+\infty} B_\infty \sin(\omega(x-t))\; d\omega +\\+
  \int\limits_0^a q(t) \; dt
  \int\limits_0^{+\infty}[B_h(\omega)-B_\infty]\Bigl(
  \sin(\omega(x-t))\; dt+\sin(\omega(x+t)\Bigr)\; d\omega =\\=
  -\pi k_1^2 E f'(x), \qquad x\in (-a,a),
\end{multline*}

\begin{lemm}\label{lemme2}
In the sense of distributions, we have the following formula
  $$\int\limits_0^{+\infty}\sin(x t) dx = \frac1t$$
\end{lemm}
\begin{proof}
First of all, we introduce the main set of functions
\begin{defn}
$\phi: \mathbb{R}^n \mapsto \mathbb{R}$, $\phi \in D$ if and only if
$\supp\phi\subset \mathbb{R}^n$, $\phi$ is infinitely differentiable
function.
\end{defn}
In this Lemma, we have $n=1$. We will denote by $D'$ the set of distributions on $D$. We introduce a linear functional $\mathcal{P}\frac1x$ on $D$:
\begin{equation*}
  \bigl(\mathcal{P}\frac1x,\phi \bigr) :=
  v.p.\int\frac{\phi(x)}{x}\; dx = \lim_{\eps\to+0}\Bigl(
  \int\limits_{-\infty}^{-\eps} +
  \int\limits_{\eps}^{+\infty}
  \Bigr) \frac{\phi(x)}{x}\; dx, \qquad \forall\phi\in D
\end{equation*}

Let us prove that $\mathcal{P}\frac1x\in D'$. It is sufficient to show that $\mathcal{P}\frac1x$ is bounded. For $\forall \phi\in D$ such that $\phi(x)\equiv 0$, $|x|>R$ we have
\begin{multline*}
  \abs{\bigl(\mathcal{P}\frac1x,\phi\bigr)} =
  \abs{v.p.\int\frac{\phi(x)}{x}} = \abs{v.p.\int\limits_{-R}^{R}
  \frac{\phi(0) + \phi(x) - \phi(0)}{x}\;dx} =\\=
  \abs{v.p.\int\limits_{-R}^{R}
  \frac{\phi(0) + x\phi'(\xi)}{x}\;dx} \leq
  \int\limits_{-R}^R \abs{\phi'(\xi)}\;dx \leq 2R
  \sup\limits_{x\in (-R,R)}\abs{\phi'(x)}.
\end{multline*}
So, we proved that $\mathcal{P}\frac1x\in D'$.

Now we have to prove the Sokhotski formulae
\begin{eqnarray}\label{sokhots1}
  \frac1{x+i0} &=& -i\pi\delta(x) + \mathcal{P}\frac1x, \\
  \frac1{x-i0} &=& i\pi\delta(x) + \mathcal{P}\frac1x.\label{sokhots2}
\end{eqnarray}
We need to show only the first formula; for example, the second one can be proved in a similar way.

Let us take $\forall \phi\in D$ such that $\supp\phi\subset(-R,R)$.
\begin{multline*}
  \lim_{\eps\to +0}\int\frac{\phi(x)}{x+i\eps}\;dx =
  \lim_{\eps\to +0}\int\limits_{-R}^R \frac{x-i\eps}{x^2+\eps^2}
  \phi(x)\; dx =\\
  = \phi(0)\lim_{\eps\to +0}\int\limits_{-R}^R
  \frac{x-i\eps}{x^2+\eps^2}\;dx +
  \lim_{\eps\to+0}\int\limits_{-R}^R \frac{x-i\eps}{x^2+\eps^2}
  (\phi(x) - \phi(0))\;dx =\\=
  -2i\phi(0)\lim_{\eps\to +0}\arctan\frac{R}{\eps} +
  \int\limits_{-R}^R\frac{\phi(x)-\phi(0)}{x}\;dx =
  -i\pi\phi(0) + v.p.\int\frac{\phi(x)}{x}\;dx
\end{multline*}
This finishes the proof of \ref{sokhots1}.

In the sequel, $H(x)$ stands for the Heaviside function
\begin{equation*}
  H(x) := \left\{%
\begin{array}{ll}
    1, & x \geq 0, \\
    0, & \hbox{elsewhere.} \\
\end{array}%
\right.
\end{equation*}

Let us calculate the Fourier transform of $H(x)$.
\begin{equation*}
  \mathfrak{F}[H(x)e^{-ax}] = \int\limits_0^{+\infty}
  e^{-ax + ix\xi}\;dx =
  \left.\frac{e^{-ax+ix\xi}}{-a+i\xi}\right|_0^{+\infty} =
  \frac{i}{\xi + i a}.
\end{equation*}
Since $H(x)e^{-ax}\to H(x)$ when $a\to +0$, one has
\begin{equation*}
  \mathfrak{F}[H(x)] = \frac{i}{\xi+i0},
\end{equation*}
or, using Sokhotski formula (\ref{sokhots1}) it follows immediately
\begin{equation*}
  \mathfrak{F}[H(x)] = \pi\delta(\xi) + i\mathcal{P}\frac1\xi.
\end{equation*}
In a similar way, one can prove
\begin{equation*}
  \mathfrak{F}[H(-x)] = \pi\delta(\xi) - i\mathcal{P}\frac1\xi.
\end{equation*}
It is clear that
\begin{equation*}
  \sign x = H(x) - H(-x).
\end{equation*}
Since the Fourier transform is a linear functional, now we are ready to calculate
\begin{equation}\label{fsign}
  \mathfrak{F}[\sign x] = \mathfrak{F}[H(x)] - \mathfrak{F}[H(-x)] =
  2i \mathcal{P}\frac1\xi.
\end{equation}
So, by definition of $\mathcal{P}\frac1\xi$ and from (\ref{fsign}) it follows that
\begin{equation}\label{f1}
  (\mathfrak{F}[\sign x],\phi) = 2i\;v.p.\int
  \frac{\phi(\xi)}{\xi}\; d\xi.
\end{equation}
By definition of Fourier transform for distributions
\begin{multline}\label{f2}
  (\mathfrak{F}[\sign x], \phi) = (\sign x, \mathfrak{F}[\phi]) =
  \int\limits_{-\infty}^{+\infty} \sign\xi\;d\xi
  \int\limits_{-\infty}^{+\infty} \phi(x) e^{i\xi x}\;dx =\\=
  \int\limits_{-\infty}^{+\infty} \sign\xi\;d\xi
  \int\limits_{-\infty}^{+\infty} \phi(x) (\cos\xi x + i\sin\xi
  x)\;dx =\\=
  \int\limits_{-\infty}^{+\infty} \phi(x)\;dx \left(
  \int\limits_{-\infty}^{+\infty} \sign \xi \cos\xi x\;d\xi
  +i\int\limits_{-\infty}^{+\infty} \sign \xi \sin\xi x\;d\xi
  \right) =\\=
  2i\int \phi(x)\;dx\int\limits_0^{+\infty}\sin\xi x\;d\xi.
\end{multline}
Comparison of (\ref{f1}) and (\ref{f2}) proves this Lemma.
\end{proof}

At this moment, we make use of Lemma \ref{lemme2} and obtain:
\begin{multline*}
  B_\infty\int\limits_0^a \frac{q(t)}{x+t} \; dt+
  B_\infty\int\limits_0^a \frac{q(t)}{x-t} \; dt+\\+
  \int\limits_0^a q(t) \; dt
  \int\limits_0^{+\infty}[B_h(\omega)-B_\infty]\sin(\omega(x+t))\;
  d\omega
  +\\+
  \int\limits_0^a q(t) \; dt
  \int\limits_0^{+\infty}[B_h(\omega)-B_\infty]\sin(\omega(x-t))\; d\omega
  = -\pi k_1^2 E f'(x),
\end{multline*}

In the integrals with $(x+t)$ we change the variable $t$ to $-t$:
\begin{multline*}
  B_\infty\int\limits_{-a}^a \frac{q(t)}{x-t} \; dt+
  \int\limits_{-a}^a q(t) \; dt
  \int\limits_0^{+\infty}[B_h(\omega)-B_\infty]\sin(\omega(x-t)) \;
  d\omega
  =\\= -\pi k_1^2 E f'(x),\qquad x\in (-a,a)
\end{multline*}
with obvious notation, it is easy to recognize a singular integral equation with Cauchy-type singularity:
\begin{equation}\label{singinteq}
\boxed{
  \int\limits_{-a}^a \frac{q(t)}{t-x} \; dt +
  \int\limits_{-a}^a q(t) K(x-t) \; dt = \phi(x), x\in (-a,a),}
\end{equation}
where
\begin{equation}\label{kernel}
  K(z) := \int\limits_{0}^{+\infty}
  \Bigl[1 - \frac{B_h(\omega)}{B_\infty}\Bigr]
  \sin(\omega z) \; d\omega
\end{equation}
and
\begin{equation*}
  \phi(x) := \frac{\pi k_1^2(h)E(h)}{B_\infty} f'(x).
\end{equation*}
Note that all integrals in this section should be understood in the sense of Cauchy's principal value.

Now, we can already start to solve numerically this integral equation, but we would like to achieve one more step. In the next subsection, we will transform this equation into an equivalent form.

\subsection{Singular integral equation regularization by Carleman-Vekua method}

In this section, we apply the well-known method of singular integral equation regularization, which was proposed by the Georgian mechanist Niko Vekua (1913-1993) in \cite{vekua}. To our knowledge, this problem was posed by the Swedish mathematician Tage Gills Torsten Carleman (1892-1949).

This method allows us to obtain an equivalent Fredholm second-kind integral equation. This is very useful because there are a lot of subroutines which were written precisely to solve this type of integral equation. Moreover, these equations have very good properties, which allow us to solve them without any difficulties.

Let us start. At the end of the last section, we obtained a singular integral equation with Cauchy-type singularity:
\begin{equation*}
  \int\limits_{-a}^a \frac{q(t)}{t-x} \; dt +
  \int\limits_{-a}^a q(t) K(x-t) \; dt = \phi(x), x\in (-a,a),
\end{equation*}
where
\begin{equation*}
  K(z) := \int\limits_{0}^{+\infty}
  \Bigl[1 - \frac{B_h(\omega)}{B_\infty}\Bigr]
  \sin(\omega z) \; d\omega
\end{equation*}
and
\begin{equation*}
  \phi(x) := \frac{\pi k_1^2(h)E(h)}{B_\infty} f'(x).
\end{equation*}

Let
\begin{equation*}
  \Phi(x) := -\int\limits_{-a}^a q(t) K(x-t) \; dt + \phi(x).
\end{equation*}
Then equation (\ref{singinteq}) becomes
\begin{equation}\label{fx}
  \int\limits_{-a}^a \frac{q(t)}{t-x}\; dt = \Phi(x),
  \qquad |x|<a.
\end{equation}

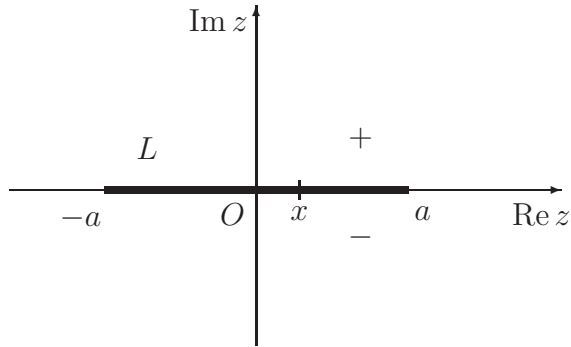
\begin{figure}[htbp]
\begin{center}
\unitlength 1mm
\begin{picture}(80.00,42.50)(0,0)

\linethickness{0.15mm}
\put(40.00,-2.50){\line(0,1){45.00}}
\put(40.00,42.50){\vector(0,1){0.12}}

\linethickness{0.15mm}
\put(7.50,18.13){\line(1,0){72.50}}
\put(80.00,18.13){\vector(1,0){0.12}}

\put(35.00,40.63){\makebox(0,0)[cc]{$\Im z$}}

\linethickness{1.00mm}
\put(20.00,18.13){\line(1,0){40.00}}

\put(16.88,14.38){\makebox(0,0)[cc]{$-a$}}

\put(61.88,15.00){\makebox(0,0)[cc]{$a$}}

\put(36.88,15.00){\makebox(0,0)[cc]{$O$}}

\put(25.63,23.75){\makebox(0,0)[cc]{$L$}}

\put(53.75,25.00){\makebox(0,0)[cc]{$+$}}

\put(53.75,11.88){\makebox(0,0)[cc]{$-$}}

\linethickness{0.15mm}
\put(45.63,16.88){\line(0,1){2.50}}

\put(45.63,15.00){\makebox(0,0)[cc]{$x$}}

\put(77.50,15.00){\makebox(0,0)[cc]{$\Re z$}}

\end{picture}
\end{center}
  \caption{Illustration for contour in complex plane.}\label{pict7}
\end{figure}

For now, we suppose that the function $\Phi(x)$ is known, and we will continue to work as if it were known. We consider the contour $L=\set{z\in \mathbb{C}: |\Re z|<a}$ in the
complex plane. This situation is illustrated in Figure \ref{pict7}.

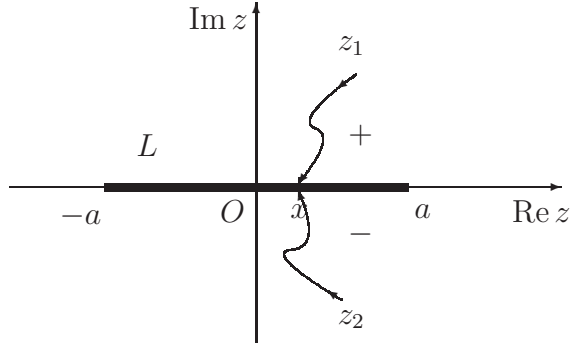
\begin{figure}[htbp]
\begin{center}
\unitlength 1mm
\begin{picture}(80.00,42.50)(0,0)

\linethickness{0.15mm}
\put(40.00,-2.50){\line(0,1){45.00}}
\put(40.00,42.50){\vector(0,1){0.12}}

\linethickness{0.15mm}
\put(7.50,18.13){\line(1,0){72.50}}
\put(80.00,18.13){\vector(1,0){0.12}}

\put(35.00,40.63){\makebox(0,0)[cc]{$\Im z$}}

\linethickness{1.00mm}
\put(20.00,18.13){\line(1,0){40.00}}

\put(16.88,14.38){\makebox(0,0)[cc]{$-a$}}

\put(61.88,15.00){\makebox(0,0)[cc]{$a$}}

\put(36.88,15.00){\makebox(0,0)[cc]{$O$}}

\put(25.63,23.75){\makebox(0,0)[cc]{$L$}}

\put(53.75,25.00){\makebox(0,0)[cc]{$+$}}

\put(53.75,11.88){\makebox(0,0)[cc]{$-$}}

\linethickness{0.15mm}
\put(45.63,16.88){\line(0,1){2.50}}

\put(45.63,15.00){\makebox(0,0)[cc]{$x$}}

\put(77.50,15.00){\makebox(0,0)[cc]{$\Re z$}}

\put(52.50,0.63){\makebox(0,0)[cc]{$z_2$}}

\put(52.50,36.88){\makebox(0,0)[cc]{$z_1$}}

\linethickness{0.95mm}

\linethickness{0.95mm}

\linethickness{0.15mm}
\qbezier(53.13,33.13)(44.86,26.95)(47.73,25.94)

\linethickness{0.15mm}
\qbezier(47.73,25.94)(50.61,24.92)(45.63,18.75)
\put(45.63,18.75){\vector(-3,-4){0.12}}

\linethickness{0.15mm}
\multiput(50.63,31.25)(0.16,0.12){16}{\line(1,0){0.16}}
\put(50.63,31.25){\vector(-4,-3){0.12}}

\linethickness{0.15mm}


\linethickness{0.15mm}
\qbezier(51.25,3.13)(40.82,9.53)(44.69,9.84)

\linethickness{0.15mm}
\qbezier(44.69,9.84)(48.55,10.16)(45.63,17.50)
\put(45.63,17.50){\vector(-1,3){0.12}}

\linethickness{0.15mm}
\multiput(49.38,4.38)(0.19,-0.13){10}{\line(1,0){0.19}}
\put(49.38,4.38){\vector(-3,2){0.12}}

\linethickness{0.15mm}


\end{picture}
\end{center}
  \caption{Illustration for the definition of
   $F^+(x)$, $F^-(x)$.}\label{pict8}
\end{figure}

We consider the complex function
\begin{equation}\label{fz}
  F(z) := \frac1{2\pi i}\int\limits_{L}
  \frac{q(t)}{t-z}\; dt
\end{equation}
and two functions linked with $F(z)$ by following definitions:
\begin{equation*}
  F^+(x) := \lim_{z\to x+0} F(z),\qquad x\in L,
\end{equation*}
\begin{equation*}
  F^-(x) := \lim_{z\to x-0} F(z),\qquad x\in L.
\end{equation*}
These definitions are illustrated in Figure \ref{pict8}.

In the Soviet literature on complex analysis (for example \cite{privalov, shabat}), we can find two formulae that were proved by Soviet mathematician Yulian Vasilievich Sokhotski (1842-1927):
\begin{equation}\label{sokh1}
  F^+(x) = \frac12 q(x) + \frac1{2\pi i}
  \int\limits_L \frac{q(t)}{t-x}\; dt,
\end{equation}
\begin{equation}\label{sokh2}
  F^-(x) = -\frac12 q(x) + \frac1{2\pi i}
  \int\limits_L \frac{q(t)}{t-x}\; dt.
\end{equation}

We will rewrite these formulae in the equivalent form:
\begin{equation}\label{sokh12}
  F^+(x) + F^-(x) = \frac1{\pi i}
  \int\limits_L \frac{q(t)}{t-x}\; dt,
\end{equation}
\begin{equation}\label{sokh22}
  F^+(x) - F^-(x) = q(x).
\end{equation}

Recall that from (\ref{fx})
$$
  \int\limits_L \frac{q(t)}{t-x}\; dt = \Phi(x).
$$
Substituting it in (\ref{sokh12}) gives us another formulation of our problem. More exactly, we have a Riemann problem for an analytical function $F(z)$:
\begin{equation}\label{riemann}
  \boxed{F^+(x) + F^-(x) = \frac1{\pi i}\Phi(x),\;\; x\in(-a,a)}
\end{equation}

Here, we would like to make a remark about the Riemann problem. In general, a Riemann problem is formulated like this. With given analytic functions $g(z)$ and $G(z)$ we have to find an analytic function $F(z)$ in $\mathbb{C}\backslash L$ which satisfies this relation on $L$:
$$
  F^+(z) + g(z)F^-(z) = G(z),\;\; z\in L.
$$

In other words, the Riemann problem consists of the reconstruction of an analytic function from its discontinuity on a contour. If $G(z)\equiv 0$, we say that the Riemann problem is homogeneous.

So, we transformed our integral equation to the Riemann problem (\ref{riemann}). Now, we want to solve this problem. For this, we need the following technical lemma:
\begin{lemm}\label{lemme3}
If a function $F(z)$ is defined by (\ref{fz}) then we have the following asymptotic behavior as $z\to\infty$:
$$
  F(z) = -\frac{Q}{2\pi i z} + O\Bigl(\frac{1}{z^3}\Bigr),
$$ 
where $Q$ is the total load applied on the block:
$$
  Q := \int\limits_{-a}^a q(t)\; dt.
$$
\end{lemm}
\begin{proof}
\begin{multline*}
  F(z) = \frac1{2\pi i}\int\limits_L \frac{q(t)}{t-z}\;dt =
  -\frac1{2\pi i z}\int\limits_{-a}^a
  \frac{q(t)}{1-\frac{t}{z}}\;dt =\\=
  -\frac1{2\pi i z}\int\limits_{-a}^a q(t)\Bigl(
  1-\frac{t}{z}\Bigr)^{-1} \;dt =\\=
  -\frac1{2\pi i z}\int\limits_{-a}^a q(t)\Bigl[
  1+\frac{t}{z} - \frac{t^2}{z^2} + O\Bigl(\frac{1}{z^3}\Bigr)
  \Bigr]\;dt =\\= -\frac1{2\pi i z}\int\limits_{-a}^a q(t)\;dt +
  O\Bigl(\frac{1}{z^3}\Bigr) =
  -\frac{Q}{2\pi i z} +
  O\Bigl(\frac{1}{z^3}\Bigr)
\end{multline*}
\end{proof}

From this Lemma, we immediately obtain a corollary:
\begin{equation*}
  F(\infty) = \lim_{z\to\infty} F(z) = 0.
\end{equation*}
Later, we will use it. For now, we will consider an auxiliary Riemann problem. It will become clear later why we do this. 

We want to find an analytical function $X(z)$ which can have singular points only on the contour $L$ and satisfies the relation:
\begin{equation}\label{riemann2}
  X^+(x) + X^-(x) = 0, \;\; x\in(-a,a)
\end{equation}
or in an equivalent form
\begin{equation}\label{riemann22}
  \frac{X^+(x)}{X^-(x)} = -1,\;\; x\in(-a,a).
\end{equation}

We consider the function
\begin{equation*}
  X(z) = \frac1{\sqrt{(z+a)(z-a)}}
\end{equation*}
and we prove that this function is the solution of the auxiliary Riemann problem (\ref{riemann2}).

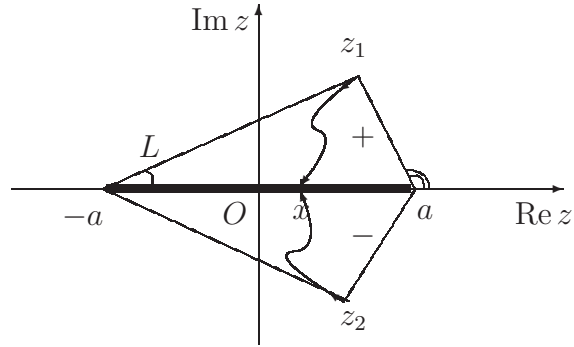
\begin{figure}[htbp]
\begin{center}
\unitlength 1mm
\begin{picture}(68.75,44.37)(0,0)

\linethickness{0.15mm}
\put(28.75,-0.63){\line(0,1){45.00}}
\put(28.75,44.37){\vector(0,1){0.12}}

\linethickness{0.15mm}
\put(-3.75,20.00){\line(1,0){72.50}}
\put(68.75,20.00){\vector(1,0){0.12}}

\put(23.75,42.50){\makebox(0,0)[cc]{$\Im z$}}

\linethickness{1.00mm}
\put(8.75,20.00){\line(1,0){40.00}}

\put(5.63,16.25){\makebox(0,0)[cc]{$-a$}}

\put(50.63,16.87){\makebox(0,0)[cc]{$a$}}

\put(25.63,16.87){\makebox(0,0)[cc]{$O$}}

\put(14.38,25.62){\makebox(0,0)[cc]{$L$}}

\put(42.50,26.87){\makebox(0,0)[cc]{$+$}}

\put(42.50,13.75){\makebox(0,0)[cc]{$-$}}

\linethickness{0.15mm}
\put(34.38,18.75){\line(0,1){2.50}}

\put(34.38,16.87){\makebox(0,0)[cc]{$x$}}

\put(66.25,16.87){\makebox(0,0)[cc]{$\Re z$}}

\put(41.25,2.50){\makebox(0,0)[cc]{$z_2$}}

\put(41.25,38.75){\makebox(0,0)[cc]{$z_1$}}

\linethickness{0.95mm}

\linethickness{0.95mm}

\linethickness{0.15mm}
\qbezier(41.88,35.00)(33.61,28.82)(36.48,27.81)

\linethickness{0.15mm}
\qbezier(36.48,27.81)(39.36,26.79)(34.38,20.62)
\put(34.38,20.62){\vector(-3,-4){0.12}}

\linethickness{0.15mm}
\multiput(39.38,33.12)(0.16,0.12){16}{\line(1,0){0.16}}
\put(39.38,33.12){\vector(-4,-3){0.12}}

\linethickness{0.15mm}


\linethickness{0.15mm}
\qbezier(40.00,5.00)(29.57,11.40)(33.44,11.71)

\linethickness{0.15mm}
\qbezier(33.44,11.71)(37.30,12.03)(34.38,19.37)
\put(34.38,19.37){\vector(-1,3){0.12}}

\linethickness{0.15mm}
\multiput(38.13,6.25)(0.19,-0.13){10}{\line(1,0){0.19}}
\put(38.13,6.25){\vector(-3,2){0.12}}

\linethickness{0.15mm}


\linethickness{0.15mm}
\multiput(41.88,35.00)(0.12,-0.24){63}{\line(0,-1){0.24}}

\linethickness{0.15mm}
\multiput(8.13,20.00)(0.27,0.12){125}{\line(1,0){0.27}}

\linethickness{0.15mm}
\multiput(40.00,5.00)(0.12,0.19){78}{\line(0,1){0.19}}

\linethickness{0.15mm}
\multiput(8.13,20.00)(0.26,-0.12){125}{\line(1,0){0.26}}

\linethickness{0.15mm}
\multiput(14.79,21.71)(0.15,-1.19){1}{\line(0,-1){1.19}}
\multiput(13.85,22.44)(0.16,-0.12){6}{\line(1,0){0.16}}

\linethickness{0.15mm}
\multiput(50.35,21.20)(0.14,-0.52){2}{\line(0,-1){0.52}}
\multiput(49.48,21.82)(0.17,-0.12){5}{\line(1,0){0.17}}
\multiput(48.42,21.73)(1.06,0.08){1}{\line(1,0){1.06}}

\linethickness{0.15mm}
\multiput(51.15,20.85)(0.08,-0.72){1}{\line(0,-1){0.72}}
\multiput(50.74,21.51)(0.14,-0.22){3}{\line(0,-1){0.22}}
\multiput(50.04,22.04)(0.17,-0.13){4}{\line(1,0){0.17}}
\multiput(49.14,22.38)(0.30,-0.11){3}{\line(1,0){0.30}}
\multiput(48.13,22.50)(1.01,-0.12){1}{\line(1,0){1.01}}

\end{picture}
\end{center}
  \caption{Auxiliary Riemann problem solution.}\label{pict9}
\end{figure}

We compute explicitly the limits $X^+(x)$ and $X^-(x)$:
\begin{equation*}
  X^+(x) = \lim_{z\to x+0} =
  \frac1{\sqrt{|x+a|}e^{i\frac02}\sqrt{|x-a|}e^{i\frac{\pi}2}} =
  \frac1{i\sqrt{a^2-x^2}},
\end{equation*}
\begin{equation*}
  X^-(x) = \lim_{z\to x-0} =
  \frac1{\sqrt{|x+a|}e^{i\frac02}\sqrt{|x-a|}e^{-i\frac{\pi}2}} =
  -\frac1{i\sqrt{a^2-x^2}}
\end{equation*}
and if we divide $X^+(x)$ by $X^-(x)$:
\begin{equation*}
  \frac{X^+(x)}{X^-(x)} = \frac{\frac1{i\sqrt{a^2-x^2}}}
  {-\frac1{i\sqrt{a^2-x^2}}} = -1.
\end{equation*}
Thus, we proved that the function $X(z)$ is a solution of the problem (\ref{riemann2}).

Let us return to the main Riemann problem. Its definition is
\begin{equation*}
  F^+(x) + F^-(x) = \frac1{\pi i} \Phi(x).
\end{equation*}
We perform some simple transformations and use the second definition (\ref{riemann22}) of the function $X(z)$:
\begin{equation*}
  F^+(x) - (-1)F^-(x) = \frac1{\pi i}\Phi(x),
\end{equation*}
\begin{equation*}
  F^+(x) - \frac{X^+(x)}{X^-(x)} F^-(x) =
  \frac1{\pi i} \Phi(x), \;\; x\in(-a,a)
\end{equation*}
\begin{equation*}
  \frac{F^+(x)}{X^+(x)} - \frac{F^-(x)}{X^-(x)} =
  \frac1{\pi i} \frac{\Phi(x)}{X^+(x)}, \;\; x\in(-a,a).
\end{equation*}
Introducing a new complex function
\begin{equation*}
  \Psi(z) = \frac{F(z)}{X(z)}
\end{equation*}
we have
\begin{equation}\label{psi}
  \Psi^+(x) - \Psi^-(x) = \frac1{\pi i}\frac{\Phi(x)}{X^+(x)}.
\end{equation}
In other words, we obtain a third Riemann problem for the function $\Psi(z)$. Fortunately, we know its solution because it is given by Sokhotski formula (\ref{sokh22}):
\begin{equation*}
  F^+(x) - F^-(x) = q(x).
\end{equation*}
Now we are able to write the solution of the third Riemann problem (\ref{psi}):
\begin{equation*}
  \Psi(z) = \frac1{2\pi i}\int\limits_{-a}^a
  \frac1{\pi i}
  \frac{\Phi(t)}{X^+(t)(t-z)}\; dt + P(z),
\end{equation*}
where $P(z)\in \mathbb{C}[z]$ is an arbitrary polynomial.

If we remember the definition of function $\Psi(z) := \frac{F(z)}{X(z)}$ we can rewrite our solution:
\begin{equation*}
  F(z) = \frac{X(z)}{2\pi i\cdot i\pi}\int\limits_{-a}^a
  \frac{\Phi(t)}{X^+(t)(t-z)}\; dt + P(z)X(z).
\end{equation*}

Earlier, we obtained a corollary that
$$
  F(\infty)=0.
$$
This condition implies that polynomial $P(z)$ is a polynomial of the first degree, i.e. $P(z) = c_0$:
\begin{equation*}
  F(z) = \frac{X(z)}{2\pi i\cdot i\pi}\int\limits_{-a}^a \frac{\Phi(t)}{X^+(t)(t-z)}\; dt + c_0 X(z).
\end{equation*}
To compute this unknown constant $c_0$, we will write one more time the asymptotic behaviour of the function $F(z)$ for $z\gg1$ from the last formula\footnote{Since $X(z) = O\left(\frac1z\right)$ and $\frac{1}{2\pi i\cdot i\pi}\int\limits_{-a}^a \frac{\Phi(t)}{X^+(t)(t-z)}\; dt = O\left(\frac1z\right)$ we have that $\frac{X(z)}{2\pi i\cdot i\pi}\int\limits_{-a}^a \frac{\Phi(t)}{X^+(t)(t-z)}\; dt =  O\left(\frac1{z^2}\right)$}:
\begin{multline*}
  F(z) = c_0(z^2-a^2)^{-\frac12} + O\Bigl(\frac1{z^2}\Bigr) =\\=
  \frac{c_0}{z}\left(1-\frac{a^2}{z^2}\right)^{-\frac12}
  + O\Bigl(\frac1{z^2}\Bigr) = \frac{c_0}{z} +
  O\Bigl(\frac1{z^2}\Bigr).
\end{multline*}
In Lemma~\ref{lemme3} we established that
\begin{equation*}
  F(z) = -\frac{Q}{2\pi i z} + O\Bigl(\frac1{z^3}\Bigr)
\end{equation*}
Comparing the last two formulae, one can conclude that
\begin{equation*}
    c_0 = -\frac{Q}{2\pi i}
\end{equation*}
So, we have
\begin{equation*}
  F(z) = \frac{X(z)}{2\pi i\cdot i\pi}\int\limits_{-a}^a
  \frac{\Phi(t)}{X^+(t)(t-z)}\; dt - \frac{Q}{2\pi i} X(z).
\end{equation*}

Now, we consider one more complex function
\begin{equation*}
  \Omega(z) = \frac1{2\pi i}\int\limits_{-a}^a
  \frac{\Phi(t)}{X^+(t)(t-z)}\; dt
\end{equation*}
We apply Sokhotski formulae to this function:
\begin{equation*}
  \Omega^+(x) = \frac12 \frac{\Phi(x)}{X^+(x)} +
  \frac1{2\pi i}\int\limits_{-a}^a
  \frac{\Phi(t)}{X^+(t)(t-x)}\; dt, \;\; x\in(-a,a),
\end{equation*}
\begin{equation*}
  \Omega^-(x) = -\frac12 \frac{\Phi(x)}{X^+(x)} +
  \frac1{2\pi i}\int\limits_{-a}^a
  \frac{\Phi(t)}{X^+(t)(t-x)}\; dt, \;\; x\in(-a,a).
\end{equation*}

At this moment, we are able to find the limits $F^+(x)$ and $F^-(x)$:
\begin{equation*}
  F^+(x) = \frac{X^+(x)}{i\pi}\left[
  \frac12 \frac{\Phi(x)}{X^+(x)} +
  \frac1{2\pi i}\int\limits_{-a}^a
  \frac{\Phi(t)}{X^+(t)(t-x)}\; dt
  \right] - \frac{Q}{2\pi i} X^+(x),
\end{equation*}
\begin{equation*}
  F^-(x) = \frac{X^-(x)}{i\pi}\left[
  -\frac12 \frac{\Phi(x)}{X^+(x)} +
  \frac1{2\pi i}\int\limits_{-a}^a
  \frac{\Phi(t)}{X^+(t)(t-x)}\; dt
  \right] - \frac{Q}{2\pi i} X^-(x).
\end{equation*}
Recall that the function $X(z)$ is a solution of the Riemann problem (\ref{riemann2}). By definition
\begin{equation*}
  X^-(x) = -X^+(x)
\end{equation*}
and the last formula can be rewritten as
\begin{equation*}
  F^+(x) = \frac{X^+(x)}{i\pi}\left[
  \frac12 \frac{\Phi(x)}{X^+(x)} +
  \frac1{2\pi i}\int\limits_{-a}^a
  \frac{\Phi(t)}{X^+(t)(t-x)}\; dt
  \right] - \frac{Q}{2\pi i} X^+(x),
\end{equation*}
\begin{equation*}
  F^-(x) = -\frac{X^+(x)}{i\pi}\left[
  -\frac12 \frac{\Phi(x)}{X^+(x)} +
  \frac1{2\pi i}\int\limits_{-a}^a
  \frac{\Phi(t)}{X^+(t)(t-x)}\; dt
  \right] + \frac{Q}{2\pi i} X^+(x)
\end{equation*}
According to Sokhotski formula (\ref{sokh22}) we have
\begin{equation*}
  F^+(x) - F^-(x) = q(x).
\end{equation*}
If we subtract expression of $F^-(x)$ from $F^+(x)$ we obtain:
\begin{equation*}
  q(x) = \frac{X^+(x)}{(\pi i)^2}
  \int\limits_{-a}^a \frac{\Phi(t)}{X^+(t)(t-x)} -
  \frac{Q}{\pi i} X^+(x).
\end{equation*}
Earlier, we performed a simple computation and established that
\begin{equation*}
  X^+(x) = \frac{1}{i\sqrt{a^2-x^2}}.
\end{equation*}
We put this expression in the integral equation:
\begin{equation}\label{lastlast}
  q(x) = -\frac{1}{\pi^2\sqrt{a^2-x^2}}
  \int\limits_{-a}^a \frac{\sqrt{a^2-t^2}\Phi(t)}{t-x}\; dt
  + \frac{Q}{\pi\sqrt{a^2-x^2}}.
\end{equation}
The last step is to recall the definition of the function $\Phi(t)$ that we considered to be known:
\begin{equation*}
  \Phi(t) = -\int\limits_{-a}^a q(s)K(t-s)\;ds + \pi k_1^2E f'(t).
\end{equation*}
We put this expression in equation (\ref{lastlast}):
\begin{multline*}
  q(x) = -\frac{1}{\pi^2\sqrt{a^2-x^2}}
  \int\limits_{-a}^a \frac{\sqrt{a^2-t^2}}{t-x}\Bigl[
  -\int\limits_{-a}^a q(s)K(t-s)\;ds +\\+ \pi k_1^2E f'(t)
  \Bigr] + \frac{Q}{\pi\sqrt{a^2-x^2}}, \;\; x\in(-a,a).
\end{multline*}
It is now very easy to obtain an equivalent Fredholm second-kind integral equation:
\begin{multline*}
  q(x) - \frac1{\pi^2\sqrt{a^2-x^2}}\int\limits_{-a}^a
  \overline{K}(x,t) q(t)\; dt = \frac{Q}{\pi\sqrt{a^2-x^2}} -\\-
  \frac{k_1^2E}{\pi\sqrt{a^2-x^2}}
  \int\limits_{-a}^a\frac{\sqrt{a^2-t^2}}{t-x}f'(t)\;dt,
\end{multline*}
where
\begin{equation*}
  \overline{K}(x,t) = \int\limits_{-a}^a
  \frac{\sqrt{a^2-s^2}K(s-t)}{s-x} \; ds
\end{equation*}

So, at the end of this section, we obtained an equivalent Fredholm second-kind integral equation, which can be solved by standard subroutines without any problem. But in the present work, we will take another way. Personally, we find it more interesting to solve numerically the singular integral equation. The next subsection is devoted to this topic.

\subsection{Numerical algorithm for solving the singular integral equation
with Cauchy type singularity}

A. A. Korneychuk obtained a numerical integration formula for singular integrals with Cauchy type kernel:
\begin{equation}\label{numquadr}
  \int\limits_{-1}^1\frac{\omega(t)u(t)\;dt}{t-x_r} =
  \sum_{m=1}^M\frac{a_m u(t_m)}{t_m - x_r}.
\end{equation}

In this formula, $u(t)$ is a regular function, and $\omega(t)$ is an integrable function which contains the singularity. The integration points $t=t_m$, $(m=1,2,\ldots,M)$ are the roots of the orthogonal polynomial $P_M(t)$ of degree $M$ with weight function $\omega(t)$, $t\in [-1,1]$. The points $x=x_r$, $r=1,2,\ldots,R$ are the roots of the function
\begin{equation}\label{qm}
  Q_M(x) = -\frac12\int\limits_{-1}^1\frac{\omega(t)P_M(t)\;dt}{t-x}.
\end{equation}

The coefficients $a_m$, $(m=1,2,\ldots,M)$ of the numerical integration formula (\ref{numquadr}) are determined by the following formula
\begin{equation*}
  a_m = -2\frac{Q_M(t_m)}{P_M'(t_m)}.
\end{equation*}
Let us notice that the formula (\ref{numquadr}) is necessarily exact if $u(t)$ is a polynomial of degree $2M$ or less. One can notice that the coefficients $a_m$ and the points $t_m$ in the formula (\ref{numquadr}) coincide with the Gauss numerical integration rule for the same weight $\omega(t)$:
\begin{equation*}
   \int\limits_{-1}^1 \omega(t) u(t)\; dt = \sum_{m=1}^M a_m u(t_m).
\end{equation*}
Therefore, the formula (\ref{numquadr}) can be considered as a Gauss numerical integration formula for a singular integral that is true for a discreet system of points $x_r$ such that $Q_M(x_r) = 0$.

Let us consider the singular integral equation
\begin{equation}\label{eq:sing}
  \int\limits_{-1}^1\left[
    \frac1{t-x} + K(t,x)
  \right]g(t)\;dt = \pi p(x), \qquad |x|<1
\end{equation}
where the functions $K(t,x)$ and $p(x)$ are continuous at $[-1,1]^2$ and $[-1,1]$ respectively.

The singular integral equations of this type arise very often in the problems of elasticity, hydro and aerodynamics.

We will make a structural assumption about the solution of the integral equation (\ref{eq:sing})
\begin{equation}\label{repr}
  g(t) = \omega(t) u(t)
\end{equation}
where $u(t)$ is a regular unknown function and $\omega(t)$ is a known, non-negative weight function.

Let us put the representation (\ref{repr}) into the integral equation (\ref{eq:sing})
\begin{equation}\label{eq:omega}
  \int\limits_{-1}^1\frac{\omega(t)u(t)\;dt}{t-x} +
  \int\limits_{-1}^1\omega(t)u(t)K(t,x)\;dt = \pi p(x),
  \qquad|x|<1.
\end{equation}
Since the functions $u(t)$, $K(x,t)$ are regular, we can approximate the integrals in (\ref{eq:omega}) at the points $x_r$, $r=1,\ldots,R$ and obtain a system of $R$ linear algebraic equations:
\begin{equation}\label{eq:alg}
  \sum_{m=1}^M a_m u(t_m)\left[
    \frac1{t_m-x_r} + k(t_m,x_r)
  \right]=\pi p(x_r), \qquad{(r=1,2,\ldots,R)}.
\end{equation}
There are two possibilities:
\begin{enumerate}
    \item $R\geq M$,
    \item $R<M$.
\end{enumerate}
In the first case, one should choose exactly $M$ equations from (\ref{eq:alg}) and, after solving them, find an unknown function at the points $t=t_m$. If $R<M$, then we should use some complementary conditions in order to obtain a complete system of equations. Now, let us consider a particular case which is particularly interesting in the context of this work. Moreover, this is a rare case when we can find the coefficients $a_m$ and the points $t_m$, $x_r$ analytically. In our problem, the solution has the form
\begin{equation*}
    g(x) = \frac{u(x)}{\sqrt{1-x^2}}.
\end{equation*}
It can be unclear why we have chosen this particular form of the solution. In fact, this can be established by solving a simple problem about the contact between a rigid block and an elastic half-space. The solution to this problem can be obtained in analytic form.

In this case $\omega(x) = (1-x^2)^{-\frac12}$ and the system of orthogonal polynomial is the system of Tchebychev polynomials of the first kind
\begin{equation}\label{tcheb1}
    T_n(x) = \cos(n\arccos x).
\end{equation}
As we established above, the roots of the polynomials (\ref{tcheb1}) are the integration points:
\begin{equation*}
  t_m=\cos\frac{2m-1}{2M}\pi
\end{equation*}
and the coefficients
\begin{equation*}
    a_m = \frac{\pi}{M}, \qquad m=1,2,\ldots,M.
\end{equation*}
Using the formula
\begin{equation*}
  \frac1{\pi}\int\limits_{-1}^1
  \frac{T_M(t)\;dt}{\sqrt{1-t^2}(t-x)}=\left\{%
\begin{array}{ll}
  0, & M=0, \\
  u_{M-1}(x), & M>0.\\
\end{array}%
\right.\qquad |x|<1
\end{equation*}
and the formula (\ref{qm}) one obtains
\begin{equation}\label{qm2}
    Q_M(x) = -2\pi U_{M-1}(x) = -2\pi
    \frac{\sin(M\arccos x)}{\sqrt{1-x^2}}
\end{equation}
where we introduced a notation
$$
  U_{M-1}(x) = \frac{\sin(M\arccos x)}{\sqrt{1-x^2}}.
$$
One can recognize that $U_{M-1}(x)$ is the Tchebychev polynomial of the second kind. As the points $x_r$ are the roots of $Q_M(x)$ we can easily find from (\ref{qm2})
\begin{equation*}
  x_r = \cos\frac{\pi r}{M}, \qquad r=1,\ldots,M-1.
\end{equation*}

In this case $R = M-1 < M$, therefore we have $M$ unknowns $u(t_m)$ and $M-1$ equations (\ref{eq:alg}). Fortunately, we have one more condition to obtain a complete system of equations
\begin{equation}\label{cond:supl}
  \int\limits_{-1}^1\frac{u(t)\;dt}{\sqrt{1-t^2}} = Q
\end{equation}
where $Q$ is the known constant with physical meaning. It represents the total load applied on the moving block. It is easy to obtain a finite-dimensional analogue of the condition (\ref{cond:supl}) applying the Gauss numerical integration rule
\begin{equation}\label{findimens}
  \int\limits_{-1}^1\frac{u(t)\;dt}{\sqrt{1-t^2}} =
  \sum_{m=1}^{M} a_m u(t_m) =
  \sum_{m=1}^{M} \frac{\pi}{M} u(t_m) = Q
\end{equation}

So, putting together equations (\ref{eq:alg}) and (\ref{findimens}) gives us a system of linear algebraic equations
\begin{eqnarray*}
  \sum_{m=1}^M a_m u(t_m)\left[
    \frac1{t_m-x_r} + K(t_m,x_r)
  \right] &=& \pi p(x_r), \qquad{(r=1,2,\ldots,M-1)} \\
  \sum_{m=1}^M\frac{\pi}{M}u(t_m) &=& Q.
\end{eqnarray*}

\subsection{Dimensionless form of the integral equation}

In order to perform numerical computations, we need to transform the integral equation (\ref{singinteq}) into dimensionless form. To do this, we change the variables:
\begin{equation*}
  x = \overline{x} a,\qquad t = \overline{t} a, \qquad
  \omega = \overline{\omega} h
\end{equation*}
where $2a$ is the block length and $h$ is the ice thickness. Putting new variables in the equation (\ref{singinteq}) gives
\begin{equation*}
  \int\limits_{-1}^1\frac{q(\overline{t}a)\;d\overline{t}}
  {\overline{t}-\overline{x}} + a\int\limits_{-1}^1
  q(\overline{t}a)K(a(\overline{x}-\overline{t}))\;d\overline{t} =
  \frac{\pi k_1^2(h)E(h)}{B_\infty}f'(x)\left|_{x=\overline{x}a}\right.
\end{equation*}
We are looking for a solution of the form
\begin{equation}\label{partsolform}
  q(x) = \frac{u(x)}{\sqrt{a^2-x^2}}.
\end{equation}
We know that the function $q(x)$ satisfies the condition
\begin{equation*}
  \int\limits_{-a}^a q(x) \; dx =
  \int\limits_{-a}^{a} \frac{u(x)}{\sqrt{a^2-x^2}}\; dx = Q,
\end{equation*}
or, in new variables
\begin{equation}\label{condnew}
  \int\limits_{-1}^1 \frac{u(\overline{x} a)}
  {\sqrt{1-\overline{x}^2}} \; d \overline{x} = Q.
\end{equation}
If we introduce a new function
\begin{equation*}
  \overline{u}(\overline{x}) = Q u(\overline{x})
\end{equation*}
we will have the condition (\ref{condnew}) in the dimensionless form
\begin{equation}\label{condnewd}
  \int\limits_{-1}^1 \frac{\overline{u}(\overline{x} a)}
  {\sqrt{1-\overline{x}^2}} \; d \overline{x} = 1.
\end{equation}
Let us transform slightly the kernel $K(x,t)$ of the integral equation
\begin{multline*}
  K(a(\overline{x}-\overline{t})) = \int\limits_0^{+\infty}
  \left(1 - \frac{B_h(\omega)}{B_{\infty}}\right) \sin(\omega
  a(\overline{x}-\overline{t}))\; d\omega =\\=
  \frac1h
  \int\limits_0^{+\infty}
  \left(1 - \frac{B(h,\frac{\omega}{h})}{B_{\infty}}\right) \sin(
  \frac{a}{h}\overline{\omega}(\overline{x}-\overline{t}))\;
  d\overline{\omega}.
\end{multline*}
In order to simplify the notation, we omit all the dashes below.

Finally we can write the dimensionless singular integral equation:
\begin{equation*}
  \int\limits_{-1}^1\frac{\frac{u(t)}{\sqrt{1-t^2}}\;dt}{t-x} +
  \kappa \int\limits_{-1}^1 \frac{u(t)}{\sqrt{1-t^2}} K(x-t)\; dt =
  \frac{a\pi k_1^2(h)E(h)}{Q B_{\infty}}\left. f'(x)
  \right|_{x=a\overline{x}}
\end{equation*}
where
\begin{equation*}
    \kappa := \frac{a}{h}.
\end{equation*}
Together with a complementary condition (\ref{condnewd}), we have a problem that is ready to be solved by the numerical algorithm described in the previous section.

\section{Appendix}

\subsection{Lam\'{e}'s equations solution for an homogeneous
ice layer}\label{app2}

\begin{verbatim}
% This code determines elastic deformations of an
% homogeneous ice plate of finite thickness h lying on a
% compressible water. Water depth is H=constant. The load
% is assumed to be a finite block that moves along
% x-axis with velocity c0.
%%%%%%%%%%%%%%%%%%%%%%%%%%%%%%%%%%%%%%%%%%%%%%%%%%%
% Author: Denys Dutykh

tic

sprintf ('Initialisation....')
%%%%%%%%%%%%%%%%%%%%%%%%%%%%%%%%%%%%%%%%%%%%%%%%%%%
% CONSTANT DECLARATION
%%%%%%%%%%%%%%%%%%%%%%%%%%%%%%%%%%%%%%%%%%%%%%%%%%%
N = 65536;      % number of discretisation subintervals
delta = 0.001;  % frequency discretisation step
a = 1.0;        % a half-length of the moving block
h = 2.0*a;      % depth of ice plate
H = 5.0*a;      % water depth
nu = 0.33;      % Poisson ratio
E = 9.5e9;      % Young module
g = 9.80665;    % The standard gravitational acceleration
c0 = 15.0;      % block velocity
gamma = 1500.0; % sound velocity in the water
rhog = 926.0;   % ice density
rhow = 1027.0;  % seawater density
P0 = 17500.0;   % block load
mu = E/(2*(1+nu));               % Lame's constant
%%%%%%%%%%%%3%%%%%%%%%%%%%%%%%%%%%%%%%%%%%%%%%%%%%%%
% useful constants
%%%%%%%%%%%%%%%%%%%%%%%%%%%%%%%%%%%%%%%%%%%%%%%%%%%
chi = sqrt(1-(c0/gamma)^2);
k1 = sqrt(1-c0^2*rhog/mu);
k2 = sqrt(1-c0^2*rhog*(1-nu-2*nu^2)/(E*(1-nu)));
P = k2*k1^2;
B1 = 1;
B2 = ((1-nu)*k2^2-nu)/(1-2*nu);
A1 = k1 + 1/k1; A2 = 2*k2;

sprintf ('Linear system solving....')
c = zeros (4,2*N-1);  % unknown constants
% Fourier transforms of displacements
uf = zeros (4,2*N-1);
% discretisation in the Fourier domain
range = (-N+1)*delta:delta:(N-1)*delta;
A = zeros(4); b = zeros(4,1); j = 0;
for omega = range, j = j+1;
    if (abs(omega) < 3)
        if (abs(omega) < 40*eps)
        % right member of linear system
            b(1) = -i*(P0/E)*(a/omega);
        else
        % right member of linear system
            b(1) = -i*P0/E*sin(omega*a)/omega^2;
        end
        % calculation of the matrix
        A(1,1) = omega*B1*cosh(k1*omega*h);
        A(1,2) = omega*B1*sinh(k1*omega*h);
        A(1,3) = omega*B2*sinh(k2*omega*h);
        A(1,4) = omega*B2*cosh(k2*omega*h);

        A(2,1) = A1*sinh(k1*omega*h);
        A(2,2) = A1*cosh(k1*omega*h);
        A(2,3) = A2*cosh(k2*omega*h);
        A(2,4) = A2*sinh(k2*omega*h);

        A(3,2) = A1;
        A(3,3) = A2;

        s = (rhog*omega*c0^2/chi*cosh(chi*omega*H)+...
            rhow*g*sinh(chi*omega*H));

        A(4,1) = omega*B1*sinh(chi*omega*H);
        A(4,2) = -s/(k1*E);
        A(4,3) = -P*s/(k1^2*E);
        A(4,4) = omega*B2*sinh(chi*omega*H);
    else
        % right member of linear system
        b(1) = -i*P0*sin(omega*a)/...
        (E*omega^3*cosh(k1*omega*h));

        % calculation of the matrix
        A(1,1) = B1;
        A(1,2) = B1*tanh(k1*omega*h);
        A(1,3) = B2*(sinh(k2*omega*h)/cosh(k1*omega*h));
        A(1,4) = B2*(cosh(k2*omega*h)/cosh(k1*omega*h));

        A(2,1) = A1*tanh(k1*omega*h);
        A(2,2) = A1;
        A(2,3) = A2*(cosh(k2*omega*h)/cosh(k1*omega*h));
        A(2,4) = A2*(sinh(k2*omega*h)/cosh(k1*omega*h));

        A(3,2) = A1;
        A(3,3) = A2;

        s = (rhog*omega*c0^2/chi*coth(chi*omega*H)+...
        rhow*g);

        A(4,1) = B1;
        A(4,2) = -s/(omega*k1*E);
        A(4,3) = -P*s/(omega*k1^2*E);
        A(4,4) = B2;
    end

    % we find unknown constants
    c(:,j) = inv(A)*b;

    for k=1:4
        if (isnan(c(k,j))|abs(c(k,j))>1e-3) % 8e-1
            c(k,j) = c(k,j-1);
        end
    end

    I = diag([1,i/k1^2,1,i/k1^2]);
    aux = [1,0,0,1; 0,k1,P,0; cosh(k1*omega*h),...
      sinh(k1*omega*h),sinh(k2*omega*h),cosh(k2*omega*h);
      k1*sinh(k1*omega*h),k1*cosh(k1*omega*h),...
      P*cosh(k2*omega*h),P*sinh(k2*omega*h)];

    % Here we find displacements Fourier transform
    uf(:,j) = I*aux*c(:,j);
end

Delta = delta/(2*pi);
deltax = 1/(N*Delta); % discretisation step
% x-variable discretisation
rangex = -1/(2*Delta):deltax:1/(2*Delta);

sprintf('Fast Fourier Transform...')

% to understand this command type: doc fftw
fftw ('planner', 'patient');

un1 = uf(:,N-(0:N-1)); un2 = uf(:,N:2*N-1);

% integral calculation by DFT
u0=(Delta*(N*ifft(un2')-fft(un1')))';

u = zeros(4,N+1);

% We form the solution
u(:,1:N/2) = real (u0(:,(N/2+1):N));
u(:,(N/2+1):(N+1)) = real(u0(:,1:(N/2+1)));

% interface between water and ice
dzeta0 = [rangex+u(1,:); u(2,:)];
% interface between air and ice
dzetah = [rangex+u(3,:); u(4,:)];

toc

sprintf('Plotting....')

subplot (2,1,1)
plot([dzetah(1,(127*N/256):(129*N/256))],
[dzetah(2,(127*N/256):(129*N/256))]),
grid on, title '\zeta(h)', xlabel 'x/a', ylabel '(y-h)/a';
subplot (2,1,2)
plot ([dzeta0(1,(127*N/256):(129*N/256))],
[dzeta0(2,(127*N/256):(129*N/256))]),
grid on, title '\zeta(0)', xlabel 'x/a', ylabel 'y/a';

sprintf('Done.')
\end{verbatim}

\subsection{Numerical computations for one homogeneous
layer}\label{app3}

\begin{figure}
  \includegraphics[width=\linewidth]{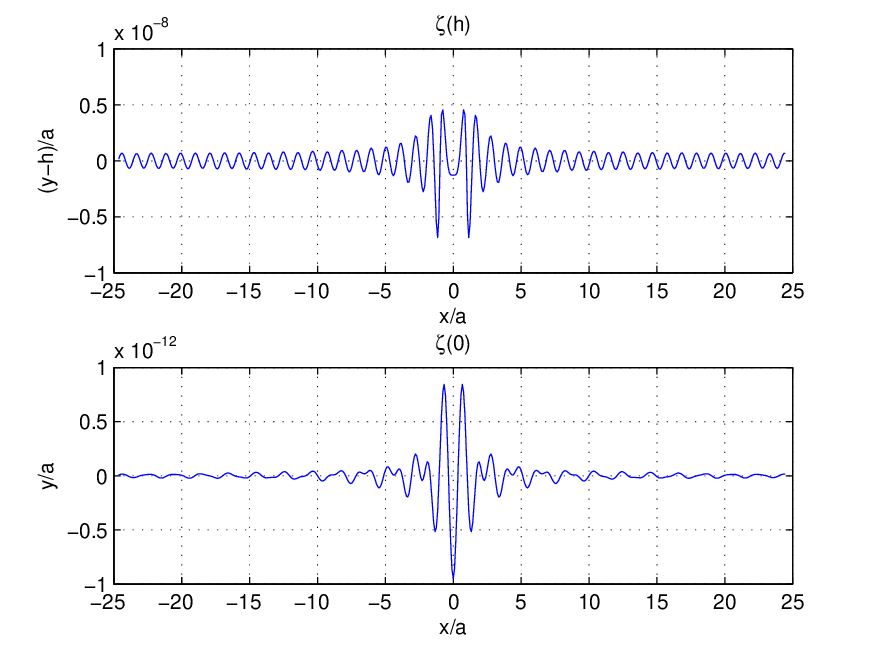}\\
  \caption{Ice deflection for $c_0=0.1m/s$}
  \label{c01:fig}
\end{figure}

\begin{figure}
  \includegraphics[width=\linewidth]{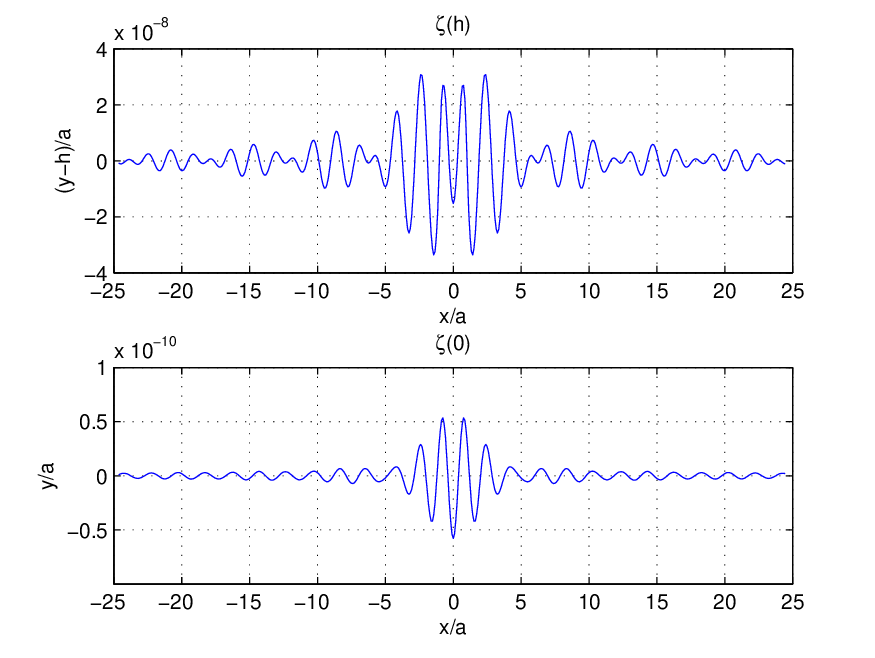}\\
  \caption{Ice deflection for $c_0=1.0m/s$}
  \label{c10:fig}
\end{figure}

\begin{figure}
  \includegraphics[width=\linewidth]{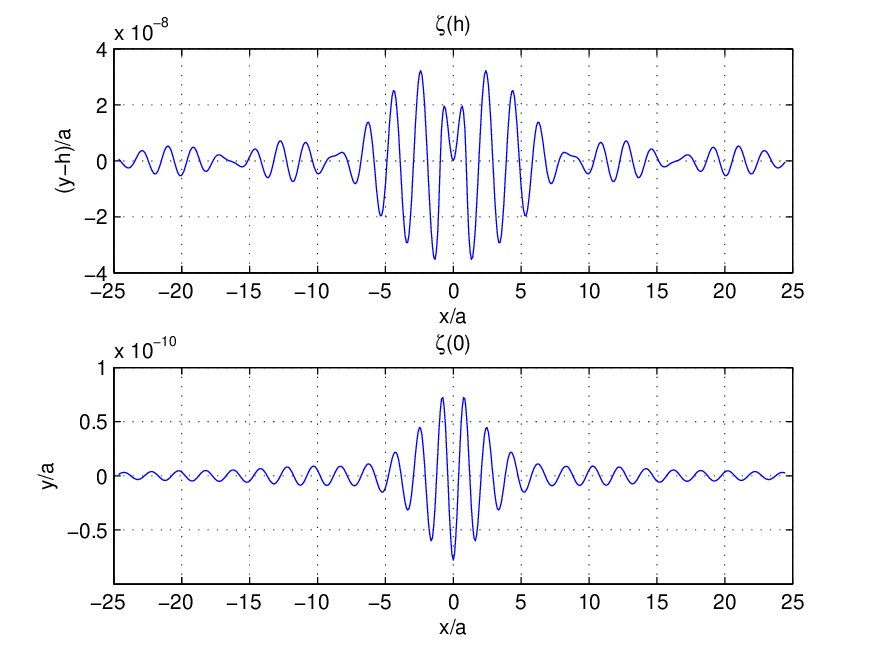}\\
  \caption{Ice deflection for $c_0=1.3m/s$}
  \label{c13:fig}
\end{figure}

\begin{figure}
  \includegraphics[width=\linewidth]{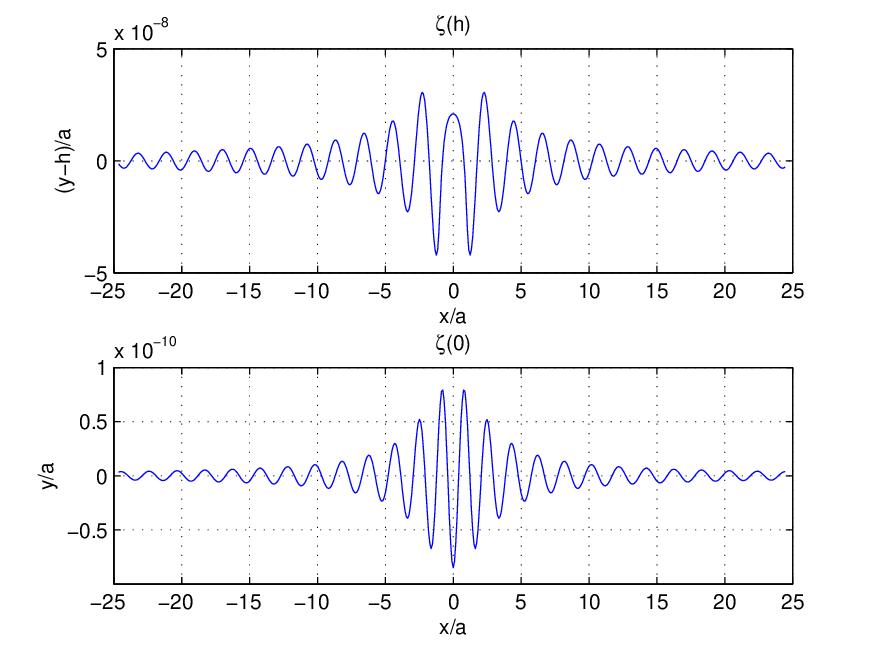}\\
  \caption{Ice deflection for $c_0=1.5m/s$}
  \label{c15:fig}
\end{figure}

\begin{figure}
  \includegraphics[width=\linewidth]{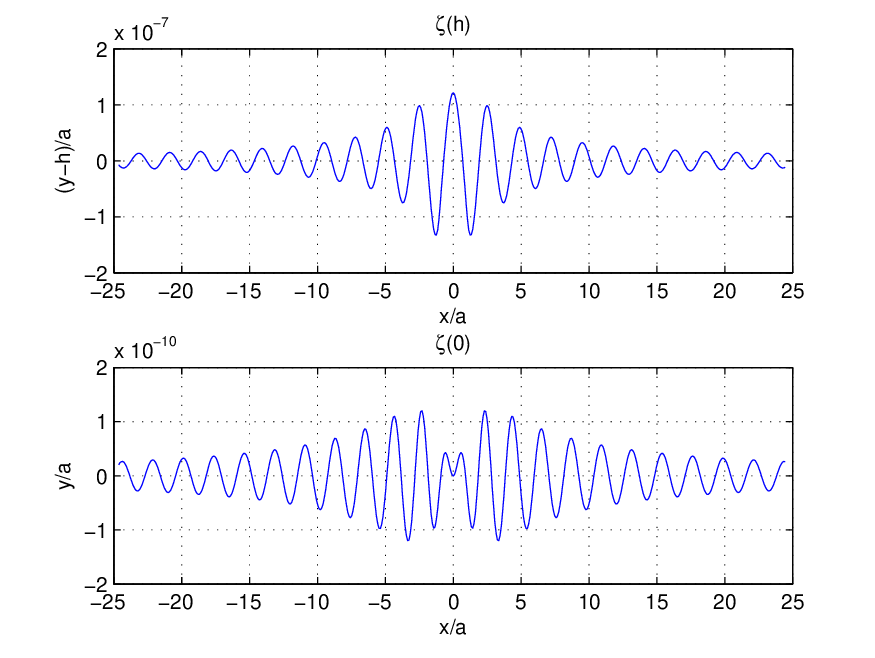}\\
  \caption{Ice deflection for $c_0=4.0m/s$}
  \label{c40:fig}
\end{figure}

\begin{figure}
  \includegraphics[width=\linewidth]{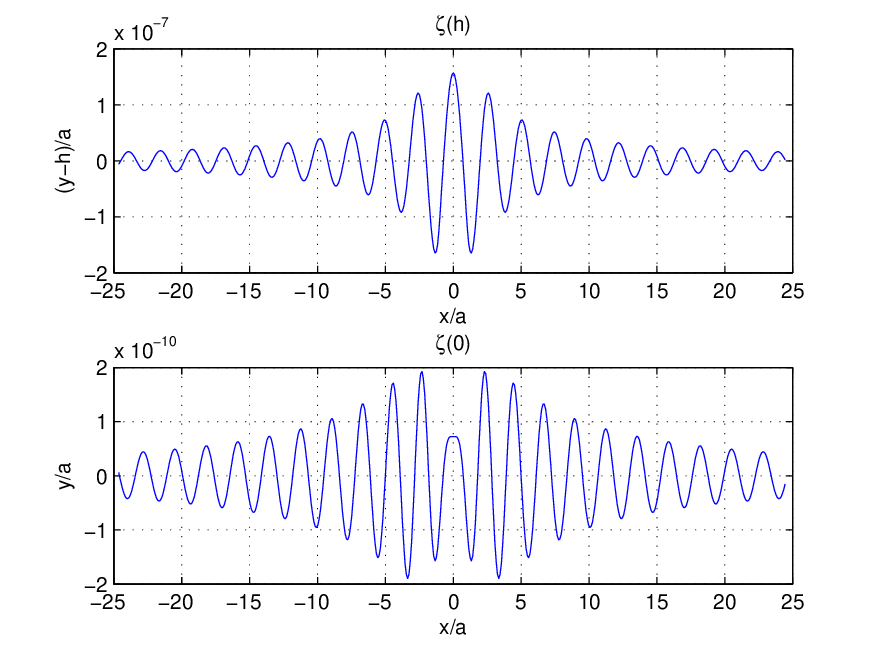}\\
  \caption{Ice deflection for $c_0=5.0m/s$}
  \label{c50:fig}
\end{figure}

\begin{figure}
  \includegraphics[width=\linewidth]{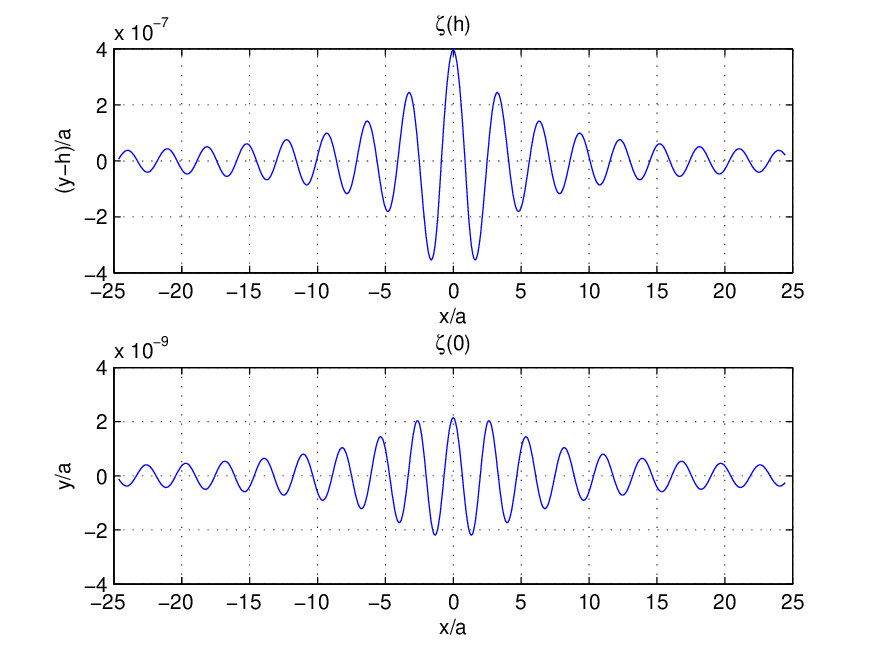}\\
  \caption{Ice deflection for $c_0=15.0m/s$}
  \label{c150:fig}
\end{figure}

\begin{figure}
  \includegraphics[width=\linewidth]{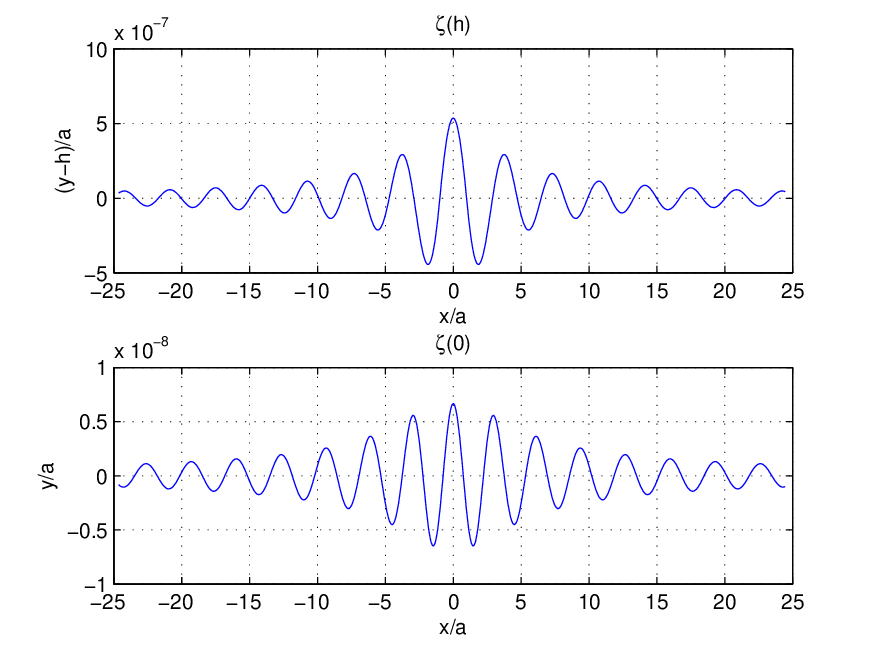}\\
  \caption{Ice deflection for $c_0=25.0m/s$}
  \label{c250:fig}
\end{figure}

In this Appendix, you can find the results of numerical computations that were performed with MatLab-code from Appendix \ref{app2}. We tried to show the different types of ice deflection for different values of the moving block velocity. The results are given in Figures \ref{c01:fig} -- \ref{c250:fig}. These graphics represent the function $\zeta(x,h)$ and $\zeta(x,0)$ of ice-air and ice-water interfaces, respectively.

\subsection{Numerical solution of the contact problem}\label{app4}

\begin{figure}
  \includegraphics[width=\linewidth]{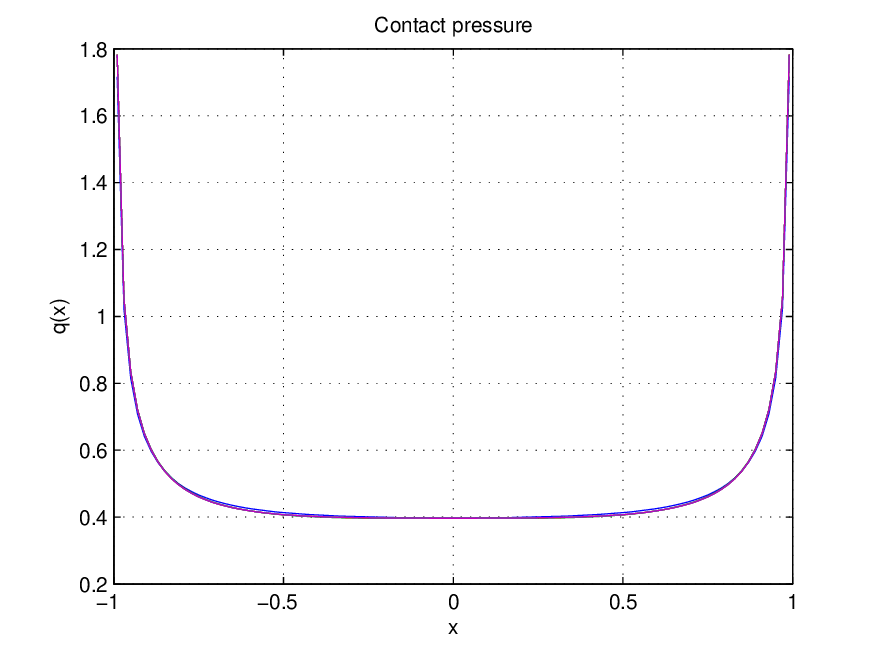}\\
  \caption{Contact pressure under the moving block for $c_0=15.0m/s$}
  \label{Cont15:fig}
\end{figure}

In this Appendix, we cite the listings of two MatLab functions that we use to solve the contact problem. In Figure \ref{Cont15:fig}, the reader can see a typical profile of the contact pressure under the rigid block. On the same image, we plotted the numerical solutions with $3$, $5$, $7$, $9$ and $21$ collocation points. The relative error of the numerical solution with three collocation points is about $1\%$, with five points about $0.01\%$.

\begin{verbatim}
function assemble
% This code performs the linear system assembling, which
% is solved later by another code contact.m
% -----------------------------------------------------
% Author: Denys Dutykh, CMLA, ENS de Cachan

sprintf ('Initialisation...')

h = 1.0;     % ice depth
kappa = 1.0; % kappa := a/h
T0 = -4.0;   % water temperature
T1 = -12.0;  % air temperature
s = 33;      % water salinity
c0 = 15.0;    % block velocity

% This number is the limit B(omega) when omega -> +inf.
% We use a Maple code to determine it.
Binf = -1.114861886;

M = 5;
t = cos((2*(1:M)-1)*pi/(2*M));
x = cos(pi*(1:M-1)/M);

sprintf ('Matrix assembling...')
tic
A = zeros (M);
for r=1:M-1
    for m=1:M
        A(r,m) = (1/(t(m)-x(r)) +...
            kappa*K(x(r)-t(m),Binf,kappa,h,s,T0,T1,c0))/M;
    end
end A(M,:) = pi/M; toc

sprintf ('Writing to file...') save a5c3_2 A

sprintf ('Done.')

function rho = rho (y,h,T0,T1) % ice density
  rho = 916.5 - 0.14*(T0*(1-y/h) + T1*y/h);

% --------------------------------------------

function nu = nu(y) % Poisson ratio
  nu = 0.33;

% --------------------------------------------

function E = E(y,h,s,T0,T1) % Young's modulus
  T = abs(T0*(1-y/h) + T1*y/h);
  if (T < 0.5) | (T > 22.9)
      error ('Function E: T is out of range');
      exit;
  end
  if (T <= 2.06)
      nub = s/1000*(52.56/T - 2.28);
  elseif (T <= 8.2)
      nub = s/1000*(45.917/T + 0.93);
  else
      nub = s/1000*(43.795/T + 1.189);
  end
  E = (10 - 3.5*nub)*1e9;

% --------------------------------------------

function dE = dE(y,h,s,T0,T1)
  dE = (log(E(y+0.00001,h,s,T0,T1)) -...
      log(E(y-0.00001,h,s,T0,T1)))/...
      0.00002;

% --------------------------------------------

function k1 = k1(y,h,s,T0,T1,c0)
  k1 = sqrt(1 - 2*c0^2*rho(y,h,T0,T1)*(1+nu(y))/...
      E(y,h,s,T0,T1));

% --------------------------------------------

function dkE = dkE(y,h,s,T0,T1,c0)
  dkE = (log(k1(y+0.00001,h,s,T0,T1,c0)^2*...
      E(y+0.00001,h,s,T0,T1)) -...
  log(k1(y-0.00001,h,s,T0,T1,c0)^2*...
  E(y-0.00001,h,s,T0,T1)))/0.00002;

% --------------------------------------------

function k2 = k2(y,h,s,T0,T1,c0)
  k2 = sqrt(1 - c0^2*rho(y,h,T0,T1)*(1-nu(y)-2*nu(y)^2)/
      (E(y,h,s,T0,T1)*(1-nu(y))));

% --------------------------------------------

function B1 = B1(y)
  B1 = 1/(1+nu(y));

% --------------------------------------------

function B2 = B2(y,h,s,T0,T1,c0)
  B2 = ((1-nu(y))*k2(y,h,s,T0,T1,c0)^2 - nu(y))/...
      ((1+nu(y))*(1-2*nu(y)));

% --------------------------------------------

function A = A(y,h,s,T0,T1,c0)
  A = k1(y,h,s,T0,T1,c0) + 1/k1(y,h,s,T0,T1,c0);

% --------------------------------------------

function mu = mu(y,h,s,T0,T1)
  mu = E(y,h,s,T0,T1)/(2*(1+nu(y)));

% --------------------------------------------

function dmu = dmu(y,h,s,T0,T1)
  dmu = (log(mu(y+0.00001,h,s,T0,T1)) -...
      log(mu(y-0.00001,h,s,T0,T1)))/0.00002;

% --------------------------------------------

function dkmu = dkmu(y,h,s,T0,T1,c0)
  dkmu = (log(k1(y+0.00001,h,s,T0,T1,c0)*...
      mu(y+0.00001,h,s,T0,T1)) -...
  log(k1(y-0.00001,h,s,T0,T1,c0)*...
  mu(y-0.00001,h,s,T0,T1)))/0.00002;

% --------------------------------------------

% Differential equations system
function dA = lamzpr(y,L,omega,h,s,T0,T1,c0) dA =
[dE(y,h,s,T0,T1)*L(1) - omega*(...
1/k1(y,h,s,T0,T1,c0)^2*L(2) +...
(k2(y,h,s,T0,T1,c0)*(2*k2(y,h,s,T0,T1,c0) -...
A(y,h,s,T0,T1,c0)*k1(y,h,s,T0,T1,c0)))/...
(B1(y)-B2(y,h,s,T0,T1,c0))*L(3) +...
(A(y,h,s,T0,T1,c0)*B2(y,h,s,T0,T1,c0)-...
2*B1(y)*k1(y,h,s,T0,T1,c0))/(k1(y,h,s,T0,T1,c0)*...
(1-k1(y,h,s,T0,T1,c0)^2))*L(1)*L(2) +...
(A(y,h,s,T0,T1,c0)*B2(y,h,s,T0,T1,c0)*k1(y,h,s,T0,T1,c0)-
2*B1(y)*k2(y,h,s,T0,T1,c0))/...
(B1(y)-B2(y,h,s,T0,T1,c0))*L(1)*L(3))

dkE(y,h,s,T0,T1,c0)*L(2) - omega*(...
(k1(y,h,s,T0,T1,c0)^2*(1-k2(y,h,s,T0,T1,c0)^2))/...
(B1(y)-B2(y,h,s,T0,T1,c0)) +...
(k1(y,h,s,T0,T1,c0)^2*(B2(y,h,s,T0,T1,c0)-...
B1(y)*k2(y,h,s,T0,T1,c0)^2))/(B1(y)-B2(y,h,s,T0,T1,c0))
*L(1)+(A(y,h,s,T0,T1,c0)*k1(y,h,s,T0,T1,c0)-...
2*k2(y,h,s,T0,T1,c0)^2)/(B1(y)-B2(y,h,s,T0,T1,c0))*L(4)+
(A(y,h,s,T0,T1,c0)*B2(y,h,s,T0,T1,c0)*k1(y,h,s,T0,T1,c0)-
2*B1(y)*k2(y,h,s,T0,T1,c0)^2)/(B1(y)-B2(y,h,s,T0,T1,c0))
*L(1)*L(4)+(A(y,h,s,T0,T1,c0)*B2(y,h,s,T0,T1,c0)-...
2*B1(y)*k1(y,h,s,T0,T1,c0))/(k1(y,h,s,T0,T1,c0)*...
(1-k1(y,h,s,T0,T1,c0)^2))*L(2)^2)

dmu(y,h,s,T0,T1)*L(3) - omega*(...
1 + (B2(y,h,s,T0,T1,c0)-B1(y)*k1(y,h,s,T0,T1,c0)^2)/...
(1-k1(y,h,s,T0,T1,c0)^2)*L(1) + 1/k1(y,h,s,T0,T1,c0)^2
*L(4)+(A(y,h,s,T0,T1,c0)*B2(y,h,s,T0,T1,c0)-...
2*B1(y)*k1(y,h,s,T0,T1,c0))/(k1(y,h,s,T0,T1,c0)*...
(1-k1(y,h,s,T0,T1,c0)^2))*L(1)*L(4) +...
(A(y,h,s,T0,T1,c0)*B2(y,h,s,T0,T1,c0)*k1(y,h,s,T0,T1,c0)-
2*B1(y)*k2(y,h,s,T0,T1,c0)^2)/(B1(y)-...
B2(y,h,s,T0,T1,c0))*L(3)^2)

dkmu(y,h,s,T0,T1,c0)*L(4) - omega*(...
(B2(y,h,s,T0,T1,c0)-B1(y)*...
k1(y,h,s,T0,T1,c0)^2)/(1-k1(y,h,s,T0,T1,c0)^2)*L(2) +...
(k1(y,h,s,T0,T1,c0)^2*(B2(y,h,s,T0,T1,c0)-...
B1(y)*k2(y,h,s,T0,T1,c0)^2))/(B1(y)-B2(y,h,s,T0,T1,c0))
*L(3)+(A(y,h,s,T0,T1,c0)*B2(y,h,s,T0,T1,c0)-...
2*B1(y)*k1(y,h,s,T0,T1,c0))/(k1(y,h,s,T0,T1,c0)*...
(1-k1(y,h,s,T0,T1,c0)^2))*L(2)*L(4) +...
(A(y,h,s,T0,T1,c0)*B2(y,h,s,T0,T1,c0)*k1(y,h,s,T0,T1,c0)-
2*k2(y,h,s,T0,T1,c0)^2*B1(y))/...
(B1(y)-B2(y,h,s,T0,T1,c0))*L(3)*L(4))];

% --------------------------------------------

function B = B(omega,h,s,T0,T1,c0)
  [y A] = ode23t(@lamzpr,[0 h],[0 0 0 0],[],omega,...
      h,s,T0,T1,c0);

  [n m] = size(A);
  B = A(n,2);

% --------------------------------------------

function K = K(z,Binf,kappa,h,s,T0,T1,c0)
  aux = 0; N = 5; A = 10; hh = A/N;
  t1 = 0.42264973;
  t2 = 1.577350269;
  omi = 0:hh:A-hh;
  for omega=omi
      o1 = omega + 0.5*hh*t1;
      o2 = omega + 0.5*hh*t2;
      aux = aux + (1-B(o1,h,s,T0,T1,c0)/Binf)*...
          sin(kappa*z*o1) +(1-B(o2,h,s,T0,T1,c0)/Binf)*...
          sin(kappa*z*o2);
  end
  K = 0.5*hh*aux;
\end{verbatim}

\begin{verbatim}
function contact
% This code solves the contact problem for an inhomogeneous layer
% with block moving on it with constant velocity c0.
% The layer is called inhomogeneous because of its mechanical
% properties depend upon the depth.
% -------------------------------------------------------
% Author: Denys Dutykh, CMLA, ENS de Cachan

load a5c1_2 A;

[M M] = size (A);
b = zeros (M,1);
b(M) = 1;

u3 = A\b;
x3 = cos((2*(1:M)'-1)*pi/(2*M));

xx = -0.99:0.02:0.99;
uu3 = spline (x3,u3,xx);
q3 = uu3./sqrt(1-xx.^2);

plot (xx, q3)
\end{verbatim}

\section{Maple code to determine the asymptotic behaviour of Lamzyuk-Privarnikov functions}\label{app5}

\pagestyle{empty}
\begin{maplegroup}
\begin{mapleinput}
\mapleinline{active}{1d}{restart:}{%
}
\end{mapleinput}
\end{maplegroup}

\begin{maplegroup}
\mapleresult
\begin{maplelatex}
\mapleinline{inert}{2d}{h := 1.0;}{%
\[
h := 1.0
\]
}
\end{maplelatex}

\begin{maplelatex}
\mapleinline{inert}{2d}{T0 := -4;}{%
\[
\mathit{T0} := -4
\]
}
\end{maplelatex}

\begin{maplelatex}
\mapleinline{inert}{2d}{T1 := -12;}{%
\[
\mathit{T1} := -12
\]
}
\end{maplelatex}

\begin{maplelatex}
\mapleinline{inert}{2d}{s := 33;}{%
\[
s := 33
\]
}
\end{maplelatex}

\begin{maplelatex}
\mapleinline{inert}{2d}{c0 := 15.0;}{%
\[
\mathit{c0} := 15.0
\]
}
\end{maplelatex}

\end{maplegroup}
\begin{maplegroup}
\mapleresult
\begin{maplelatex}
\mapleinline{inert}{2d}{T := proc (y) options operator, arrow; T0*(1-y/h)+T1*y/h end proc;}{%
\[
T := y\rightarrow \mathit{T0}\,(1 - {\displaystyle \frac {y}{h}} ) +
{\displaystyle \frac {\mathit{T1}\,y}{h}}
\]
}
\end{maplelatex}

\end{maplegroup}
\begin{maplegroup}
\mapleresult
\begin{maplelatex}
\mapleinline{inert}{2d}{rho := proc (y) options operator, arrow; -.14*T(y)+916.5 end proc;}{%
\[
\rho  := y\rightarrow  - 0.14\,\mathrm{T}(y) + 916.5
\]
}
\end{maplelatex}

\end{maplegroup}
\begin{maplegroup}
\mapleresult
\begin{maplelatex}
\mapleinline{inert}{2d}{nu := proc (y) options operator, arrow; .33 end proc;}{%
\[
\nu  := y\rightarrow 0.33
\]
}
\end{maplelatex}

\end{maplegroup}
\begin{maplegroup}
\mapleresult
\begin{maplelatex}
\mapleinline{inert}{2d}{nub := proc (y) options operator, arrow;
piecewise(T(y) < -.5 and -2.06 <=
T(y),1/1000*s*(52.56/abs(T(y))-2.28),T(y) < -2.06 and -8.2 <=
T(y),1/1000*s*(45.917/abs(T(y))+.93),T(y) < -8.2 and -22.9 <=
T(y),1/1000*s*(43.795/abs(T(y))+1.189)) end proc;}{%
\maplemultiline{ \mathit{nub} := y\rightarrow
\mathrm{piecewise}(\mathrm{T}(y) <
 - 0.5\ \textbf{and}\  - 2.06\leq \mathrm{T}(y),  \\
{\displaystyle \frac {1}{1000}} \,s\,({\displaystyle \frac {52.56 }{
\left|  \! \,\mathrm{T}(y)\, \!  \right| }}  - 2.28), \,
\mathrm{T}(y) <  - 2.06\ \textbf{and}\  - 8.2\leq \mathrm{T}(y),
 \\
{\displaystyle \frac {1}{1000}} \,s\,({\displaystyle \frac {
45.917}{ \left|  \! \,\mathrm{T}(y)\, \!  \right| }}  + 0.93), \,
\mathrm{T}(y) <  - 8.2\ \textbf{and}\  - 22.9\leq \mathrm{T}(y),
 \\
{\displaystyle \frac {1}{1000}} \,s\,({\displaystyle \frac {
43.795}{ \left|  \! \,\mathrm{T}(y)\, \!  \right| }}  + 1.189))
 }
}
\end{maplelatex}

\end{maplegroup}
\begin{maplegroup}
\mapleresult
\begin{maplelatex}
\mapleinline{inert}{2d}{E := proc (y) options operator, arrow;
10000000000-1000000000*3.5*nub(y) end proc;}{%
\[
E := y\rightarrow 10000000000 - 1000000000\,(3.5)\,\mathrm{nub}(y )
\]
}
\end{maplelatex}

\end{maplegroup}
\begin{maplegroup}
\mapleresult
\begin{maplelatex}
\mapleinline{inert}{2d}{B1 := proc (y) options operator, arrow; 1/(1+nu(y)) end proc;}{%
\[
\mathit{B1} := y\rightarrow {\displaystyle \frac {1}{1 + \nu (y)} }
\]
}
\end{maplelatex}

\end{maplegroup}
\begin{maplegroup}
\mapleresult
\begin{maplelatex}
\mapleinline{inert}{2d}{k1 := proc (y) options operator, arrow;
sqrt(1-2*c0^2*rho(y)*(1+nu(y))/E(y)) end proc;}{%
\[
\mathit{k1} := y\rightarrow \sqrt{1 - {\displaystyle \frac {2\,
\mathit{c0}^{2}\,\rho (y)\,(1 + \nu (y))}{\mathrm{E}(y)}} }
\]
}
\end{maplelatex}

\end{maplegroup}
\begin{maplegroup}
\mapleresult
\begin{maplelatex}
\mapleinline{inert}{2d}{k2 := proc (y) options operator, arrow;
sqrt(1-c0^2*rho(y)*(1-nu(y)-2*nu(y)^2)/E(y)/(1-nu(y))) end proc;}{%
\[
\mathit{k2} := y\rightarrow \sqrt{1 - {\displaystyle \frac {
\mathit{c0}^{2}\,\rho (y)\,(1 - \nu (y) - 2\,\nu (y)^{2})}{
\mathrm{E}(y)\,(1 - \nu (y))}} }
\]
}
\end{maplelatex}

\end{maplegroup}
\begin{maplegroup}
\mapleresult
\begin{maplelatex}
\mapleinline{inert}{2d}{B2 := proc (y) options operator, arrow;
((1-nu(y))*k2(y)^2-nu(y))/(1+nu(y))/(1-2*nu(y)) end proc;}{%
\[
\mathit{B2} := y\rightarrow {\displaystyle \frac {(1 - \nu (y))\,
\mathrm{k2}(y)^{2} - \nu (y)}{(1 + \nu (y))\,(1 - 2\,\nu (y))}}
\]
}
\end{maplelatex}

\end{maplegroup}
\begin{maplegroup}
\mapleresult
\begin{maplelatex}
\mapleinline{inert}{2d}{A := proc (y) options operator, arrow; k1(y)+1/k1(y) end proc;}{%
\[
A := y\rightarrow \mathrm{k1}(y) + {\displaystyle \frac {1}{
\mathrm{k1}(y)}}
\]
}
\end{maplelatex}

\end{maplegroup}
\begin{maplegroup}
\mapleresult
\begin{maplelatex}
\mapleinline{inert}{2d}{eq1 :=
1.000058209*B0+1.985059695*C0+.3436171886e-4*A0*B0-3.492445777*A0*C0
=
0;}{%
\maplemultiline{ \mathit{eq1} := 1.000058209\,\mathit{B0} +
1.985059695\,\mathit{
C0} + 0.00003436171886\,\mathit{A0}\,\mathit{B0} \\
\mbox{} - 3.492445777\,\mathit{A0}\,\mathit{C0}=0 }
}
\end{maplelatex}

\end{maplegroup}
\begin{maplegroup}
\mapleresult
\begin{maplelatex}
\mapleinline{inert}{2d}{eq2 :=
.6748914792-.4925039384*A0-1.310175580*D0-2.984992094*A0*D0+.343617188
6e-4*B0^2 = 0;}{%
\maplemultiline{ \mathit{eq2} := 0.6748914792 -
0.4925039384\,\mathit{A0} -
1.310175580\,\mathrm{D0} \\
\mbox{} - 2.984992094\,\mathit{A0}\,\mathrm{D0} +
0.00003436171886\,\mathit{B0}^{2}=0 }
}
\end{maplelatex}

\end{maplegroup}
\begin{maplegroup}
\mapleresult
\begin{maplelatex}
\mapleinline{inert}{2d}{eq3 :=
1+.3759423427*A0+1.000058209*D0+.3436171886e-4*A0*D0-2.984992094*C0^2
= 0;}{%
\maplemultiline{ \mathit{eq3} := 1 + 0.3759423427\,\mathit{A0} +
1.000058209\,
\mathrm{D0} \\
\mbox{} + 0.00003436171886\,\mathit{A0}\,\mathrm{D0} -
2.984992094\,\mathit{C0}^{2}=0 }
}
\end{maplelatex}

\end{maplegroup}
\begin{maplegroup}
\mapleresult
\begin{maplelatex}
\mapleinline{inert}{2d}{eq4 :=
.3759423427*B0-.4925039384*C0+.3436171886e-4*B0*D0-2.984992094*D0*C0
=
0;}{%
\maplemultiline{ \mathit{eq4} := 0.3759423427\,\mathit{B0} -
0.4925039384\,
\mathit{C0} + 0.00003436171886\,\mathit{B0}\,\mathrm{D0} \\
\mbox{} - 2.984992094\,\mathrm{D0}\,\mathit{C0}=0 }
}
\end{maplelatex}

\end{maplegroup}
\begin{maplegroup}
\mapleresult
\begin{maplelatex}
\mapleinline{inert}{2d}{\{C0 = 0., A0 =
-1.330029533+.3682545932e-2*I, D0 = -.4999793286-.1384344364e-2*I,
B0 = 0.\}, \{C0 = 0., A0 = -1.330029533-.3682545932e-2*I, D0 =
-.4999793286+.1384344364e-2*I, B0 = 0.\}, \{C0 = -146.8271028, D0 =
-32561.86908, A0 = -130446.4524, B0 = -19208927.65\}, \{C0 =
146.8271028, B0 = 19208927.65, D0 = -32561.86908, A0 =
-130446.4524\}, \{D0 = -.8111804628, A0 = -.9008595176, C0 =
-.2240739028*I, B0 = 1.149750727*I\}, \{D0 = -.8111804628, A0 =
-.9008595176, C0 = .2240739028*I, B0 = -1.149750727*I\}, \{B0 =
-1.114861886, D0 = .3818699670e-1, A0 = 1.030356050, C0 =
-.6910649695\}, \{C0 = .6910649695, D0 = .3818699670e-1, A0 =
1.030356050, B0 = 1.114861886\}, \{D0 = 21623.35164, C0 =
-146.8443305, B0 = -8470576.270, A0 = 38196.73964\}, \{C0 =
146.8443305, D0 = 21623.35164, B0 = 8470576.270, A0 =
38196.73964\};}{%
\maplemultiline{ \{\mathit{C0}=0., \,\mathit{A0}=-1.330029533 +
0.003682545932\,I
,  \\
\mathrm{D0}=-0.4999793286 - 0.001384344364\,I, \,\mathit{B0}=0.\}
, \{\mathit{C0}=0.,  \\
\mathit{A0}=-1.330029533 - 0.003682545932\,I,  \\
\mathrm{D0}=-0.4999793286 + 0.001384344364\,I, \,\mathit{B0}=0.\}
, \{ \\
\mathit{C0}=-146.8271028, \,\mathrm{D0}=-32561.86908, \,\mathit{
A0}=-130446.4524,  \\
\mathit{B0}=-0.1920892765\,10^{8}\}, \{\mathit{C0}=146.8271028,
 \\
\mathit{B0}=0.1920892765\,10^{8}, \,\mathrm{D0}=-32561.86908, \,
\mathit{A0}=-130446.4524\},  \\
\{\mathrm{D0}=-0.8111804628, \,\mathit{A0}=-0.9008595176, \,
\mathit{C0}=-0.2240739028\,I,  \\
\mathit{B0}=1.149750727\,I\}, \{\mathrm{D0}=-0.8111804628, \,
\mathit{A0}=-0.9008595176,  \\
\mathit{C0}=0.2240739028\,I, \,\mathit{B0}=-1.149750727\,I\}, \{
\mathit{B0}=-1.114861886,  \\
\mathrm{D0}=0.03818699670, \,\mathit{A0}=1.030356050, \,\mathit{
C0}=-0.6910649695\}, \{ \\
\mathit{C0}=0.6910649695, \,\mathrm{D0}=0.03818699670, \,\mathit{
A0}=1.030356050,  \\
\mathit{B0}=1.114861886\}, \{\mathrm{D0}=21623.35164, \,\mathit{
C0}=-146.8443305,  \\
\mathit{B0}=-0.8470576270\,10^{7}, \,\mathit{A0}=38196.73964\},
\{\mathit{C0}=146.8443305,  \\
\mathrm{D0}=21623.35164, \,\mathit{B0}=0.8470576270\,10^{7}, \,
\mathit{A0}=38196.73964\} }
}
\end{maplelatex}
\end{maplegroup}

\pagebreak

\bibliographystyle{unsrt}
\bibliography{elast}
\end{document}